\RequirePackage{fix-cm}

\let\oldvec\vec
\documentclass[smallextended]{svjour3}       
\let\vec\oldvec

\smartqed  

\usepackage{breakcites}
\usepackage{booktabs} 
\usepackage{amsmath}
\usepackage{amsfonts}
\usepackage{graphicx}
\usepackage{color}
\usepackage{xspace}
\usepackage{pdfpages}
\usepackage{ifdraft}
\usepackage{xspace}
\usepackage{relsize}  
\usepackage{hyperref} 
\usepackage{listings} 
\usepackage{array}              
\usepackage{pifont}             
\usepackage{MnSymbol}
\usepackage{wasysym}
\usepackage{enumitem}

\usepackage{tikz}

\usepackage{supertabular}
\usepackage[framemethod=tikz]{mdframed}
\usetikzlibrary{patterns}

\usepackage[english]{babel}
\usepackage{natbib}

\usepackage{multirow}
\usepackage{multicol}

\usepackage{mdframed}
\usepackage{xcolor}

\usepackage{boxedminipage}

\usepackage{siunitx}
\usepackage{siunitx}

\newmdenv[innerlinewidth=0.5pt, roundcorner=4pt,linecolor=gray,innerleftmargin=4pt,innerrightmargin=4pt,innertopmargin=4pt,innerbottommargin=4pt]{note1}

\newenvironment{result}%
{\medskip\begin{note1}\centering\em}%
{\end{note1}\medskip}

\definecolor{Grey}{rgb}{0.5,0.5,0.5}
\definecolor{LightGrey}{rgb}{0.9,0.9,0.9}
\definecolor{Green}{rgb}{0.0,0.6,0.0}
\definecolor{Red}{rgb}{0.6,0.0,0.0}
\definecolor{Blue}{rgb}{0.0,0.0,0.6} 

\newcommand{\PASS}{\text{\color{Green}\ding{52}}\xspace}
\newcommand{\FAIL}{\text{\color{Red}\ding{56}}\xspace}

\newmdenv[innerlinewidth=0.5pt, roundcorner=4pt,linecolor=gray,innerleftmargin=0pt,
innerrightmargin=1pt,innertopmargin=2pt,innerbottommargin=0pt]{quotes}
            
\newif\ifextended
\extendedfalse

\newif\ifoldresults
\oldresultsfalse

\newcommand{\todo}[1]{}%
\renewcommand{\todo}[1]{{\color{red} TODO: {#1}}}%
\newcommand{\Coreb}{CoREBench\xspace}
\newcommand{\adslice}{approximate dynamic slice\xspace}
\newcommand{\adslicing}{approximate dynamic slicing\xspace}

\newif\ifblinded
\blindedfalse

\begin{document}

\title{
Locating Faults with Program~Slicing: An~Empirical~Analysis
}


\author{Ezekiel Soremekun \and Lukas Kirschner 
	\and Marcel B{\"o}hme   \and Andreas Zeller 
}


\institute{Ezekiel Soremekun \at
				SnT, University of Luxembourg, Luxembourg \\
              \email{ezekiel.soremekun@uni.lu} 
           \and
           Lukas Kirschner \at
              CISPA – Helmholtz Center for Information Security, Saarbr{\"u}cken, Germany \\
               \email{s8lukirs@stud.uni-saarland.de} 
           \and
           Marcel B{\"o}hme \at
              Monash University, Melbourne, Australia \\
               \email{marcel.boehme@acm.org}  
           \and
           Andreas Zeller \at
               CISPA – Helmholtz Center for Information Security, Saarbr{\"u}cken, Germany \\
               \email{zeller@cispa.saarland} 
}

\date{Received: date / Accepted: date}

\maketitle

\begin{abstract}
Statistical fault localization is an easily deployed technique for quickly determining candidates for faulty code locations.  If a human programmer has to search the fault beyond the top candidate locations, though, more traditional techniques of following dependencies along dynamic slices may be better suited. In a large study of 
457 bugs (369 single faults and 88 multiple faults) in 46 open source C programs, 
we compare the effectiveness of statistical fault localization against dynamic slicing. For single faults, we find that dynamic slicing was eight percentage points more effective than the best performing statistical debugging formula; for 66\% of the bugs, dynamic slicing finds the fault earlier than the best performing statistical debugging formula. In our evaluation, dynamic slicing is more effective for programs with single fault, but statistical debugging performs better on multiple faults. Best results, however, are obtained by a \emph{hybrid approach}: If programmers first examine at most the top five most suspicious locations from statistical debugging, and then switch to dynamic slices, on average, they will need to examine 15\% (30 lines) of the code. These findings hold for 18~most effective statistical debugging formulas and our results are independent of the number of faults (i.e. single or multiple faults) and error type (i.e. artificial or real errors).
\end{abstract}

\section{Introduction}
\label{intro}
In the past 20~years, the field of \emph{automated fault localization} (AFL) has found considerable interest among researchers in Software Engineering.  Given a program failure, the aim of fault localization is to suggest locations in the program code where a fault in the code causes the failure at hand.  Locating a fault is an obvious prerequisite for removing and fixing it; and thus, \emph{automated} fault localization brings the promise of supporting programmers during arduous debugging tasks.  Fault localization is also an important prerequisite for \emph{automated program repair,} where the identified fault locations serve as candidates for applying the computer-generated patches \citep{genprog,semfix,par,rsrepair}.

The large majority of recent publications on automated fault localization fall into the category of \emph{statistical debugging} (also called \textit{spectrum-based fault localization} (SBFL)), an approach pioneered 18~years ago~\citep{tarantula,statistical1,statistical2}. A recent survey~\citep{tse} lists more than 100~publications on statistical debugging. The core idea of statistical debugging is to take a set of passing and failing runs, and to record the program lines which are executed (``covered'') in these runs. The stronger the correlation between the execution of a line and failure (say, because the line is executed only in failing runs, and never in passing runs), the more we consider the line as ``suspicious''.

\begin{figure}[!htbp]
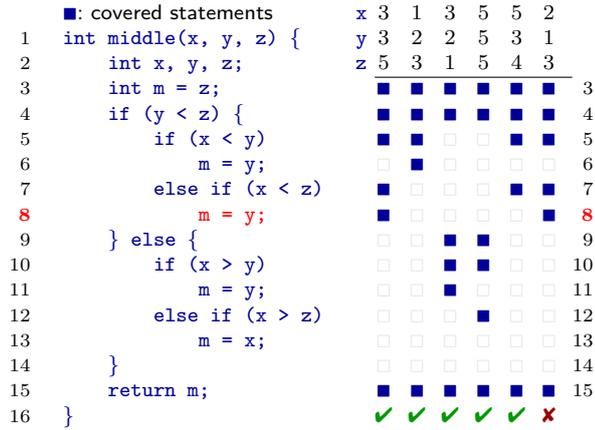
\small
\centering
\def\*{{\color{Blue}$\blacksquare$}}
\def\+{{\color{Blue}$\blacksquare$}}
\def\-{{\color{LightGrey}$\Box$}}
\def\ind{\qquad}
\begin{tabular}{@{}>{\scriptsize}r>{\tt\color{Blue}}ll@{\ }c@{\ \ }c@{\ \ }c@{\ \  }c@{\ \  }c@{\ \ }c@{\ \ }>{\scriptsize}r}
\\
& \textsf{\color{black}\*: covered statements} & \color{Blue}\texttt{x} &  3   & 1   & 3   & 5   & 5  & 2  \\
1  & int middle(x, y, z) \{                        & \color{Blue}\texttt{y} & 3   & 2   & 2   & 5   & 3  & 1  \\
2  & \ind  int x, y, z;                            & \color{Blue}\texttt{z} & 5   & 3   & 1   & 5   & 4  & 3 
 \\ \cline{4-9}
 3  & \ind  int m = z;            & & \*  & \*  & \*  & \*  & \*  & \+ & 3 \\ 
4  & \ind  if (y < z) \{         & & \*  & \*  & \*  & \*  & \*  & \+ & 4 \\ 
5  & \ind \ind  if (x < y)       & & \*  & \*  & \-  & \-  & \*  & \+ & 5 \\ 
6  & \ind \ind \ind  m = y;      & & \-  & \*  & \-  & \-  & \-  & \- & 6 \\ 
7  & \ind \ind  else if (x < z)  & & \*  & \-  & \-  & \-  & \*  & \+ & 7 \\ 
{\color{red}\textbf{8}}  & \ind \ind \ind  {\color{red}\textbf{m = y;} }     & & \*  & \-  & \-  & \-  & \-  & \+ & {\color{red}\textbf{8}} \\ 
9  & \ind  \} else \{            & & \-  & \-  & \*  & \*  & \-  & \- & 9 \\ 
10 & \ind \ind  if (x > y)       & & \-  & \-  & \*  & \*  & \-  & \- & 10 \\ 
11 & \ind \ind \ind  m = y;      & & \-  & \-  & \*  & \-  & \-  & \- & 11 \\ 
12 & \ind \ind  else if (x > z)  & & \-  & \-  & \-  & \*  & \-  & \- & 12 \\ 
13 & \ind \ind \ind  m = x;      & & \-  & \-  & \-  & \-  & \-  & \- & 13 \\ 
14 & \ind  \}                    & & \-  & \-  & \-  & \-  & \-  & \- & 14 \\ 
15 & \ind  return m;             & & \*  & \*  & \*  & \*  & \*  & \+ & 15 \\ 
 16 & \}                          & & \PASS & \PASS & \PASS & \PASS & \PASS & \FAIL \\
\end{tabular}
\caption{Statistical debugging illustrated~\citep{statEval1}: The \texttt{middle} function takes three values and returns that value which is greater than or equals the smallest and less than or equals the biggest value; however, on the input (2, 1, 3), it returns 1 rather than 2.  Statistical debugging reports the faulty Line~8 (in {\color{red}\textbf{bold red}}) as the most suspicious one, since the correlation of its execution with failure is the strongest.}
\label{fig:middle-coverage}
\end{figure}

As an example, let us have a look at the function \texttt{middle}, used in~\citep{tarantula} to introduce the technique (\emph{see \autoref{fig:middle-coverage}}). The \texttt{middle} function computes the middle of three numbers \texttt{x, y, z}; \autoref{fig:middle-coverage} shows its source code as well as statement coverage for few sample inputs. On most inputs, \texttt{middle} works as advertised; but when fed with \texttt{x} = 2, \texttt{y} = 1, and \texttt{z} = 3, it returns~1 rather than the middle value~2. Note that the statement in Line~8 is incorrect and should read \texttt{m = x}. Given the runs and the lines covered in them, statistical debugging assigns a  \emph{suspiciousness score} to each program statement---a function on the number of times it is (not) executed by passing and failing test cases. The precise function it uses differs for each statistical debugging technique. Since the statement in Line~8 is executed most often by the failing test case and least often by any passing test case, it is reported as most suspicious fault location.

\begin{figure}[!htbp]
\centering
\includegraphics*[width=0.5\linewidth,trim={0.6cm 5cm 1cm 5cm},clip]{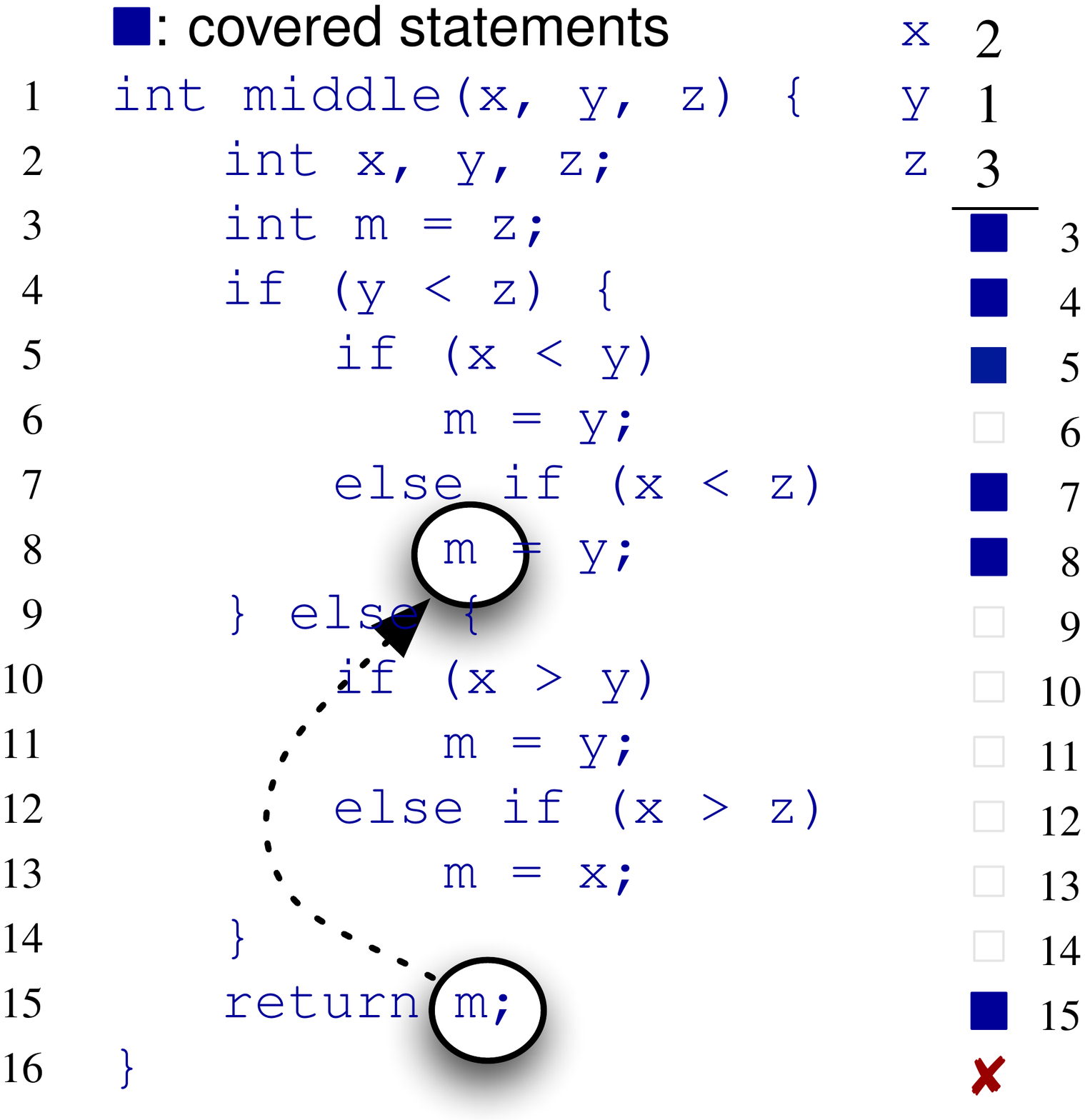}
\caption{Dynamic slicing illustrated~\citep{statEval1}: The \texttt{middle} return value in Line~15 can stem from any of the assignments to~\texttt{m}, but only those in Lines 3~and~8 are executed in the failing run.  Following back the dynamic dependency immediately gets the programmer to Line~8, the faulty one.}
\label{fig:middle-dependencies}
\vspace{-0.6cm}
\end{figure}

Statistical debugging, however, is not the first technique to automate fault localization.  In his seminal paper titled ``Programmers use slices when debugging''~\citep{programmerSlice}, Mark Weiser introduced the concept of a \emph{program slice} composed of data and control dependencies in the program. Weiser argued that during debugging, programmers would start from the location where the error is observed, and then proceed backwards along these dependencies to find the fault. In a debugging setting, programmers would follow \emph{dynamic} dependencies to find those lines that actually impact the location of interest in the \emph{failing run}.  In our example (\autoref{fig:middle-dependencies}), they could simply follow the dynamic dependency of Line~15 where the value of \texttt{m} is unexpected, and immediately reach the faulty assignment in Line~8. Consequently, on the \emph{example originally introduced to show the effectiveness of statistical debugging} (\autoref{fig:middle-coverage}), the older technique of dynamic slicing is just as effective (\textit{see \autoref{fig:middle-dependencies}}). 

Thus, we investigate the fault localization effectiveness of \textit{the most effective} statistical debugging formulas against dynamic program slicing. A few researchers have empirically evaluated the fault localization effectiveness of different slicing algorithms~\citep{zhang2007study,zhang2005experimental}. However, they did not compare the effectiveness of slicing to that of statistical debugging. To the best of our knowledge, this is the \emph{first empirical study} to evaluate the fault localization effectiveness of program slicing versus (one of) the most effective statistical debugging formulas. This is also one of the \emph{largest empirical studies} of fault localization techniques, evaluating hundreds of faults (707) in C programs.

In this paper, we use four benchmarks with 35 tools, 46 programs and 457 bugs to compare fault localization techniques against each other. This set of bugs comprises of 295 real single faults, 74 injected single faults, and 88 injected multiple faults containing about four faults per program, on average. In total, we had 707 program faults. Our takeaway findings are as follows:

\begin{enumerate}
\item \textbf{Top ranked locations in statistical debugging can pinpoint the fault.} If one is only interested in a \emph{small set of candidate locations,} statistical debugging frequently pinpoints the faults, it correctly localizes 33\% of faults after inspecting \textit{only the single most suspicious code location}. It outperforms dynamic slicing in the top 5\% of the most suspicious locations, by localizing faults in \textit{twice as many} buggy programs as dynamic slicing. In our experiments, looking at only the top 5\% of the most suspicious code locations, statistical debugging would reveal faults for 6\% of all buggy programs, twice as many as slicing (3\% of buggy programs). This result is important for \emph{automatic program repair} (APR) techniques, as the search for possible repairs can only consider a limited set of candidate locations; also, the repair attempt is not necessarily expected to succeed.

\item \textbf{If one \emph{must} fix a (single-fault) bug, dynamic slicing is more effective.}\footnote{In our evaluation, dynamic slicing is more effective than SBFL on single faults. However, other factors such as multiple faults (\emph{see RQ7}), test generation~\citep{yang2017better}, test reduction~\citep{yu2008empirical} and program sizes may influence its effectiveness (\textit{see RQ6 and \autoref{sec:threats}}).} In our experiments, dynamic slicing is 62\% more likely to find the fault location earlier than statistical debugging, for single faults. In absolute terms, locating faults along dynamic dependencies requires programmers to examine on average 21\% of the code (40 LoC); whereas the most effective statistical debugging techniques require~26\% (51 LoC).  Not only is the average better; the effectiveness of dynamic slicing also has a much lower standard deviation and thus is more predictable.  Both features are important for \emph{human debuggers,} as they eventually must find and fix the fault: If they follow the dynamic slice from the failing output, they will find the fault quicker than if they examine locations whose execution correlates with failure. Moreover, dynamic slicing needs only the failing run, whereas statistical debugging additionally requires multiple similar passing runs. Although dynamic slicing is more effective on single faults, statistical debugging performs better on multiple faults (\textit{see RQ7}).

\item \textbf{Programmers can start with statistical debugging, but should quickly switch to dynamic slicing after a few locations.}  In our experiments, it is a \emph{hybrid} strategy that works best: First consider the top locations of statistical debugging (if applicable), and then proceed along the dynamic slice. In our experiments, \emph{the hybrid approach is significantly more effective than both slicing and statistical debugging.} For most errors (98\%), the hybrid approach localizes the fault within the top-20 most suspicious statements, in contrast, both slicing and statistical debugging will localize faults for most errors (98\%) after inspecting about five times as many statements (100 LoC). Notably, the hybrid approach is more effective than statistical debugging and dynamic slicing, regardless of the error type (real/artificial) and the number of faults (single/multiple) in a buggy program (\textit{see RQ6 and RQ7, respectively}).

\end{enumerate}


%

\noindent
The remainder of this paper is organized as follows. After introducing dynamic slicing and statistical debugging in \autoref{sec:background}, this paper makes the following key contributions: 
\begin{enumerate}
\item \autoref{sec:hybrid} presents a \textit{hybrid approach} that merges both dynamic slicing and statistical debugging into a strategy, where the developer switches to slicing after investigating a handful of the most suspicious statements reported by statistical debugging. 
\item We describe our evaluation setup (\autoref{sec:eval-setup}) and empirically evaluate 
the fault localization effectiveness of dynamic slicing, statistical debugging and our hybrid approach (\textit{RQ1 to 5 in \autoref{sec:eval-results})}. 
\item We conduct an empirical study on the effect of error type and the number of faults on the effectiveness of AFL techniques. We examine the difference between evaluating an AFL technique on \textit{real vs. artificial faults} (\textit{RQ6 in \autoref{sec:eval-results}}), as well as \textit{single vs. multiple faults} (\textit{RQ7 in \autoref{sec:eval-results}}). 
\end{enumerate}
In \autoref{sec:threats}, we discuss the limitations and threats to the validity of this work. \autoref{sec:future-work} and \autoref{sec:related-work} present future work and related work, respectively. 
Finally, \autoref{sec:conclusion} closes with conclusion and consequences.


The contributions and findings of this paper are important for debugging and repair stakeholders. 
Programmers, debugging tools and automated program repair (APR) tools 
need \textit{effective} fault localization techniques, in order to reduce the amount of time and effort spent (automatically) debugging and fixing errors. 
These findings enable APR tools, debuggers and programmers to be effective and efficient in bug diagnosis and bug fixing. 

\section{Background}
\label{sec:background}

In this section, we provide background on the two main AFL techniques evaluated in this paper, namely \emph{program slicing} and \emph{statistical debugging}. 

\subsection{Program Slicing}
\label{sec:slicing}
More than three decades ago, Mark Weiser~\citep{programmerSlice,slicing} noticed that developers localize the root cause of a failure by following chains of statements starting from where the failure is  observed. Starting from the symptomatic statement~$s$, where the error is observed, developers would identify those program locations that directly influence the variable values or execution of~$s$. 
This traversal continues transitively, until the root cause of the failure (i.e., the \emph{fault}) is found. This procedure allows developers to investigate those parts of the program involved in the infected information-flow in reversed order towards the location where the failure is first observed.

\subsection{Static Slicing}
Weiser developed \emph{program slicing}  as the first automated fault localization technique~\citep{programmerSlice}. A programmer marks the statement where the failure is observed (i.e., the failure's symptom) as \emph{slicing criterion}~$C$. To determine the potential impact of one statement onto another, the program slicer first computes the Program Dependence Graph (PDG) for the buggy program. 

The \emph{PDG} is a directed graph with nodes for each statement and an edge from a node~$s$ to a node~$s'$ if
\begin{enumerate}
  \setlength{\itemsep}{0pt}%
  \setlength{\parsep}{0pt}
  \setlength{\parskip}{0pt}
  \item statement~$s'$ is a conditional (e.g., an \texttt{if}-statement) and~$s$ is executed in a branch of~$s'$ (i.e., the values in~$s'$ \emph{control} whether or not~$s$ is executed), or
  \item statement~$s'$ defines a variable~$v$ that is used at~$s$ and~$s$ may be executed after~$s'$ without~$v$ being redefined at an intermediate location (i.e., the values in~$s'$ directly \emph{influence the value} of the variables in~$s$).  
\end{enumerate}  
The first condition elicits \emph{control dependence} while the second elicits \emph{data dependence}. The PDG essentially captures the inform\-ation-flow among all statements in the program. If there is no path from node~$n$ to node~$n'$, then the values of the variables at~$n$ have definitely no impact on the execution of~$n'$ or its variable values. 

The \emph{static program slice}~\citep{slicing,slicingSurvey} computed w.r.t.~$C$ consists of all statements that are reachable from~$C$ in the PDG. In other words, it contains all statements that potentially impact the execution and program states of the slicing criterion. Note that static slicing only removes those statements that are \emph{definitely not} involved in observing the failure at~$C$. The statements in the static slice may or may not be involved. Static program slices are often very large
~\citep{sliceSize}.

\subsection{Dynamic, Relevant, and Execution Slicing}
A \emph{dynamic program slice}~\citep{dynamicSl1,dynamicSl2} is computed for a specific failing input~$t$ and is thus much smaller than a static slice. It is able to capture all statements that are \emph{definitely} involved in computing the values that are observed at the location where the failure is observed for failing input~$t$. Specifically, the dynamic slice computed w.r.t. slicing criterion~$C$ for input~$t$ consists of all statements whose instances are reachable from~$C$ in the Dynamic Dependence Graph (DDG) for~$t$. The \emph{DDG} for~$t$ is computed similarly as the PDG, but the nodes are the statement \emph{instances} in the execution trace~$\pi(t)$. The DDG contains a separate node for each occurrence of a statement in~$\pi(t)$ with outgoing dependence edges to only those statement instances on which this statement instance depends in~$\pi(t)$~\citep{dynamicSl2}. However, an error is not only explained by the actual information-flow towards~$C$. 
It is important to also investigate statements that could have contributed towards an alternative, potentially correct information-flow.

The \emph{relevant slice}~\citep{relevantSl1,relevantSl2} computed for a failing input~$t$ subsumes the dynamic slice for~$t$ and also captures the fact that the fault may be in \emph{not} executing an alternative, correct path. It adds conditional statements (e.g., \texttt{if}-statements) that were executed by~$t$ and if evaluated differently may have contributed to a different value for the variables at~$C$. It requires computing (static) potential dependencies. In the execution trace~$\pi(t)$, a statement instance~$s$ \emph{potentially depends} on conditional statement instance~$b$ if there exists a variable~$v$ used in~$s$ such that (i)~$v$ is not defined between~$b$ and~$s$ in trace~$\pi(t)$, (ii)~there exists a path~$\sigma$ from $\varphi(s)$~to~$\varphi(b)$ in the PDG along which~$v$ is defined, where $\varphi(b)$ is the node in the PDG corresponding to the instance~$b$, and (iii) evaluating~$b$ differently may cause this untraversed path~$\sigma$ to be exercised. 
\cite{relevantProof} proved that the relevant slice w.r.t.~$C$ for~$t$ contains \emph{all} statements required to explain the value of~$C$ for~$t$.

The \emph{approximate dynamic slice}~\citep{adslicing,dynamicSl1} 
is computed w.r.t. slicing criterion~$C$ for failing input~$t$ as the set of \emph{executed} statements in the static slice w.r.t.~$C$.
The approximate dynamic slice subsumes the dynamic slice because there can be an edge from an instance~$s$ to an instance~$s'$ in the DDG for~$t$ only if there is an edge from statement~$\varphi(s)$ to statement~$\varphi(s')$ in the PDG.
The approximate dynamic slice subsumes the relevant slice because it also accounts for potential dependencies: Suppose instance~$s$ potentially depends on instance~$b$ in execution trace~$\pi(t)$. Then, by definition there exists a path~$\sigma$  from $\varphi(s)$~to~$\varphi(b)$ in the PDG along at least one control- and one data-dependence edge (via the node defining~$v$); and if $\varphi(s)$ is in the static slice, then $\varphi(b)$ is as well.
Note that the approximate dynamic slice is 
(1)~\emph{easier to compute} than dynamic slices (static analysis),
(2)~\emph{significantly smaller} than the static slice, and still
(3)~\emph{as ``complete''} as the relevant slice.
In summary, \emph{dynamic slice} $\subseteq$ \emph{relevant slice} $\subseteq$ \emph{approximate dynamic slice} $\subseteq$ \emph{static slice}.

\begin{figure}[!tp]\centering
\includegraphics[width=0.7\linewidth]{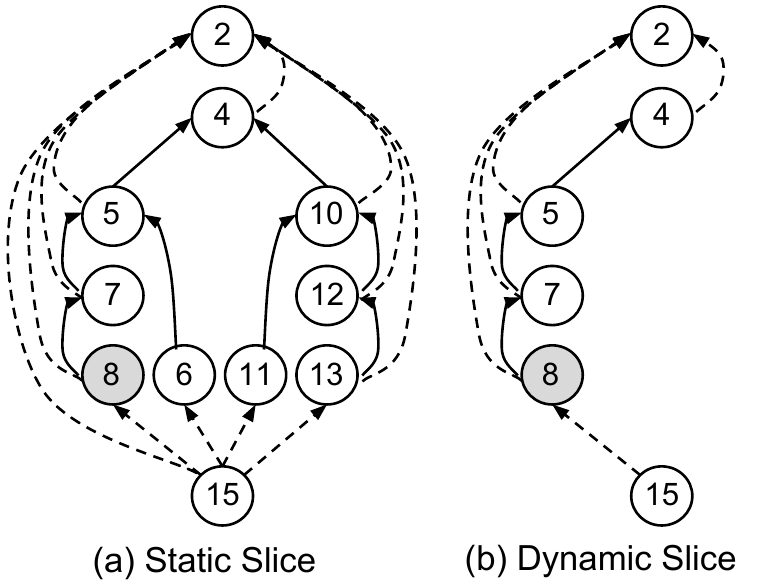}
\caption{Slicing Example: Nodes are statements in each line of the \texttt{middle} program (see \autoref{fig:middle-coverage}). Control-dependencies are shown as dashed lines while data dependencies are shown as concrete lines.}
\label{fig:middle-slicing}
\vspace{-\baselineskip}
\end{figure}

\autoref{fig:middle-slicing} (a) and (b) show the static and the dynamic slice for the \texttt{middle} program, respectively. The slicing criterion was chosen as the return statement of the program---that statement where the failure is observed. As test case, we chose the single failing test case \texttt{x} = 2, \texttt{y} = 1, and \texttt{z} = 3. 
In this example, the \adslice matches exactly the dynamic slice. For our evaluation, we implemented \adslicing, and in our evaluation results and discussions \emph{we refer to \adslicing as ``dynamic slicing"}.

\subsection{Statistical Debugging}
\label{sec:sfl}
Almost two decades ago, \cite{tarantula} introduced the first statistical debugging technique -- \textsc{Tarantula}, quickly followed by \cite{statistical1} and \cite{statistical2}.
The main idea of \emph{statistical debugging} is to associate the execution of a particular program element with the occurrence of failure using so-called \emph{suspiciousness measures}. 
Program elements (like statements, basic blocks, functions, components, etc.) that are observed more often in failed executions than in correct executions are deemed as more suspicious. A program element with a high suspiciousness score is more likely to be related to the root cause of the failure. An important property of statistical debugging is that apart from measuring coverage, it requires no specific static or dynamic program analysis.  This made it easy to implement and deploy, in particular as part of several \emph{automated program repair techniques}~\citep{genprog,semfix,par,rsrepair}, which first consider the highest ranked, most suspicious elements as patch location. Using a more effective debugging technique thus directly increases the effectiveness of such repair techniques.

\autoref{fig:middle-statistical} shows the scores computed for the executable lines in our motivating example. The statement in Line~8 is incorrect and should read \texttt{m = x;} instead. This statement is also the most suspicious according to all three statistical fault localization techniques in the example. Notice that only twelve (12) lines are actually executable. Evidently, in this example from Jones and Harrold~\citep{statEval1}, the faulty statement is also the most suspicious for these three statistical fault localization techniques.\footnote{The scores for the faulty statement in Line~8 are \texttt{tarantula}$(s_8)=\frac{1}{1}/\left(\frac{1}{1}+\frac{1}{5}\right)$, \texttt{ochiai}$(s_8)=\frac{1}{\sqrt{1(1+1)}}$, and \texttt{naish2}$(s_8)=1 - \frac{1}{1+4+1}$.} 

In this paper, we focus on four sets of measures consisting of 18 statistical fault localization formulas; namely seven human-generated optimal measures, three most popular measures, four Genetic Programming (GP) evolved measures and four  measures targeted at \emph{single bug optimality}. 

\begin{enumerate}

\item \textbf{Human-generated measures:} The first set of measures includes two \texttt{DStar ($D^*$)} formulas and five formulas which have been theoretically proven to be optimal and found to be the most effective in existing studies~\citep{xie2013provably}. These formulas include \texttt{Wong1}, \texttt{Russel\_Rao}, \texttt{Binary}, \texttt{Naish1} and \texttt{Naish2}~\citep{wong2012software,wong2013dstar, wong2007effective, russell1940habitat, naish2011model}. 
For the \texttt{DStar} algorithm, we have selected ``\textit{star}" ($^*$) values two and three (i.e., $D^*=\{D^2$,$D^3\}$) which have been demonstrated to be the most effective values for single and multiple faults, respectively~\citep{wong2012software,wong2013dstar}. 
The other five measures were selected in a theoretical evaluation of over 50 formulas and recommended as the only optimal formulas to be applied for statistical fault localization~\citep{xie2013theoretical, xie2013provably}. 

\item \textbf{Popular measures:} These measures are the most popular statistical fault localization measures, \texttt{Tarantula}, \texttt{Ochiai}, and \texttt{Jaccard}~\citep{tarantula, ochiai,jaccard,statEval12}. They have been used in recent automated program repair (APR) techniques and have been shown to improve the effectiveness of program repair~\citep{semfix, flrepair}. 

\item \textbf{Genetic Programming (GP) evolved measures:} These measures are GP-evolved formulas, which have been found to be human-competitive (comparable to human-generated measures) and theoretically maximal (i.e., the best performing measures), namely \texttt{GP02}, \texttt{GP03}, \texttt{GP13} and \texttt{GP19}~\citep{yoo2012evolving}. 

\item \textbf{Single Bug Optimal measures: } These statistical formulas are optimized for programs containing a single bug, based on the observation that if a program contains only a single bug, then all failing traces cover that bug~\citep{naish2013duals}. These measures have been empirically demonstrated to be optimal in a large-scale comparison of 157 measures. The single bug optimal measures in our study include \texttt{m9185}, \texttt{Kulczynski2} \texttt{LexOchiai} and \texttt{Pattern-Similarity}. In particular, \texttt{LexOchiai} and \texttt{Patt-\\ern-Similarity} measures perform best overall \citep{landsberg2016methods,landsberg2015evaluation}.
\end{enumerate}

\begin{figure}[!tp]\footnotesize
\def\ind{\qquad}
\begin{tabular}{@{}>{\tiny}r>{\tt\color{Blue}}ll@{\ \ }r@{\ \ }r@{\ \ }r}
1  & int middle(x, y, z) \{     & &\textbf{Tarantula} &\textbf{Ochiai} &\textbf{Naish2}  \\
2  & \ind  int x, y, z;         & &0.500  &0.408  &0.167 \\ 
3  & \ind  int m = z;           & &0.500  &0.408  &0.167 \\ 
4  & \ind  if (y < z) \{        & &0.500  &0.408  &0.167 \\ 
5  & \ind \ind  if (x < y)      & &0.625  &0.500  &0.500 \\ 
6  & \ind \ind \ind  m = y;     & &0.000  &0.000  &$-$0.167 \\ 
7  & \ind \ind  else if (x < z) & &0.714  &0.578  &0.667 \\ 
{\color{red}\textbf{8}}  & \ind \ind \ind  {\color{red}\textbf{m = y;}}     & & {\color{red}\textbf{0.833}}  & {\color{red}\textbf{0.707}}  & {\color{red}\textbf{0.833}} \\ 
9  & \ind  \} else \{           & &0.000  &0.000  &$-$0.333 \\ 
10 & \ind \ind  if (x > y)      & &0.000  &0.000  &$-$0.333 \\ 
11 & \ind \ind \ind  m = y;     & &0.000  &0.000  &$-$0.167 \\ 
12 & \ind \ind  else if (x > z) & &0.000  &0.000  &$-$0.167 \\ 
13 & \ind \ind \ind  m = x;     & &0.000  &0.000  &0.000 \\ 
14 & \ind  \}                   & &0.000  &0.000  &0.000 \\ 
15 & \ind  return m;            & &0.500  &0.408  &0.167 \\ 
16 & \}                         & & & &          \\ 
\end{tabular}
\caption{Statistical Fault Localization Example: Scores for the faulty line 8 are in {\color{red}\textbf{bold red}}} 
\label{fig:middle-statistical}
\vspace{-\baselineskip}
\end{figure}

\section{A Hybrid Approach}
\label{sec:hybrid}

Even though dynamic slicing is generally more effective than statistical debugging, we observe that statistical debugging can be highly effective for some bugs, especially when inspecting only the top most suspicious statements. For instance, statistical debugging can pinpoint a single faulty statement as the most suspicious statement for about 40\% of the errors in \texttt{IntroClass} and \texttt{SIR}, i.e. a developer can find a faulty statement after inspecting only one suspicious statement (\textit{see \autoref{fig:effectiveness-by-benchmark}}). This is further illustrated by the clustering of some points in the rightmost corner below the diagonal line of the comparison charts (\textit{see \autoref{fig:meandiff}}).

In this paper, we assume that a programmer in the end \emph{has} to fix a bug, and a viable ``alternative'' method is following the dependencies by (dynamic) slicing.  To this end, we investigate a \emph{hybrid} fault localization approach which leverages the strengths of both dynamic slicing and statistical debugging. The goal is to improve on the effectiveness of both approaches by harnessing the power of statistical correlation and dynamic program analysis. The hybrid approach first reports the \emph{top most suspicious statements} (e.g. top five statements) before it reports the statements in the dynamic slice computed w.r.t. the symptomatic statement.

The concept of examining only the top most suspicious statements is also backed by user studies on statistical fault localization.
In a recent survey~\citep{kochhar2016practitioners}, Kochhar et~al. found that three quarter of surveyed practitioners would investigate \emph{no more than the top-5 ranked statements}---which should contain the faulty statement at least three out of four times---before switching to alternative methods.  This is also confirmed by the study of Parnin and Orso~\citep{parnin}, who observed that programmers tend to transition to traditional debugging (i.e., finding those statements that impact the value of the symptomatic statement) after failing to locate the fault within the first~$N$ top-ranked most suspicious statements.  This transition is exactly what the hybrid approach provides.

Specifically, the hybrid approach proceeds in two phases. In the first phase, it reports the top $N$ (e.g. $N = 5$) most suspicious statements, obtained from the ordinal ranking\footnote{In ordinal ranking, lines with the same score are ranked by line number.} of a statistical fault localization technique. Then, if the fault is not found, it proceeds to the second phase where it reports the symptom's dynamic backward dependencies. In the second phase, we only report statements that have not already been reported in the first phase.\footnote{This is to avoid duplication of inspected statements, i.e. avoid double inspection.}

\section{Evaluation Setup}
\label{sec:eval-setup}
Let us evaluate the effectiveness of all three fault localization techniques and 
the influence of the number of faults 
and error type 
on the effectiveness of these AFL techniques. Specifically, we ask the following research questions:
\begin{itemize} [label=$\ast$] 
\item \textbf{RQ1: Effectiveness of Dynamic Slicing}: How effective is dynamic slicing in fault localization, i.e. localizing fault locations in buggy programs?

\item \textbf{RQ2: Effectiveness of Statistical Debugging}: Which statistical formula is the most effective at fault localization? 

\item \textbf{RQ3: Comparing Statistical Debugging and Dynamic Slicing}: How effective is the most effective statistical formula in comparison to dynamic slicing?

\item \textbf{RQ4: Sensitiveness of the Hybrid Approach}: How many suspicious statements (reported by statistical debugging, i.e. Kulczynski2) should a tool or developer inspect before switching to slicing?

\item \textbf{RQ5: Effectiveness of the Hybrid Approach}: Which technique is the most effective in fault localization? Which technique is more likely to find fault locations earlier?

\item \textbf{RQ6: Real Errors vs. Artificial Errors}: Does the type of error influence the effectiveness of AFL techniques? Is there a difference between evaluating an AFL technique on real or artificial errors?

\item \textbf{RQ7: Single Fault vs Multiple Faults}: What is the effect of the number of faults on the effectiveness of AFL techniques? Is there a difference between evaluating an AFL technique on single or multiple fault(s)?

\end{itemize}

In this paper, we evaluate the performance of statistical debugging, dynamic slicing and the hybrid approach in the framework of Steimann, Frenkel, and Abreu~\citep{framework} where we fix the granularity of fault localization at \emph{statement level} and the fault localization mode at \emph{one-at-a-time} (except for multiple faults in RQ7). In this setting with real errors and real test suites, the provided test suites \emph{may not be coverage adequate}, e.g. they may not execute all program statements. Fault localization effectiveness is evaluated as \emph{relative wasted effort} based on the ranking of units in the order they are suggested to be examined (see \autoref{measures} for more details). 

\subsection{Implementation}
Let us provide implementation details for each AFL technique in this paper. 

\subsubsection{Dynamic Slicing Implementation}
The \adslice is computed using \texttt{Frama-C},\footnote{http://frama-c.com/} \texttt{gcov}, \texttt{git-diff}, \texttt{gdb}, and several Python libraries. Given the preprocessed source files of a C program, \texttt{Frama-C} computes the static slices for each function and their call graphs as DOT files. The \texttt{gcov}-tool determines the executed/covered statements in the program. The \texttt{git-diff}-tool determines the changed statements in the patch and thus the faulty statements in the program. The \texttt{gdb}-tool allows to derive coverage information even for crashing inputs and to determine the slicing criterion as the last executed statement.
Our Python script intersects the statements in the static slice and the set of executed statements to derive the \adslice. 
We use the Python libraries \texttt{pygraphviz}\footnote{https://pygraphviz.github.io/}, \texttt{networkx},\footnote{https://networkx.github.io/} and \texttt{matplotlib}\footnote{http://matplotlib.org/} to process the DOT files and compute the \emph{score} for the \adslice.

\subsubsection{Statistical Debugging Implementation}
The statistical debugging tool was implemented using two bash scripts with several standard command line tools, notably \texttt{gcov},\footnote{https://gcc.gnu.org/onlinedocs/gcc/Gcov.html} \texttt{git-diff}\footnote{https://git-scm.com/docs/git-diff} and \texttt{gdb}\footnote{https://www.gnu.org/software/gdb/documentation/}. The differencing tool \texttt{git-diff} identifies those lines in the buggy program that were changed in the patch. If the patch only added statements, we cannot determine a corresponding faulty line. Some errors were thus excluded from the evaluation. The code coverage tool, \texttt{gcov} identifies those lines in the buggy program that are covered by an executed test case. When the program crashes, \texttt{gcov} does not emit any coverage information. If the crash is \emph{not} caused by an infinite loop, it is sufficient to run the program under test in \texttt{gdb} and force-call the \texttt{gcov}-function from \texttt{gdb} to write the coverage information once the segmentation fault is triggered. This was automated as well. However, for some cases, no coverage information could be generated due to infinite recursion. \texttt{Gcov} also gives the number of \emph{executable} statements in the buggy program (i.e.,~$|P|$).\footnote{The executable statements refers to statements for which coverage information are obtainable by \texttt{Gcov}, in particular, all program statements except spaces, blanks and comments.} Finally, our Python implementation of the scores is used to compute the fault localization effectiveness. 

\subsubsection{Hybrid Approach Implementation} 
The hybrid approach is implemented simply as a combination of both tools. If the top-$N$ most suspicious statements do not contain the fault, the dynamic slicing component is informed about the set of statements already inspected in the first phase. Given the unranked suspiciousness score of every executable statement in the program, the hybrid fault localizer performs an ordinal ranking of all statements. It then determines the proportion of the top~$N$ rank of suspicious statements, based on the~$N$ value of the hybrid approach. For instance, a hybrid approach with~$N =5$ takes the five topmost suspicious statements. Then, it determines the highest ranked faulty statement in the rank of all suspicious statements. If the faulty statement is in the top~$N$ suspicious positions (e.g. third position), then it  reports the number of statements in the top ranked positions up till the faulty statement, as a proportion of all executable program statements. 

In the case that the suspicious statement is not in the top~$N$ suspicious positions (e.g. seventh position), then it proceeds to slicing and reports the cardinality of the set union of all~$N$ top ranked statements and the number of inspected statements in the slice before the first faulty statement. 

\subsection{Metrics and Measures}
\emph{Odds Ratio}~$\psi$. To establish the superiority of one technique~$A$ over another technique~$B$, it is common to measure the effect size of~$A$ w.r.t.~$B$. A standard measure of effect size and widely used is the \emph{odds ratio}~\citep{oddsratio}. It ``is a measure of how many times greater the odds are that a member of a certain population will fall into a certain category than the odds are that a member of another population will fall into that category''~\citep{oddsratio}. In our case, let ``$A$ is \emph{successful}'' mean that fault localization technique~$A$ is more effective than fault localization technique~$B$ and let~$a$ be the number of successes for~$A$, $b$~the number of successes for~$B$, and $n=a+b$ the total number of successes. Then, the odds ratio~$\psi$ is calculated as  
$$
\psi = \left.\left(\cfrac{a+\rho}{n+\rho-a}\right)\right/\left(\cfrac{b+\rho}{n+\rho-b}\right)
$$
where~$\rho$ is an arbitrary positive constant (e.g.,~$\rho=0.5$) used to avoid problems with zero successes. There is no difference between the two algorithms when $\psi=1$, while $\psi>1$ indicates that technique~$A$ has higher chances of success. For example, an odds ratio of five means that fault localization technique~$A$ is five times more likely to be successful (i.e., more effective as compared to~$B$) at fault localization than~$B$.

\emph{The Mann-Whitney $U$\textit{-test}} is used to show whether there is a statistical difference between two techniques~\citep{mann1947test}. In general, it is a non-parametric test of the null hypothesis that two samples come from the same population against an alternative hypothesis, especially that a particular population tends to have larger values than the other. Unlike the $t$-test it does not require the assumption that the data is normally distributed. More specifically, it shows whether the difference in performance of two  techniques is actually statistically significant.

A \emph{cumulative frequency curve} is a running total of frequencies. We use such curves to show the percentage of errors that require examining up to a certain number of program locations. The number of code locations examined is plotted on a log-scale because the difference between examining 5 to 10 locations is more important than difference between examining 1005 to 1010 locations.

\begin{table}[!htbp]\centering
\caption{Details of Subject Programs \\ 
}
\begin{tabular}{| c | l | r | r | r | r | r |}
\hline
\textbf{Benchmark}  & \textbf{Tool} & \textbf{Avg. Size} & \textbf{\#Errors}  & \textbf{\#Fail.} & \textbf{\#Pass.} \\
(Error Type) & (Program) & (LoC) &  & \textbf{Tests} & \textbf{Tests} \\
\hline
\multirow{6}{*}{\shortstack{SIR \\(Artificial)}}  & \texttt{tcas} & 65.1 & 37 & 1356 & 58140 \\
 & \texttt{print\_tokens} & 199 & 3 & 184 & 12206 \\
 & \texttt{print\_tokens2} & 199.5 & 8 & 2031 & 30889 \\
 & \texttt{tot\_info} & 125 & 18 & 1528 & 17408 \\
 & \texttt{schedule} & 160.5 & 4 & 690 & 9910 \\
 & \texttt{schedule2} & 139.2 & 4 & 116 & 10724 \\
\hline
\multirow{6}{*}{\shortstack{IntroClass \\(Real; \\Students)}}  & \texttt{checksum} & 11.3 & 3 & 7 & 41 \\
 & \texttt{digits} & 17.4 & 3 & 16 & 32 \\
 & \texttt{grade} & 16.1 & 8 & 30 & 114 \\
 & \texttt{median} & 13.5 & 2 & 8 & 18 \\
 & \texttt{smallest} & 13.2 & 2 & 16 & 16 \\
 & \texttt{syllables} & 11.6 & 2 & 12 & 20 \\
\hline
\multirow{20}{*}{\shortstack{Codeflaws \\(Real; \\Competitions)}} & \texttt{WTLW (71A)} & 10.3 & 11 & 60 & 61 \\
 & \texttt{HQ9+ (133A)} & 10.9 & 18 &  270 & 1260 \\
 & \texttt{AG (144A)} & 24.5 & 13 & 302 & 202 \\
 & \texttt{IB (478A)} & 8.6 & 20 & 31 & 329 \\
 & \texttt{TN (535A)} & 61.2 & 9 & 118 & 778 \\
 & \texttt{Exam (534A)} & 17.5 & 12 & 108 & 68 \\
 & \texttt{Holidays (670A)} & 12.8 & 9 &  662 & 1118 \\
 & \texttt{DC (495A)} & 14.5 & 13 &  96 & 279 \\
 & \texttt{VBT(336A)} & 13.6 & 14 & 108 & 309 \\
 & \texttt{PP(509B)} & 21.2 & 16 & 84 & 98 \\
 & \texttt{DHHF (515B)} & 29.5 & 15 & 127 & 707 \\
 & \texttt{HVW2 (143A)} & 17.5 & 16 &  124 & 707 \\
 & \texttt{Ball Game (46A)} & 10 & 8 &  114 & 148 \\
 & \texttt{WE (31A)} & 14.4 & 14 &  187 & 200 \\
 & \texttt{LM (146B)} & 29.5 & 11 &  116 & 355 \\
 & \texttt{SG (570B)} & 7.5 & 11 &  69 & 531 \\
 & \texttt{WD (168A)} & 7.7 & 9 &  132 & 254 \\
 & \texttt{Football (417C)} & 13.2 & 13 &  64 & 352 \\
 & \texttt{MS (218A)} & 16.5 & 10 &  156 & 156 \\
 & \texttt{Joysticks (651A)} & 12.6 & 8 & 66 & 246 \\
\hline
\multirow{14}{*}{\shortstack{\Coreb \\(Real; \\Developers)}}  & \texttt{core. (cut)} & 306 & 4 & 4  & 6 \\ 
  & \texttt{core. (rm)} &  110 &  1 & 1 & 63 \\ 
  & \texttt{core. (ls)} & 1605.5 &  2  & 2 & 73 \\ 
  & \texttt{core. (du)} & 315 & 1 & 1 & 28 \\ 
  & \texttt{core. (seq)} & 219.7  & 3  & 3 & 5 \\ 
  & \texttt{core. (expr)} & 321 & 1 & 1 & 1 \\ 
  & \texttt{core. (copy)} & 897 & 1  & 1 & 59 \\ 
  & \texttt{find (parser)} & 119.3 & 3 & 3 & 286 \\ 
  & \texttt{find (ftsfind)} & 211.5  & 2 & 2 & 183 \\ 
  & \texttt{find (pred)} & 825 & 2 & 2 & 235 \\ 
  & \texttt{grep (dfasearch)} &  181.5 & 2 & 2 & 46\\ 
  & \texttt{grep (savedir)} &  64 & 1 & 1 & 15 \\ 
  & \texttt{grep (kwsearch)} &  77 & 2 & 2 & 46 \\ 
  & \texttt{grep (main)} & 853.5 & 2 & 2 & 45 \\ 
\hline
\textbf{Total} & 35 (46)  &  & 369 & 9012 &  148767 \\
\hline
\end{tabular}
\label{tab:subjects}
\vspace{-0.5cm}
\end{table}

\subsection{Objects of Empirical Analysis}
\emph{Programs and Bugs:} We evaluated each fault localization technique using 45 C programs containing hundreds of (369) errors and thousands of (9012) failing tests (\textit{cf. \autoref{tab:subjects}}). These programs were collected from four benchmarks, in particular, three benchmarks containing real world errors, namely IntroClass, Codeflaws and \Coreb, and one benchmark with artificial faults, namely the Software-artifact Infrastructure Repository (SIR). We selected these benchmarks of C programs to obtain a large variety of bugs and programs. Each benchmark contains a unique set of programs containing errors introduced from different sources such as developers, students, programming competitions and fault seeding (e.g. via code mutation). These large set of bugs allows us to rigorously evaluate each fault localization technique. The following briefly describes each benchmark used in our evaluation:

\begin{enumerate}
\item \emph{Software-artifact Infrastructure Repository (SIR)} is a repository designed for the evaluation of program analysis and software testing techniques using controlled experimentation~\citep{SIRBenchmark}. It contains small C programs, with seeded errors and test suites containing thousands of failing tests. In particular, this benchmark allows for the controlled evaluation of the effects of large test suites on debugging activities. 

\item \emph{IntroClass}  is a collection of small programs written by undergraduate students in a programming course \citep{IntroClass}. It contains six C programs, each with tens of instructor-written test suites. This benchmark allows for the evaluation of factors that affect debugging in a development scenario, especially for novice developers. 

\item \emph{Codeflaws} is a collection of programs from online programming competitions held on Codeforces.\footnote{\url{https://codeforces.com/}} These programs were collected for the comprehensive evaluation of debugging tools using different types of errors. It contains 3902 errors classified across 40 defect classes in total \citep{Codeflaws}. In particular, this benchmark allows for the evaluation of fault localization techniques on different defect types.

\item \emph{\Coreb} is a collection of 70 real errors that were systematically extracted from the repositories and bug reports of four open-source software projects: Make, Grep, Findutils, and Coreutils~\citep{corebench}.\footnote{\url{http://www.comp.nus.edu.sg/~release/corebench/}} These projects are well-tested, well-maintained, and widely-deployed open source programs for which the complete version history and all bug reports can be publicly accessed. All projects come with an extensive test suite. \Coreb allows for the evaluation of fault localization techniques on real world errors (unintentionally) introduced by developers. It has been used in several debugging studies, including a study that investigates how developers debug and fix real faults~\citep{dbgbench}.
\end{enumerate}

\autoref{tab:subjects} lists all the programs and bugs investigated in our study. We use six programs each from the SIR and IntroClass benchmarks. This  includes \texttt{tcas} -- this program is \emph{the} most well-studied subject according to a recent survey on fault localization~\citep{tse}. We selected 20 programming competitions from Codeflaws, including popular and difficult contests, such as ``Tavas and Nafas (535A)" and ``Lucky Mask (146B)". From \Coreb, we used three projects, namely the \texttt{Coreutils}, \texttt{Grep} and \texttt{Find} project. Notably, all projects in \Coreb come from the GNU open source C programs, in particular, these three projects contain a total of 103 tools. Due to code modularity, the program size for a single tool (e.g. \texttt{cut} in \texttt{coreutils}) contains a few hundred LoC (about 306 LoC), however, the entire code base for \Coreb is fairly large. For instance, \texttt{Coreutils}, \texttt{Grep} and \texttt{Find} have 83k, 18k and 11k LoC, respectively~\citep{corebench}. For each benchmark, we exempted programs where \texttt{Frama-C} \emph{could not} construct the Program Dependence Graph (PDG). For instance, because it cannot handle some recursive or variadic method calls. In addition, we excluded an error if no coverage information could be generated (e.g., infinite loops) or the faulty statement could not be identified (e.g., \emph{omission faults} where the patch only added statements).

\emph{Single Faults:} For our evaluation (all RQs except RQ7), we used buggy programs collected from four well-known benchmarks, where programs contained only a single fault. To determine single faults in our bug dataset, for each program, we executed all tests available for a project on the fixed version of the program, in order to determine if there are any failing test cases that are unrelated to the bug at hand. Our evaluation revealed that our dataset contained mostly single bugs (368/369=99.7\%). Almost all buggy program versions had exactly one fault, except for a single program -- \texttt{Codeflaws} version \texttt{DC} \texttt{495A}. For all benchmarks, only this program contained multiple faults, i.e. more than one fault. This distribution of single faults portrays the high prevalence of single faults and single-fault fixes in the wild~\citep{perez2017prevalence}.

\begin{table}[!tbp]\centering
\caption{Details of Multiple Faults}
\begin{tabular}{| c | l | r | r | r | r | r |}
\hline
\textbf{Benchmark}  & \textbf{Tool} & \textbf{\# Buggy} & \textbf{\#Faults} & \textbf{\#Failing} & \textbf{\#Passing} \\
(Error Type) &  &  \textbf{Programs} &  & \textbf{Tests} & \textbf{Tests} \\
\hline
\multirow{6}{*}{\shortstack{\texttt{SIR-MULT} \\ (Mutated)}}  & \texttt{tcas} & 37 & 144  & 19973 & 39523  \\
 & \texttt{print\_tokens} & 3 & 11 &  12074 & 316  \\
 & \texttt{print\_tokens2} & 8 & 28 & 27630 & 5290  \\
 & \texttt{tot\_info} & 18 & 64 & 16667 & 2269  \\
 & \texttt{schedule}  & 4 & 17 &  9673 & 927  \\
 & \texttt{schedule2} & 4 & 16 &  8616 & 2224 \\
\hline
\multirow{6}{*}{\shortstack{\texttt{IntroClass} \\ \texttt{-MULT} \\ (Mutated)}}  & \texttt{checksum} & 1 & 4 & 15 & 1  \\
 & \texttt{digits} & 2 & 7 & 30 & 2  \\
 & \texttt{grade} & 5 & 22 & 67 & 23  \\
 & \texttt{median} & 2 & 8 & 13 & 13  \\
 & \texttt{smallest} & 1 & 4 & 8 & 8  \\
 & \texttt{syllables} & 3 & 13 & 34 & 14  \\
 \hline
\textbf{Total} & & 88 & 338 & 94800  & 50610  \\
\hline
\end{tabular}
\label{tab:mult-bug}
\vspace{-0.5cm}
\end{table}

\emph{Multiple Faults:} To evaluate the effectiveness of all three fault localization techniques on multiple faults (\textit{see} RQ7), we automatically curated a set of multiple faults using \emph{mutation-based fault injection}. This is in line with the evaluation of multiple faults in previous works~\citep{abreu2009spectrum,zheng2006statistical,digiuseppe2011influence,wong2013dstar,wong2012software}.\footnote{To the best of our knowledge, there is no known benchmark of real-world programs containing multiple faults.} 
We automatically mutated the original passing version of each program until we have a buggy version containing between three to five faults. 
In particular, we performed logical and arithmetic operator mutation on each passing version of the programs contained in the \texttt{SIR} and \texttt{IntroClass} benchmarks. 
\autoref{tab:mult-bug} provides details of the buggy programs with multiple faults, the number of faults, as well as the number of failing and passing test cases. For each fault contained in the resulting program, we store the failing test case(s) that expose the bug, as well as the corresponding patches for each fault and all faults. 
In total, we collected 88 programs with multiple faults containing 338 injected faults, in total. Each program in this dataset contained about four unique faults, on average. Specifically, we collected 74 and 14 programs from the \texttt{SIR} and \texttt{IntroClass} benchmarks, 
and injected a total of 280 and 58 faults in each benchmark, 
respectively. The programs containing multiple faults are called \texttt{SIR-MULT} and \texttt{IntroClass-MULT}, respectively (\emph{see \autoref{tab:mult-bug}}). 

\emph{Minimal Patches:} The user-generated patches are used to identify those statements in the buggy version that are marked faulty. In fact, Renieris and Reiss~\citep{renieris} recommend identifying as faulty statements those that need to be changed to derive the (correct) program that does not contain the error. For each error, only patched statements are considered faulty. 
All bugs in our corpus are patched with at least one statement changed in the buggy program, all \emph{omission bugs} are exempted. Omission bugs require special handling since they are quite difficult to curate, localize and fix. Collecting patches and fault locations for omission bugs is difficult because their patches are similar to 
the implementation of new features. 
A faulty code location is unclear for omission 
bugs (in the failing commit), this makes them even more difficult to evaluate for typical AFL techniques, including statistical debugging and dynamic slicing~\citep{lin2018break}.

\emph{Slicing Criterion:} All aspects of dynamic slicing can be fully automated. To this end, as the slicing criterion we chose the last statement that is executed or the return statement of the last function that is executed. For instance, when the program crashes because an array is accessed out of bounds, the location of the array access is chosen as the slicing criterion. In our implementation, the slicing criterion is automatically selected by a bash script running \texttt{gdb}.

\emph{Passing and Failing Test Cases:} All programs in our dataset come with an extensive test suite which checks corner cases and that previously fixed errors do not re-emerge. For statistical debugging, we execute each of these (passing) test cases individually to collect coverage information. For dynamic slicing, we perform slicing for each failing test case.

In summary, for our automated evaluation, we used 457 errors in dozens of programs from four well-known benchmarks (\textit{see \autoref{tab:subjects} and \autoref{tab:mult-bug}}). Our corpus contained 46 different programs in 35 software tools. Each faulty program in our corpus had about 11 bugs, 257 failing test cases and thousands (4250) of passing test cases, on average. For single faults, we have 295 real faults and 74 injected faults. Meanwhile, we have 88 buggy programs containing multiple faults, each program contains about four faults, on average.

\subsection{Measure of Localization Effectiveness}
\label{measures}
We measure \emph{fault localization effectiveness} as the proportion of statements that do \emph{not} need to be examined until finding the first fault. This allows us to assign a score of 0 for the worst performance (i.e., all statements must be examined) and 1 for the best. More specifically, we measure the \emph{score} $=1-p$ where $p$~is the proportion of statements that needs to be examined before the first faulty statement is found. Not all failures are caused by a single faulty statement. In a study of B\"ohme and Roychoudhury, only about 10\% of failures were caused by a single statement, while there is a long tail of failures that are substantially more complex~\citep{corebench}. Focusing on the first faulty statement found, the \emph{score} measures the effort to find a good starting point to initiate the bug-fixing process rather than to provide the complete set of code that must be modified, deleted, or added to fix the failure. \cite{tse} motivates this measure of effectiveness and presents an overview of other measures.

\subsubsection{General Measures}
\emph{Ranking}. All three fault localization techniques presented in this paper produce a ranking. The developer starts examining the highest ranked statement and goes down the list until reaching the first faulty statement. To generate the ranking for \emph{statistical debugging}, we list all statements in the order of their suspiciousness (as determined by the technique), most suspicious first. 
To generate the ranking for \emph{\adslicing
}, given the statement~$c$ where the failure is observed, we rank first those statements in the slice that can be reached from~$c$ along one backward dependency edge. Then, we rank those statements that can be reached from~$c$ along two backward dependency edges, and so on. Generally, for all techniques, the $\textit{score}$ is computed as 
$$
\textit{score} = 1-\frac{|S|}{|P|}
$$ 
where~$S$ are all statements with the same rank or less as the highest ranked faulty statement and~$P$ is the set of all statements in the program. So, $S$~represents the statements a developer needs to examine until finding the first faulty one.\footnote{Note that all executable program statements are ranked in the suspiciousness rank, executable statements that are not contained in the dynamic slice are ranked lowest.}. 

\emph{Multiple Statements, Same Rank}. In most cases there are several statements that have the same rank as the faulty statement. For all our evaluations, we employ ordinal ranking, in order to effectively determine the top $N$ most suspicious statements for each technique. This is necessary to evaluate the fault localization effectiveness of each technique, if a developer is only willing to inspect $N$ most suspicious statements~\citep{kochhar2016practitioners}. In ordinal ranking, lines with the same score are re-ranked by line numbers.\footnote{Ranking ties are broken in ascending order, i.e. if both lines 10 and 50 have the same score, then line number 10 is ranked above line number 50.} This is in agreement with evaluations of fault localization techniques in previous work~\citep{tse, pearson2017evaluating,kochhar2016practitioners}.

\emph{Multiple Faults, Expense Score}. For \emph{multiple faults}, we measure fault localization effectiveness using the \emph{expense score}~\citep{yu2008empirical}. The expense score is the percentage of the program (statements) that must be examined to find the \textit{first fault}, in particular, the \textit{first faulty statement in the first localized fault}, using the ranking given by the fault localization technique. It is similar to the score employed for single faults, and it has been employed in previous evaluations of multiple faults, such as \citep{wong2012software,wong2013dstar}. Formally:
$$
\textit{expense~score} = \frac{|S|}{|P|} * 100
$$
where~$S$ are all statements with the same rank or less as the highest ranked faulty statement for \emph{the first fault found} and~$P$ is the set of all executable statements in the program. So, $S$~represents the statements a developer needs to examine until finding \emph{the first faulty statement, for the first localized fault}. 
The assumption is that it is the first fault that the developer would begin fixing, thus, finding the first statement suffices for the diagnosis of all faults~\citep{yu2008empirical}. In our evaluation of multiple faults, the fault localization effectiveness $\textit{score}$ is computed similarly to single faults as
$$
\textit{score}_{\text{mult}} = 1 - (\textit{expense~score}/100)
$$

\subsubsection{Dynamic Slicing Effectiveness}
\label{sec:slicing-experiment}
We define the effectiveness of \emph{\adslicing
}, the \emph{score$_{\text{ads}}$} according to Renieris and Reiss~\citep{renieris} as follows. Given a failing test case~$t$, the symptomatic statement~$c$, let $P$ be the set of all statements in the program, let $\zeta$ be the \adslice computed w.r.t. $c$~for~$t$, let $k_{min}$ be the minimal number of backward dependency edges between~$c$ and any faulty statement in $\zeta$, and let $DS_*(c,t)$ be the set of statements in $\zeta$ that are reachable from~$c$ along at most $k_{min}$ backward dependency edges.
Then,
\begin{align}
\text{\emph{score$_{\text{ads}}$}} = 1 - \frac{|DS_*(c,t)|}{|P|}\nonumber
\end{align}

Algorithmically, the \emph{score$_{\text{ads}}$} is computed by (i)~measuring the minimum distance $k_{min}$ from the statement~$c$ where the failure is observed to any faulty statement along the backward dependency edges in the slice, (ii)~marking all statements in the slice that are at distance~$k_{min}$ or less from~$c$, and (iii)~measuring the proportion of marked statements in the slice. This measures the part of code a developer investigates who follows backward dependencies of executed statements from the program location where the failure is observed towards the root cause of the failure. 

In the \adslice in our motivating example (\autoref{fig:middle-slicing}), we have \emph{score}$_{\text{ads}} = 1-\frac{1}{12} = 0.92$. The slicing criterion is $c=s_{15}$. The program size is $|P|=12$. The faulty statement~$s_8$ is ranked first. Statements $s_7$~and~$s_2$ are both ranked third according to \emph{modified competition ranking}\footnote{In this case, when several statements have the same rank as the faulty statement, we made the conservative assumption that a developer finds the faulty statement among other statements with the same rank.}. Statements $s_5$~and~$s_4$ are ranked fourth and fifth, respectively, while the remaining, not executed (but executable) statements are ranked 12$^{\text{th}}$.  

\subsubsection{Statistical Debugging Effectiveness}
\label{sec:sfl-experiment}
We define the effectiveness of a \emph{statistical fault localization} technique, the \emph{score$_{\text{sfl}}$} as follows. Given the ordinal ranking of program statements in program~$P$ for test suite~$T$ according to their suspiciousness as determined by the statistical fault localization method, let~$r_f$ be the rank of the highest ranked faulty statement and~$P$ is the set of all statements in the program. Then,
\begin{align}
\text{\emph{score$_{\text{sfl}}$}} = 1 - \frac{r_f}{|P|}\nonumber
\end{align}
Note that $\emph{score$_{\text{sfl}}$} = 1 - \emph{EXAM-score}$ where the well-known \emph{EXAM-score}~\citep{statEval1,statEval2} gives the proportion of statements that need to be examined until the first fault is found. Intuitively, the \emph{score$_{\text{sfl}}$} is its complement assigning 0 to the worst possible ranking where the developer needs to examine all statements before finding a faulty one.

For instance, \emph{score$_{\text{sfl}}$} $=1-\frac{1}{12}=0.92$ for our motivating example and all considered statistical debugging techniques. All statistical debugging techniques identify the faulty statement in Line~8 as most suspicious. So, there is only one top-ranked statement (Rank 1). But there are six statements with the lowest rank (Rank 12). If the fault was among one of these statements, the programmer might need to look at all statements of our small program \texttt{middle} before localizing the fault. 

\subsubsection{Hybrid Approach Effectiveness}
\label{sec:hybrid-experiment}
We define the effectiveness of the hybrid approach, the \emph{score$_{\text{hyb}}$} as follows. 
Let~$R$ be the set of faulty statements, $H$ be the $N$~most suspicious statements -- sorted first by suspiciousness score and then by line numbers and~$P$ is the set of all statements in the program. Given the failing test case~$t$ and a statement~$c$ that is marked as symptomatic, we have
\begin{equation}
\text{\emph{score$_{\text{hyb}}$}} = \begin{cases}
\min(\emph{score$_{\text{sfl}}$},N) & \text{ if $R\cap H\neq \emptyset$}\\
1 - |H \cup DS_*(c,t)|\ /\ |P| & \text{ otherwise}\nonumber
\end{cases} 
\end{equation}
Essentially, \emph{score$_{\text{hyb}}$} computes the score for the statistical fault localization technique if the faulty statement is within the first~$N$ most suspicious statements, and the score for approximate dynamic slicing while accounting for the statements already reported in the first phase. 
For instance, for $N=2$ we have \emph{score$_{\text{hyb}}$} $=1-\frac{1}{12}=0.92$ for the motivating example in \autoref{fig:middle-coverage} since the fault is amongst the $N$ most suspicious statements.

\subsection{Infrastructure}
We performed the experiments on a virtual machine (VM) running Arch Linux. The VM was running on a Dell Precision 7510 with a 2.7GHz Intel Core i7-6820hq CPU and 32GB of main memory. 

\begin{table}[!tp]\centering
\scriptsize
\caption{Effectiveness of Dynamic Slicing on Single Faults} 
\begin{tabular}{| c | r | r | r | r | r | r | r |} 
\hline
\multirow{4}{*}{\textbf{Benchmark}} &\multirow{4}{*}{\shortstack{ \textbf{Score}}}  &  \multicolumn{4}{c|}{\textbf{\% Errors Localized}} \\ 
 & & \multicolumn{4}{c|}{if developer inspects \textbf{N}} \\ 
& & \multicolumn{4}{c|}{most suspicious LoC}  \\ 
 &  &  \textbf{5} &  \textbf{10} &  \textbf{20} &  \textbf{30} \\ 
\hline
IntroClass  & 0.83 &  70.00 & 100 & 100 & 100 \\ 
 \hline
Codeflaws & 0.78  &   75.30 & 92.71  & 98.79 & 100  \\ 
\hline
\Coreb & 0.85  & 18.52 & 18.52  & 29.63 & 40.74 \\ 
\hline
\textbf{Real} &0.79 &  69.73 & 86.39 & 92.52 & 94.56\\ 
\hline
\textbf{Artificial} (SIR)  & 0.79 & 32.43 & 44.59 & 55.41 & 60.81 \\ 
\hline
\textbf{Avg. (Bugs)}  & 0.774 & 62.23 & 77.99 & 85.05  & 87.77  \\ 
\hline
\end{tabular}
\label{tab:slicing-effectiveness}
\end{table}

\section{Evaluation Results}
\label{sec:eval-results}
Let us discuss the results of our evaluation and their implications. All research questions (RQs) are evaluated using single faults (i.e., RQ1 to RQ6), except for RQ7, which is evaluated on both single and multiple faults.

\subsection{RQ1: Effectiveness of Dynamic Slicing.} 
\label{sec:slicing-effectiveness}
\emph{How effective is dynamic slicing in fault localization?}
To investigate the fault localization effectiveness of dynamic slicing, we examined the proportion of statements a developer would \emph{not need to} inspect after locating the faulty statement (\textit{score} in \autoref{tab:slicing-effectiveness}). Then, we examine the percentage of errors for which a developer can effectively locate the faulty statement, if she inspects only $N$ most suspicious statements reported by dynamic slicing for $N \in \{5,10,20,30\}$ (\textit{\% Errors Localized} in \autoref{tab:slicing-effectiveness}). 

Overall, a single faulty statement is ranked within the first quarter of the most suspicious program statements reported by dynamic slicing, on average (\textit{cf. \autoref{tab:slicing-effectiveness}}). 
This implies that a developer (using dynamic slicing) inspects 21\% (about 40 LoC) of the executable program statements before locating the fault, on average (i.e., equal to $1 - score$).
This performance was independent of the source or type of the errors (i.e., real or seeded errors). Dynamic slicing was particularly highly effective in locating faults for errors in \Coreb and errors in IntroClass, where it ranks the faulty statements within the top 15\% (81 LoC) and 17\% (3 LoC) of the program statements, respectively (\textit{cf. \autoref{tab:slicing-effectiveness}}). 

\begin{result}
For all programs, dynamic slicing reports the faulty statement within the top 21\% (40 LoC) of the most suspicious statements, on average.
\end{result}

A developer or tool using dynamic slicing will locate the faulty statement after inspecting a handful of suspicious statements. 
In our evaluation, for most errors, the faulty statement can be identified after inspecting only five to ten most suspicious statements reported by dynamic slicing. Specifically, the faulty statement is ranked within the top five to ten most suspicious statements for 62\% to 78\% of all errors, respectively (\textit{cf. \autoref{tab:slicing-effectiveness}}). 
Notably, a developer will locate the faulty statement for 55\% of artificial errors and 92\% of real errors if she inspects the top 20 most suspicious statements. 
Overall, most programs (85\%) can be debugged by inspecting the top 30\% (58 LoC, on average) of the statements reported by dynamic slicing. These results demonstrate the high effectiveness of dynamic slicing in fault localization.

\begin{result}
Dynamic slicing reports a single faulty statement within the top 5--10 most suspicious statements for most errors (62\%~to~78\%, respectively).  
\end{result}

\begin{table}[]\centering
\caption{Effectiveness of Statistical Debugging on Single Faults. Best scores for each (sub)category are in \textbf{bold}; higher scores are better. For instance, \texttt{Kulczynski2} is the best performing (single bug optimal) formula for \emph{all programs (0.737)}, on average.}
\begin{tabular}{| c | l | r | r | r | r | r | r |}
\hline
\textbf{SBFL} & \multirow{2}{*}{\shortstack{\textbf{Formula}}} & \multirow{2}{*}{\shortstack{\texttt{SIR}}} & \textbf{\texttt{Intro}}  & \textbf{\texttt{Code}}  & \textbf{\texttt{Core}} & \multicolumn{2}{c|}{\textbf{Average}} \\ 
\textbf{Family} & & & \textbf{\texttt{Class}}  & \textbf{\texttt{flaws}}  & \textbf{\texttt{bench}} & \textbf{\texttt{(Bugs)}} & \textbf{\texttt{(Prog.)}} \\
\hline
\multirow{3}{*}{\shortstack{Popular}}
 & \texttt{Tarantula} & 0.78 &  \textbf{0.76} & \textbf{0.70}  &\textbf{ 0.79}  &  \textbf{0.709} & 0.732 \\ 
 & \texttt{Ochiai} &  \textbf{0.83 }& \textbf{0.76} & 0.69 & \textbf{0.79}   &  \textbf{0.709}  & \textbf{0.735} \\
 & \texttt{Jaccard} & 0.80 & \textbf{0.76} & 0.69 & \textbf{0.79} & 0.702 & 0.728 \\ 
 \hline
\multirow{7}{*}{\shortstack{Human \\ Generated}} 
 & \texttt{Naish\_1} &  \textbf{0.83} & \textbf{0.74} & \textbf{0.69} & 0.79 & \textbf{0.710}   & \textbf{0.733} \\ 
 & \texttt{Naish\_2} & 0.81  & \textbf{0.74} & \textbf{0.69} & 0.79 &  0.709 & 0.731 \\
 & \texttt{Russel\_Rao} &  0.67 & 0.59 & 0.57 & 0.77 & 0.602 & 0.611 \\ 
 & \texttt{Binary} &  0.69 & 0.59 & 0.57 & 0.77 & 0.603 & 0.614 \\ 
 & \texttt{Wong\_1} & 0.67 & 0.59 & 0.57 &0.77  &  0.602  & 0.611 \\ 
  & \texttt{$D^2$} & 0.73 & 0.62 & 0.56 & \textbf{0.80}  & 0.598  & 0.618 \\ 
  & \texttt{$D^3$} & 0.75 & 0.62 & 0.56 & \textbf{0.80}  & 0.601  & 0.622 \\ 
\hline
\multirow{4}{*}{\shortstack{GP \\ Evolved}} 
 & \texttt{GP\_02} &  0.75 & 0.72 & 0.66 & 0.69 & 0.668 & 0.688 \\ 
 & \texttt{GP\_03} &  0.77 & 0.68 & 0.63 & 0.63 & 0.643 & 0.663 \\ 
 & \texttt{GP\_13} & \textbf{0.81} & \textbf{0.74} & \textbf{0.69} & \textbf{0.79} & \textbf{0.709} & \textbf{0.731} \\ 
 & \texttt{GP\_19} & 0.56 & 0.69 & 0.65 & 0.75 & 0.631 & 0.649 \\ 
 \hline
\multirow{4}{*}{\shortstack{Single Bug \\ Optimal}}  
 & \texttt{PattSim\_2} &  \textbf{0.85} & 0.68 & 0.69 & 0.76 & 0.705 & 0.721 \\ 
 & \texttt{lex\_Ochiai} & 0.83 & 0.74 & 0.69 & \textbf{0.79} & 0.710 & 0.733 \\ 
 & \texttt{m9185} & 0.83 & 0.74 & \textbf{0.70} & \textbf{0.79} & \textbf{0.715} &  0.735 \\   
 & \texttt{Kulczynski2} &  0.83 & \textbf{0.76 }& \textbf{0.70} & \textbf{0.79} & 0.713 & \textbf{0.737} \\
 \hline		
\end{tabular}
\label{tab:SBFL-Effectiveness}
\end{table}

\begin{figure}[!tp]\centering 
\begin{tabular}{@{}c@{}c@{}}
\includegraphics[width=0.5\columnwidth]{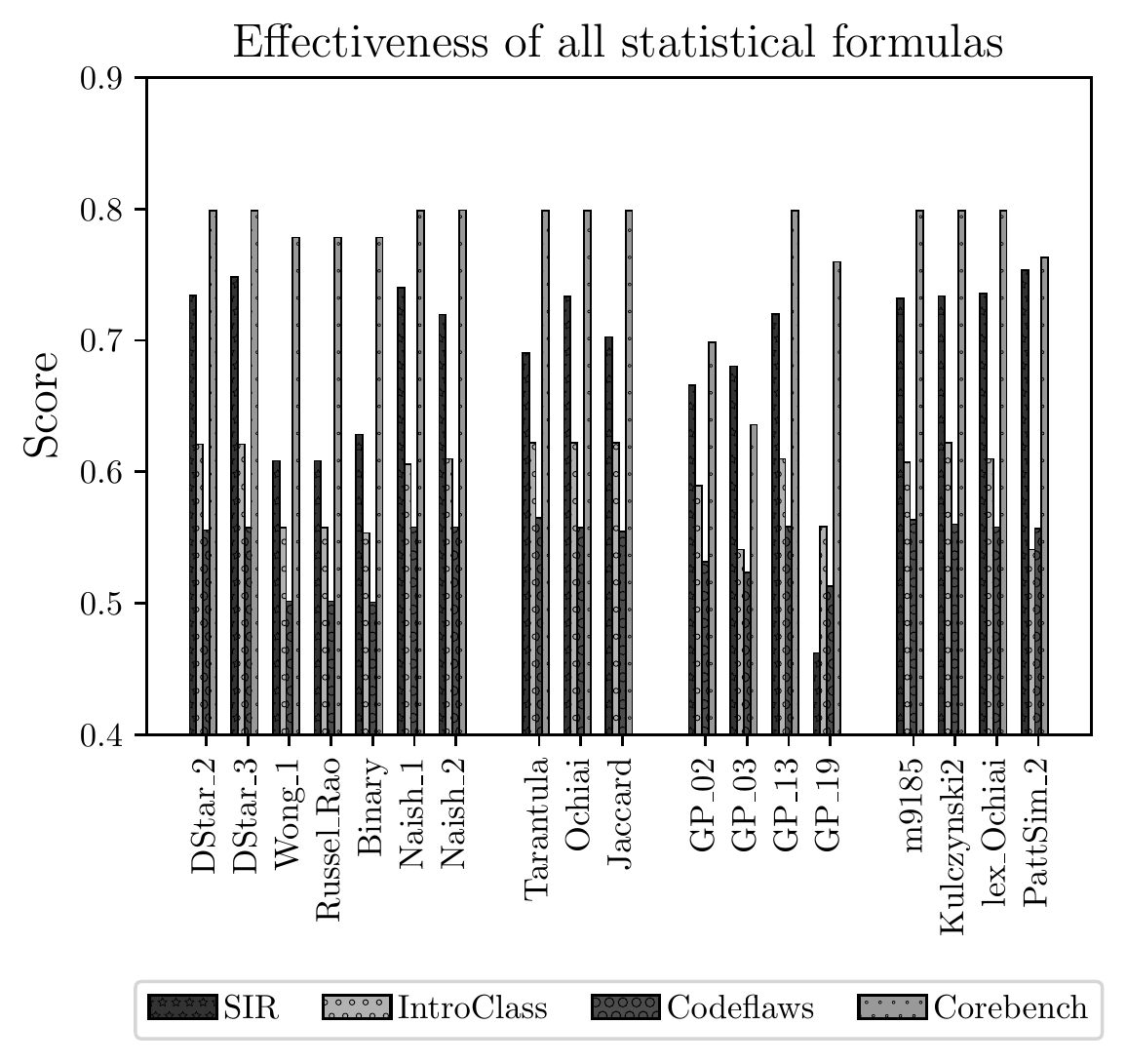} &
\includegraphics[width=0.5\columnwidth]{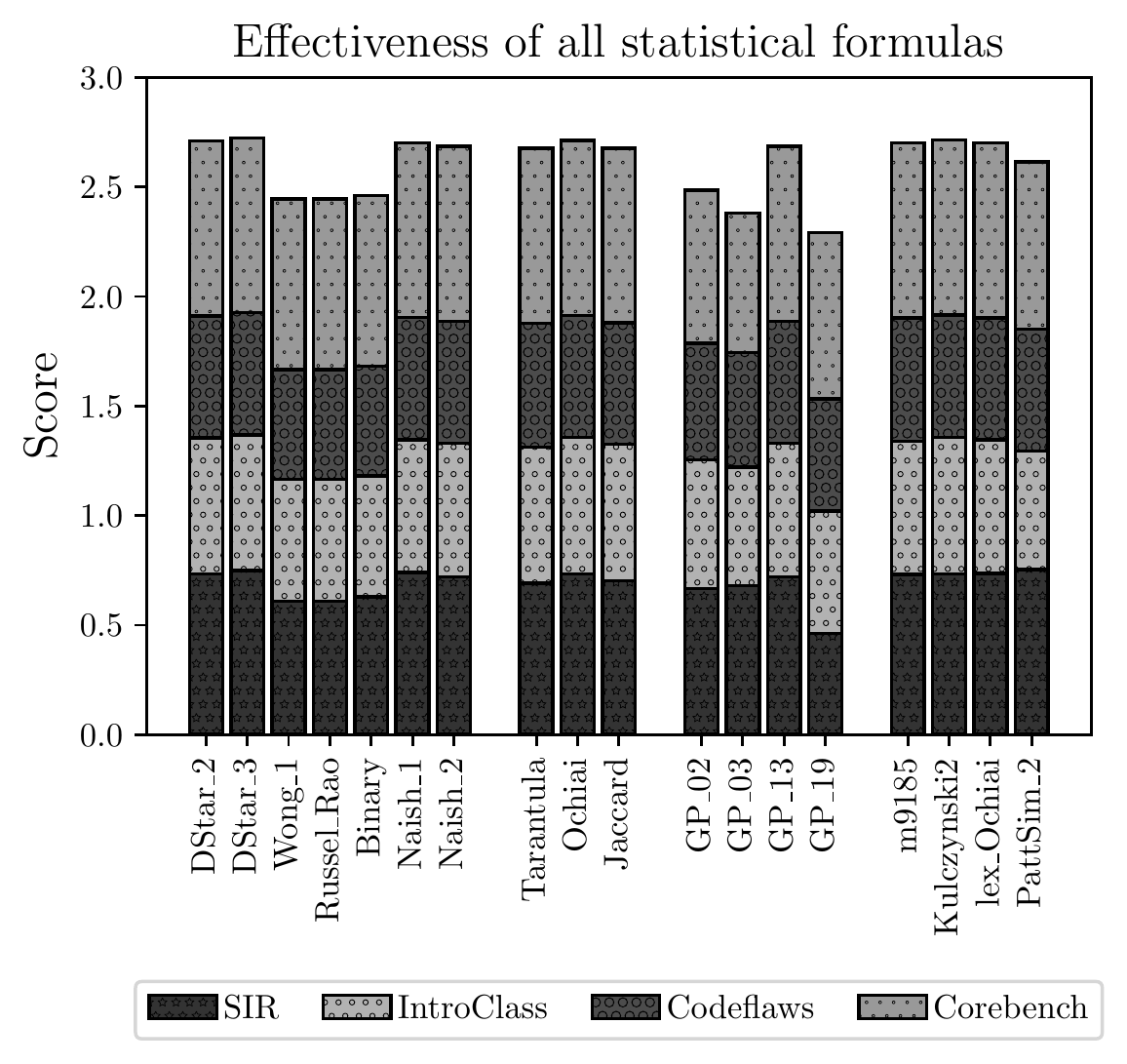} \\
\quad\quad\textbf{(a) Each Benchmark} &\quad\quad\textbf{(b) All Benchmarks }\\
\end{tabular} 
\caption{
Effectiveness of each SBFL formula on Single Faults; Results are grouped into bars for each family showing (a) the performance of each SBFL formula on each benchmark and (b) the cumulative results for all benchmarks using stacked bars
}
\label{fig:stat-dbg-effectiveness-all}
\end{figure}

\begin{figure}[!tp]\centering 
\begin{tabular}{@{}c@{}c@{}}
    \begin{minipage}[b]{0.5\textwidth}\includegraphics[scale=0.45]{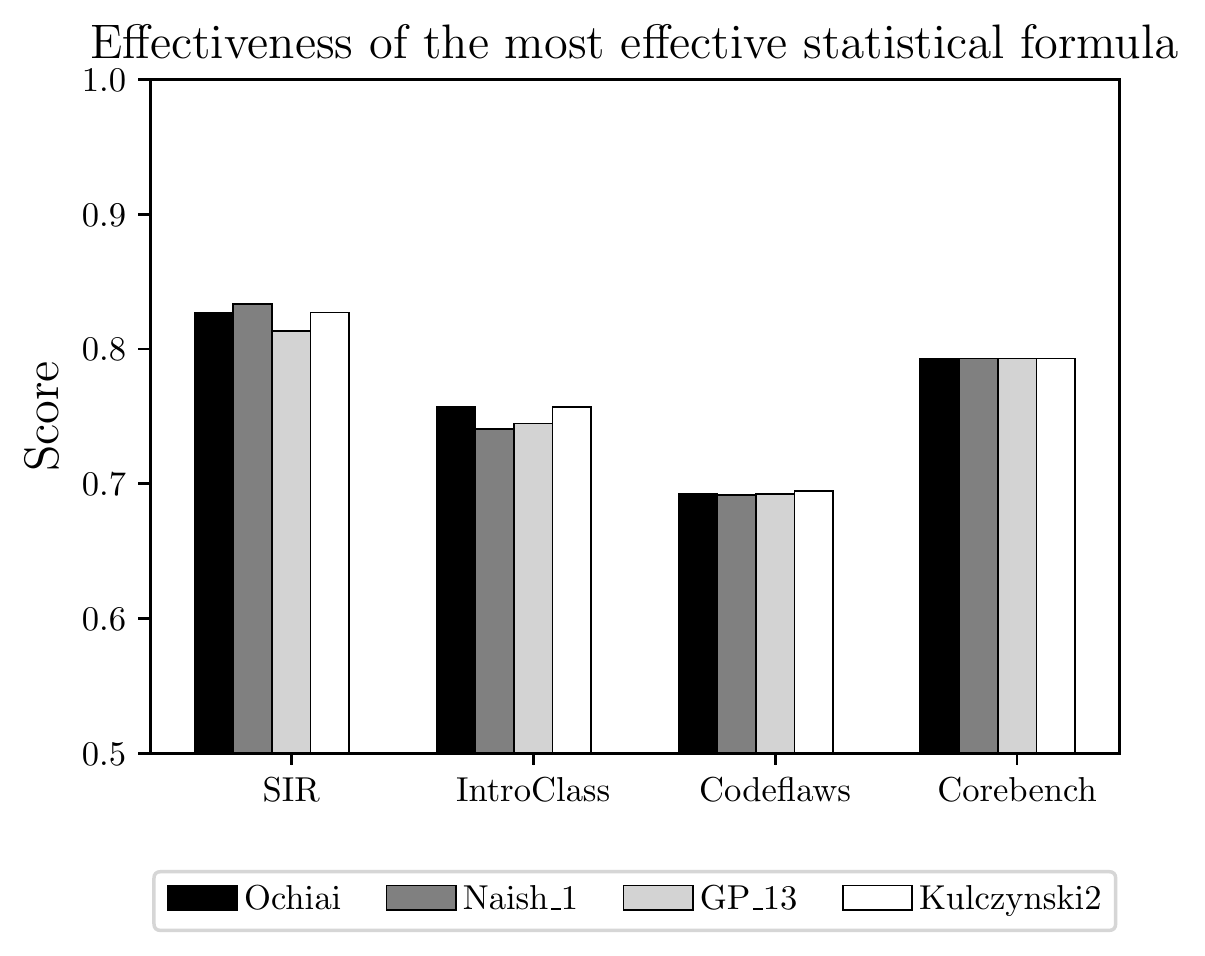}\\~\end{minipage} &
    \begin{minipage}[b]{0.5\textwidth}\raisebox{1.5mm}{\includegraphics[scale=0.46]{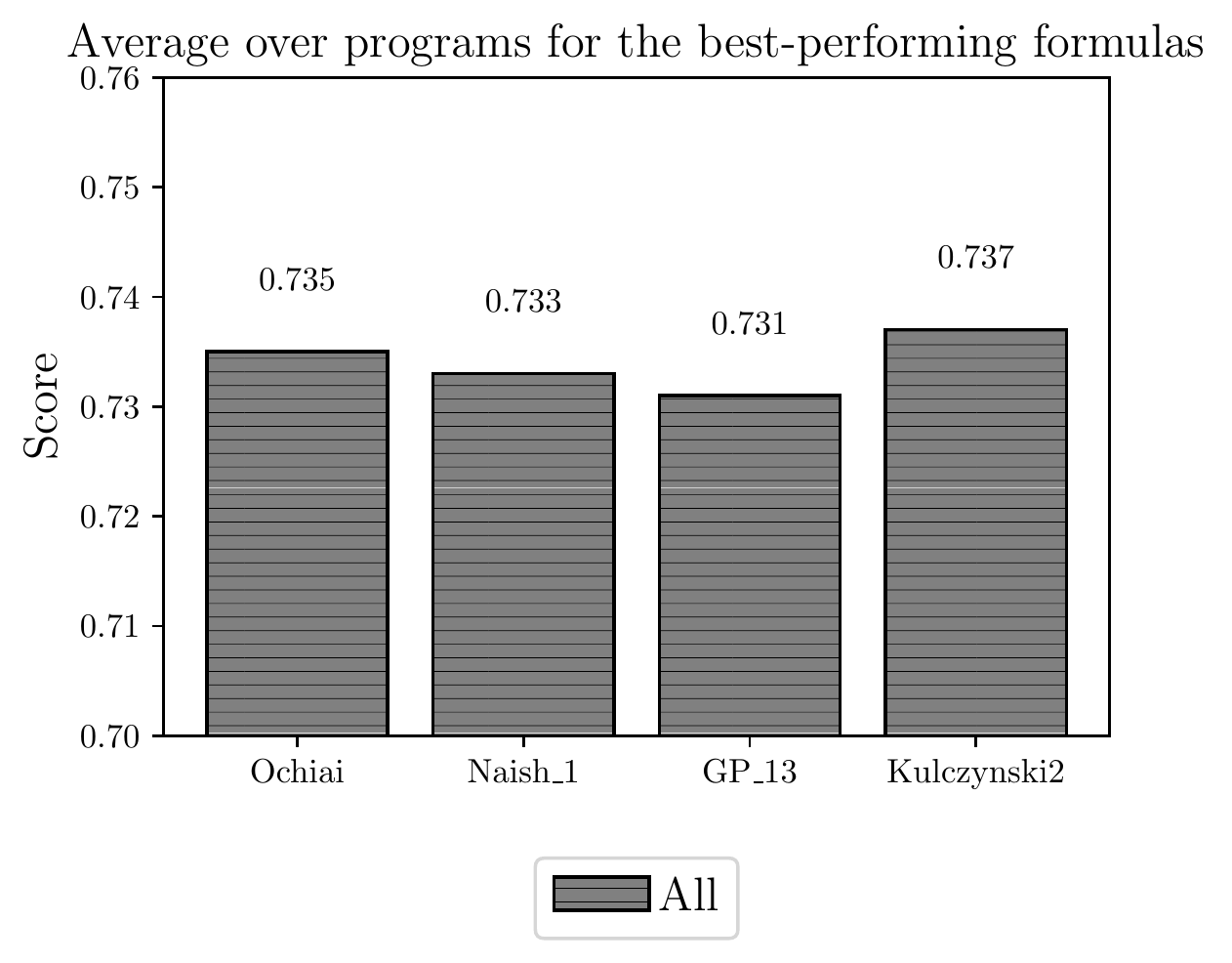}}\end{minipage} \\
\quad\quad\textbf{(a) Each Benchmark} &\quad\quad\textbf{(b) All Benchmarks }\\
\end{tabular} 
\caption{Effectiveness of the most effective statistical debugging formula in (a) each family (bars are grouped by benchmarks), and (b) the overall average (i.e., mean) for all benchmarks}
\label{fig:stat-dbg-effectiveness-best}
\end{figure}

\subsection{RQ2: Effectiveness of Statistical Debugging}
\label{sec:sfl-effectiveness}
\emph{Which statistical formula is the most effective at fault localization?} First, we investigate the effectiveness of 18 statistical formulas using four benchmarks containing 369 errors (\textit{cf. \autoref{tab:subjects}}). To determine the most effective statistical formula, for each formula, we examined the proportion of statements a developer would \emph{not need} to inspect after locating a single faulty statement (\textit{score} in \autoref{tab:SBFL-Effectiveness}). \autoref{fig:stat-dbg-effectiveness-all} (a) and (b) further illustrate the effectiveness of the SBFL formulas. Then, for the best performing statistical formula, we inspected the percentage of errors for which a developer can effectively locate the faulty statement, if she inspects only $N$ most suspicious statements for $N \in \{5,10,20,30\}$ (\textit{\% Errors Localized} in \autoref{tab:Kulczynski2-effectiveness}). 

Overall, the \textit{single bug optimal} formulas are \emph{the most effective family of statistical formulas}, they are the best performing formulas across all errors and programs. In particular, on average, \texttt{PattSim2} performed best for injected errors (i.e. \texttt{SIR}), while \texttt{Kulczynski2} outperformed all other formulas for real errors, especially for \texttt{IntroClass} (\emph{cf. \autoref{tab:SBFL-Effectiveness}}). \textbf{Bold} values in \autoref{tab:SBFL-Effectiveness} indicate the best performing formula for each family and (sub)category. For instance, \texttt{Kulczynski2} is the best performing (single bug optimal) formula for \emph{all programs (0.737)}. The performance of single-bug optimal formulas supports the results obtained in previous works~\citep{landsberg2016methods}. This family of statistical formulas are particularly effective because they are optimized for programs containing a single bug; based on the observation that if a program contains only a single bug, then all failing traces cover that bug~\citep{naish2013duals}. 

\begin{result}
The \textnormal{\texttt{single bug optimal}} statistical formulas outperformed all other SBFL formulas,
for both injected and real errors, on average.
\end{result}

The \textit{most effective statistical formula is \texttt{Kulczynski2}}, it outperformed all other formulas in our evaluation (\textit{see \autoref{tab:SBFL-Effectiveness} and \autoref{fig:stat-dbg-effectiveness-all} (a) and (b)}). The most effective statistical formula for each family are \texttt{Ochiai}, \texttt{Naish\_1},	\texttt{GP\_13} and	\texttt{Kulczynski2} for the \textit{popular}, \textit{human-generated}, \textit{genetically-evolved} and \textit{single bug optimal} families, respectively. \autoref{fig:stat-dbg-effectiveness-best} (a) and (b) compares the performance of the most effective formula in each family. For instance, in the \emph{popular} statistical family, \texttt{Ochiai} is the best performing formula, both for \emph{all errors (0.709)} and \emph{all programs (0.735)} (\textit{cf. \autoref{tab:SBFL-Effectiveness}}). Meanwhile, in the single bug optimal family, \texttt{Kulczynski2}  is the best performing formula for \emph{all programs (0.737)} (\textit{cf. \autoref{tab:SBFL-Effectiveness}}).

\begin{table}[!tp]\centering%
\scriptsize
\caption{Effectiveness of \texttt{Kulczynski2} (i.e., the most effective statistical formula) on Single faults} 
\begin{tabular}{| l | r | r | r | r | r | r | r | } 
\hline
\multirow{4}{*}{\textbf{Benchmark}} &\multirow{4}{*}{\textbf{Score}}  &  \multicolumn{4}{c|}{\textbf{\% Errors Localized}}  \\ 
 & & \multicolumn{4}{c|}{if developer inspects \textbf{N}} \\ 
& & \multicolumn{4}{c|}{most suspicious LoC}  \\
 &  &  \textbf{5} &  \textbf{10} &  \textbf{20} &  \textbf{30} \\ 
\hline
IntroClass  & 0.76  &  80.00 & 85.00 & 100 & 100 \\ 
 \hline
Codeflaws & 0.69 & 64.37 & 86.23 & 97.57 & 99.19  \\ 
\hline
\Coreb & 0.79 &  22.22 & 25.93 & 37.04 & 48.15  \\ 
\hline
\textbf{Real} & 0.72 & 61.56 & 80.61 & 92.18 & 94.56 \\ 
\hline
\textbf{Artificial} (SIR) & 0.83  &  35.14 & 41.89 & 68.92 & 71.62  \\ 
\hline
\textbf{Avg. (Bugs)}  & 0.713 &   56.25 & 72.83 & 87.50 & 89.95 \\ 
\hline
\end{tabular}
\label{tab:Kulczynski2-effectiveness}
\end{table}

Indeed, a developer using \texttt{Kulczynski2} will inspect the least number of suspicious program statements before finding the faulty statement. On average, \texttt{Kulczynski2} required a developer to inspect about 26\% (51 LoC) of the program code before finding the faulty statement. Among all statistical formulas, it has the highest suspiciousness rank for 40\% (14 out of 35) of the programs and 72\% (265 out of 369) of all errors. It is also the most effective statistical formula for localizing real errors.\footnote{Further evaluations (on single faults) in this paper use \texttt{Kulczynski2} as the default ``\textit{statistical debugging}" formula.}

\begin{result}
  \textnormal{\texttt{Kulczynski2}} is the most effective statistical formula, requiring a developer to inspect 26\% of code (51 LoC) before finding the fault, on average.
\end{result}

\begin{figure}[!tp]\centering 
\begin{tabular}{@{}c@{}c@{}}
\includegraphics[width=0.5\columnwidth]{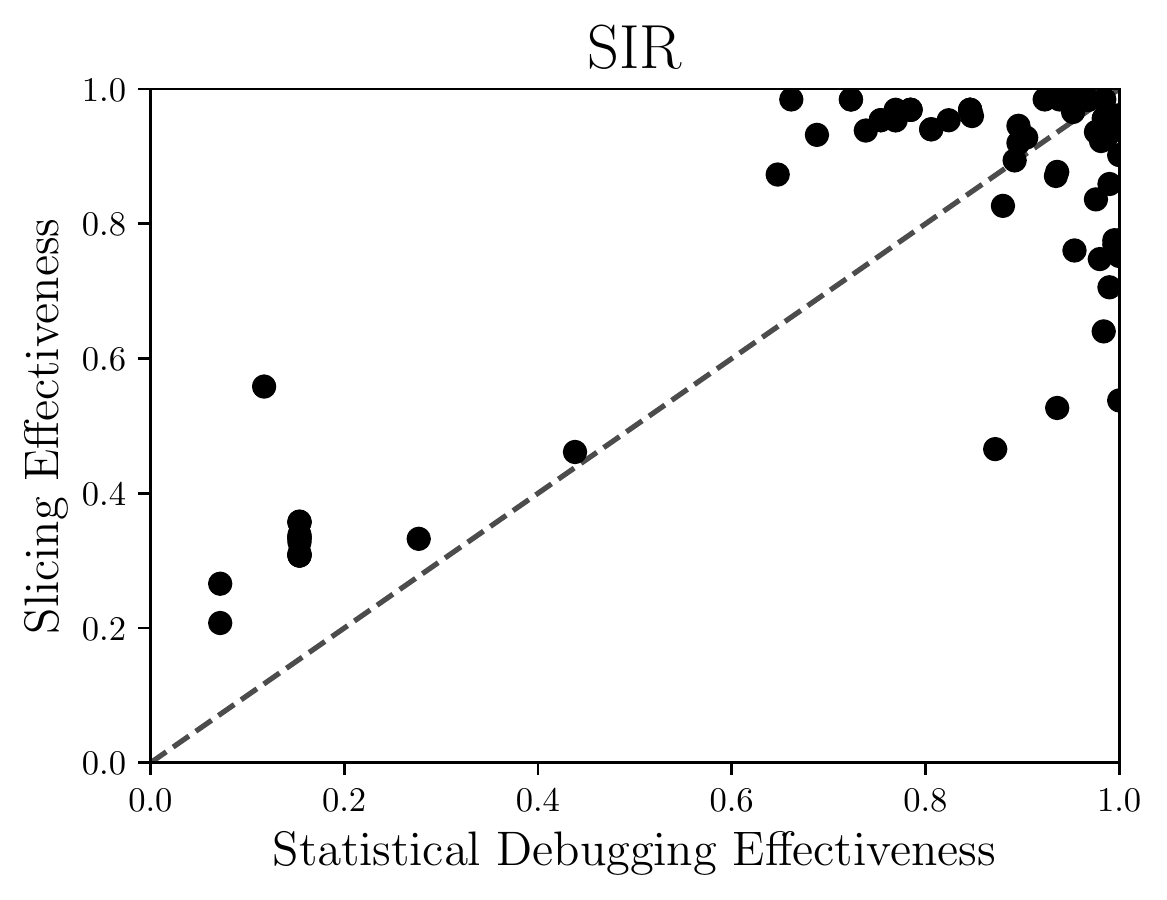} & 
\includegraphics[width=0.5\columnwidth]{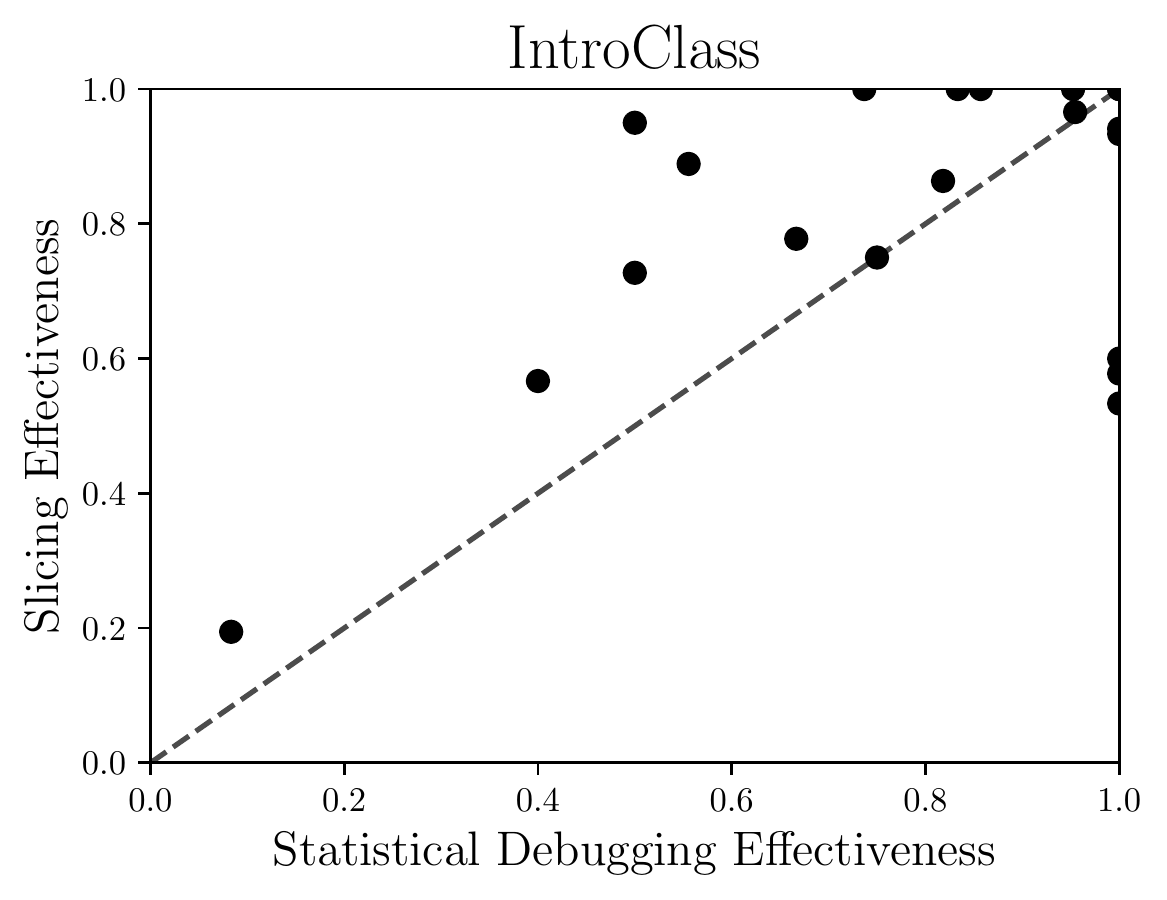} \\
\quad\quad\textbf{(a) SIR} &\quad\quad\textbf{(b) IntroClass }\\
\includegraphics[width=0.5\columnwidth]{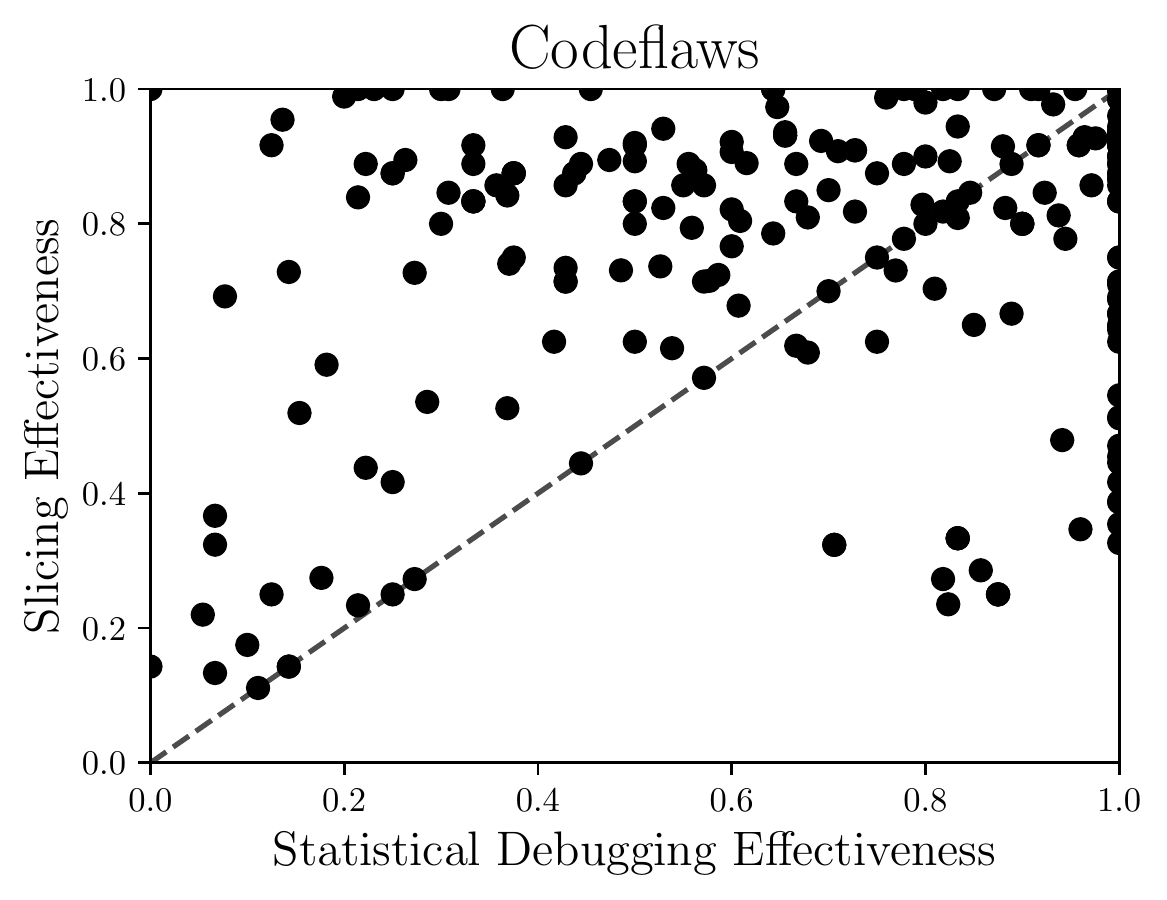} & 
\includegraphics[width=0.5\columnwidth]{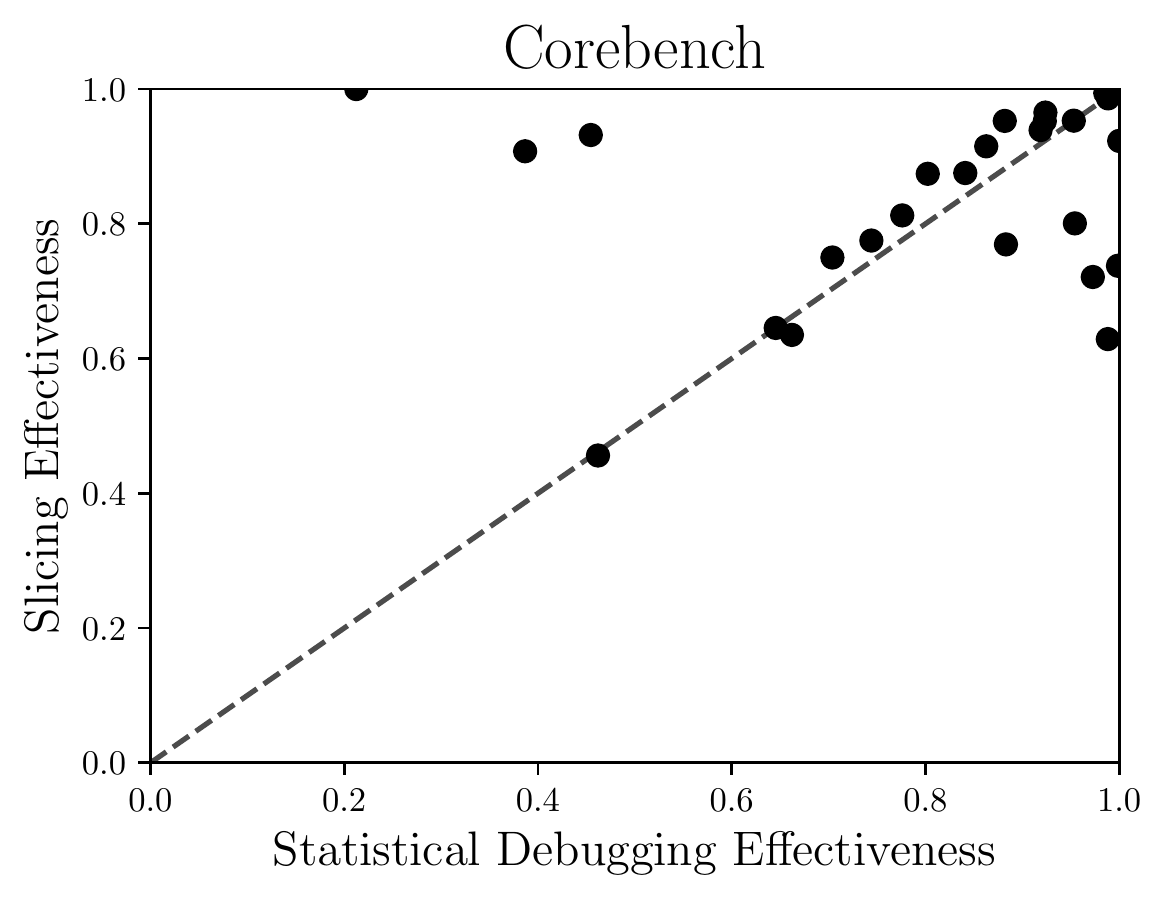} \\
\quad\quad\textbf{(c.) Codeflaws} &\quad\quad\textbf{(d.) \Coreb }\\
\end{tabular} 
\caption{Direct comparison of fault localization effectiveness between statistical debugging (Kulczynski2) and dynamic slicing (on single faults) in each benchmark}
\label{fig:meandiff}
\end{figure}

A tool or developer using \texttt{Kulczynski2} will locate the faulty statement after inspecting five to ten most suspicious statements. The faulty statement is ranked within the top five to ten most suspicious statements for most errors, i.e. 56\% to 73\% of all errors, respectively (\textit{cf. \autoref{tab:Kulczynski2-effectiveness}}). Overall, most programs (60\%) can be debugged by inspecting the top 30\% (58 LoC) of the suspicious statements reported by \texttt{Kulczynski2}. 

\begin{result}
  \textnormal{\texttt{Kulczynski2}} reports the faulty statements within the top 5--10 most suspicious statements for 56\%~to~72\% of all errors, respectively.  
\end{result}

\begin{table}[!tp]\centering 
\sisetup{round-mode=places,round-precision=4}
\caption{Statistical Tests for the most effective Statistical Debugging Formulas; \textit{Odds ratio}~$\psi$ (all ratios are statistically significant \textit{Mann-Whitney} $U$\textit{-test}$< 0.05$ for all tests)}
\begin{tabular}{|r |r r r |} \hline
&\multicolumn{3}{c|}{\textbf{Odds Ratio}~$\psi$ (Mann Whitney test score~$U$)} \\ 
\multicolumn{1}{|c|}{\textbf{Benchmark}} 
&\multicolumn{1}{c}{\texttt{Kulczynski2}} & \multicolumn{1}{c}{\texttt{Kulczynski2}} & \multicolumn{1}{c|}{\texttt{Kulczynski2}} \\ 
&\multicolumn{1}{c}{vs. \texttt{Ochiai}} & \multicolumn{1}{c}{vs. \texttt{Naish1}} & \multicolumn{1}{c|}{vs. \texttt{GP\_13}} \\ \hline  
\texttt{SIR}                & \num{0.298544} (\num{0.000158})  & \num{0.000416} (\num{0}) & \num{0.000416} (\num{0}) \\
\texttt{IntroClass}                  & \num{0.000595} (\num{0})  & \num{0.018262} (\num{0}) & \num{0.005917} (\num{0}) \\
\texttt{Codeflaws}                 & \num{0.00126} (\num{0})  & \num{0.000514} (\num{0}) & \num{0.000205} (\num{0}) \\
\texttt{\Coreb}                 & \num{0.000331} (\num{0})  & \num{0.000331} (\num{0}) & \num{0.000331} (\num{0}) \\ \hline
\textbf{All Bugs}  & \num{0.010638} (\num{0}) & \num{0.000556} (\num{0}) & \num{0.000229} (\num{0}) \\ \hline
\end{tabular}
\label{tab:stat-debug-oddstable}
\end{table}

\emph{Is the difference in the performance of \texttt{Kulczynski2} statistically significant, in comparison to the best performing formula for each statistical debugging family?} In our evaluation, the  difference in the performance of \texttt{Kulczynski2} (i.e. the best performing formula) is not statistically significant. \autoref{tab:stat-debug-oddstable} highlights the statistical tests comparing \texttt{Kulczynski2} to the best performing statistical formula in each family, i.e. \texttt{Kulczynski2} vs. \{\texttt{Ochiai}, \texttt{Naish1}, \texttt{GP13}\}. Notably, the performance of \texttt{Kulczynski2} is not statistically significant, in comparison to the best statistical formula for each family. This is evident from the fact that the \emph{odds ratio is less than one ($\psi < 1$) for all test comparisons} (\textit{see \autoref{tab:stat-debug-oddstable}}). This suggests that \texttt{Kulczynski2} has no statistically significant advantage over the best performing statistical formulas in each family; despite the fact that, in absolute terms, \texttt{Kulczynski2} outperforms the best formula in each family.

\begin{result}
\texttt{Kulczynski2} has no statistically significant advantage over the best formula in other SBFL families (i.e., \texttt{Ochiai}, \texttt{Naish1} and \texttt{GP13}). 
\end{result}

\begin{figure}[!tp]\centering 
\begin{tabular}{@{}c@{}c@{}}
\includegraphics[width=0.5\columnwidth]{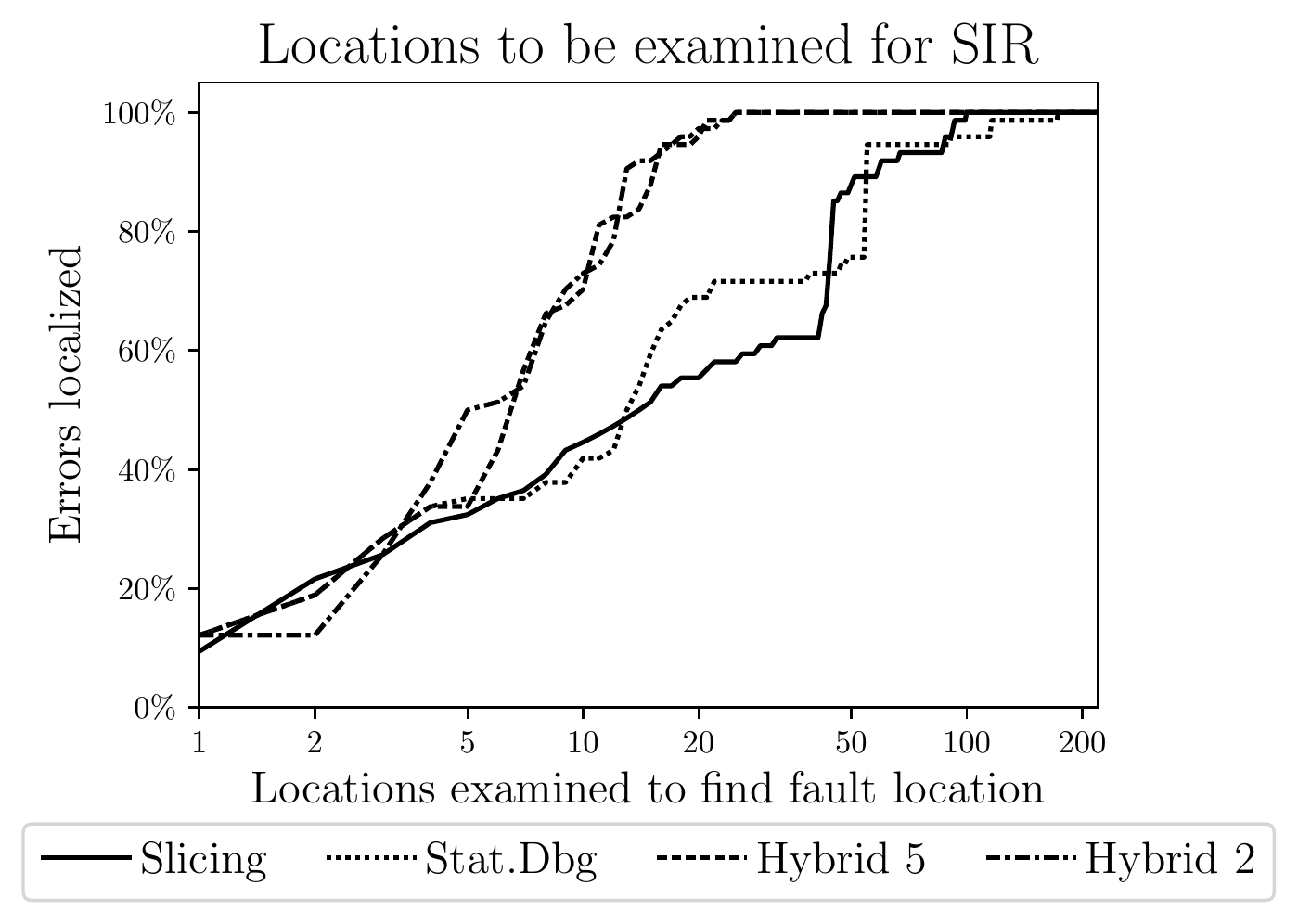} & 
\includegraphics[width=0.5\columnwidth]{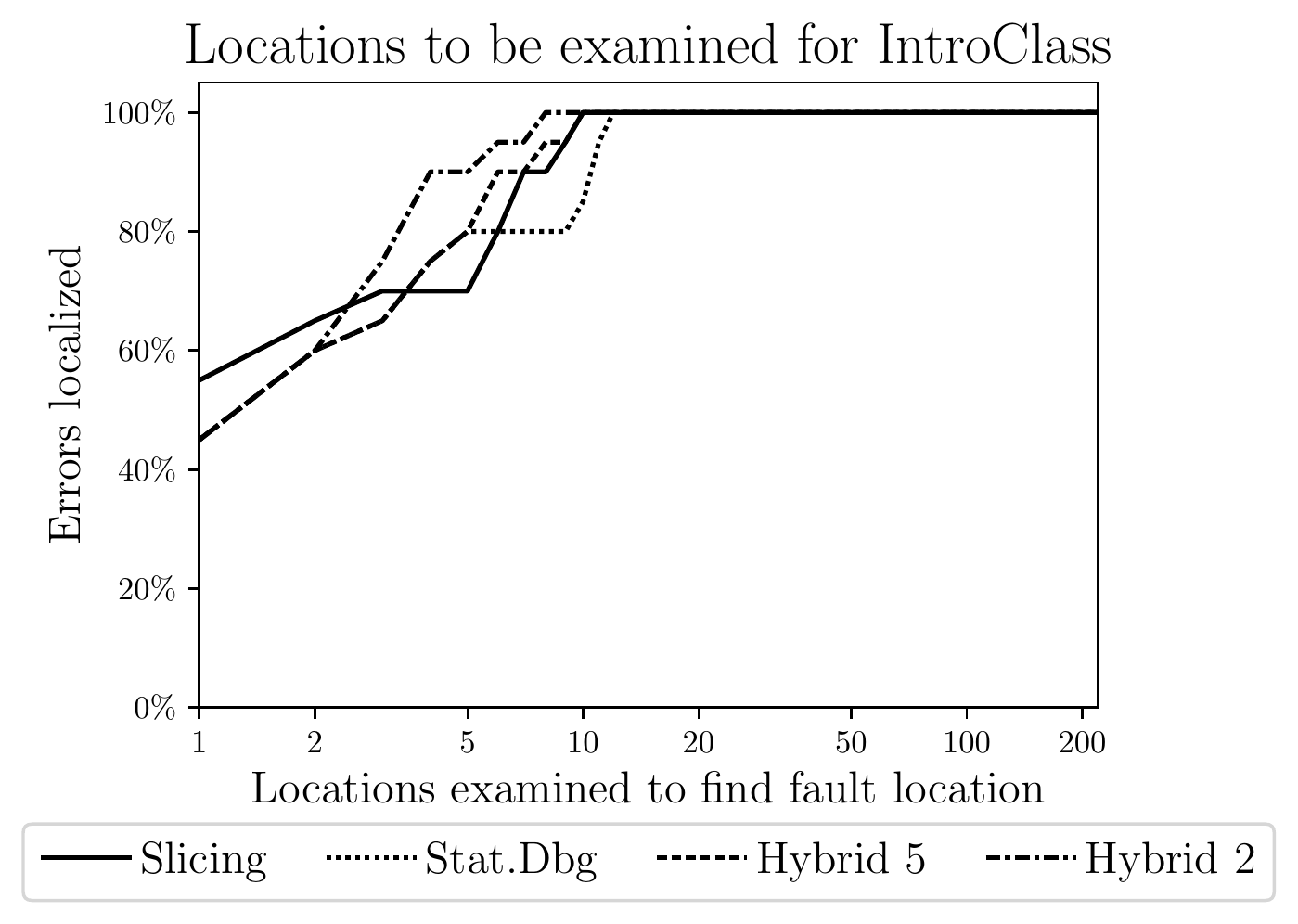} \\
\quad\quad\textbf{(a) SIR} &\quad\quad\textbf{(b) IntroClass }\\
\includegraphics[width=0.5\columnwidth]{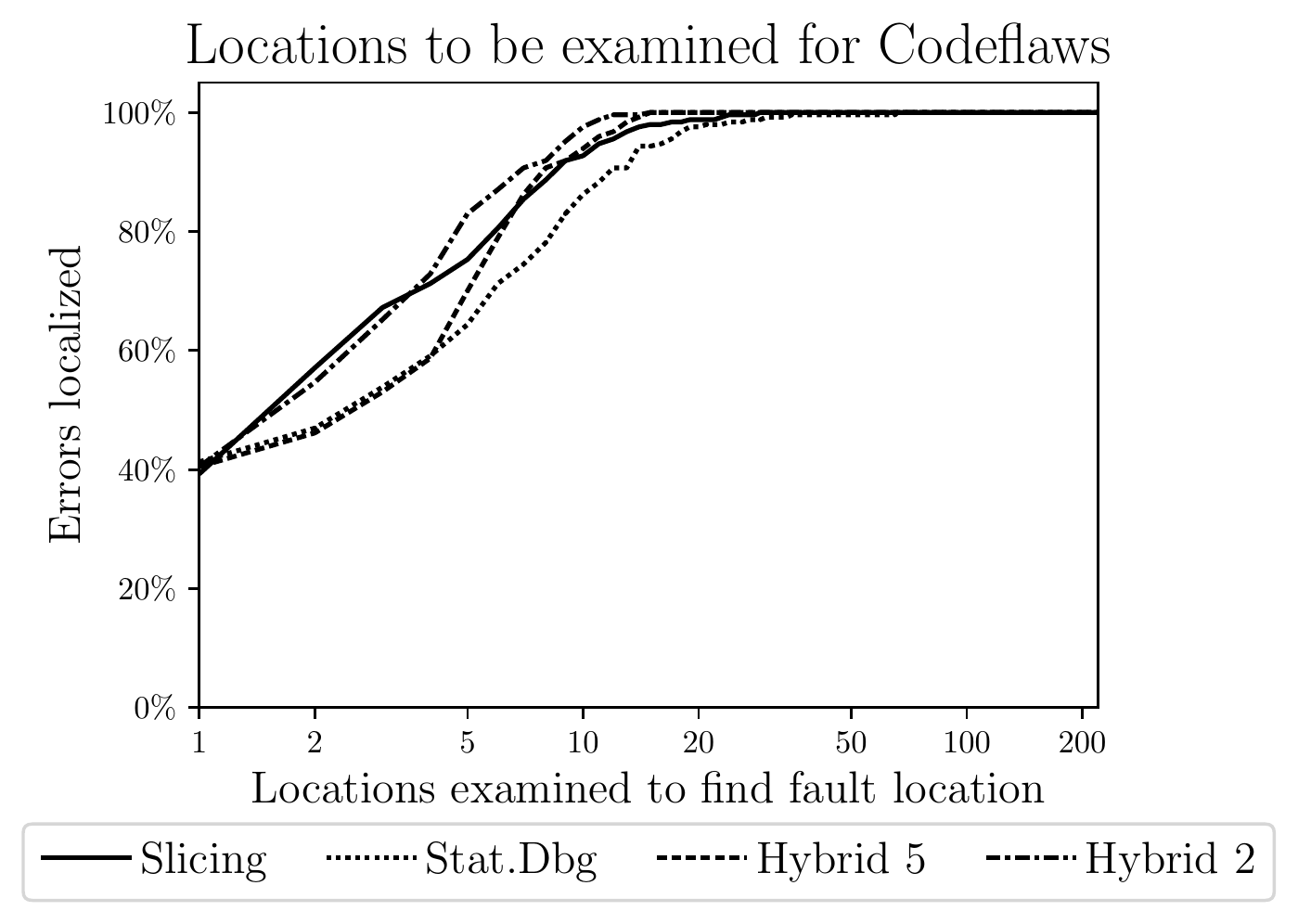} & 
\includegraphics[width=0.5\columnwidth]{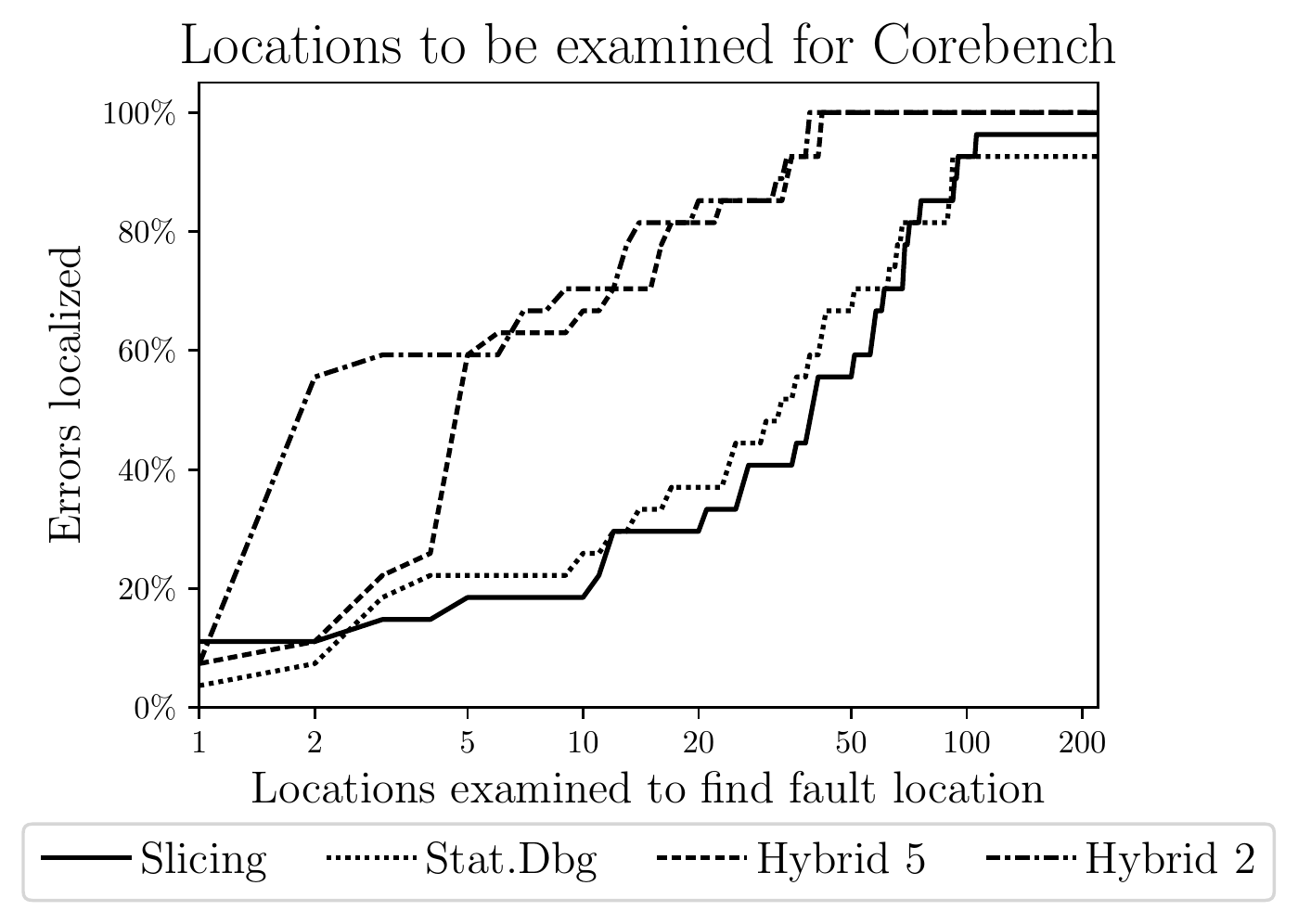} \\
\quad\quad\textbf{(c.) Codeflaws} &\quad\quad\textbf{(d.) \Coreb }\\
\end{tabular} 
\caption{Cumulative frequency of the locations to be examined, for dynamic slicing vs. statistical debugging vs. the hybrid approach (on single faults) in each benchmark }
\label{fig:effectiveness-by-benchmark}
\end{figure}


\subsection{RQ3: Comparing Statistical Debugging and Dynamic Slicing}
\emph{How effective is the most effective statistical formula in comparison to dynamic slicing?} We compare the performance of the most effective statistical formula (\texttt{Kulczynski2}) to that of dynamic slicing (\textit{cf. \autoref{fig:meandiff} and \autoref{fig:effectiveness-by-benchmark}}).

We find that, on average, \emph{dynamic slicing is more effective than statistical debugging at fault localization.} \emph{Slicing is about eight percentage points more effective than the best performing statistical formula} for all programs in our evaluation (\emph{cf. \autoref{fig:effectiveness-by-benchmark}, \autoref{tab:slicing-effectiveness} and \autoref{tab:Kulczynski2-effectiveness}}). For all errors in our study, a programmer using dynamic slicing 
needs to examine about three-quarters (78\%) of those statements that she would need to examine if she used statistical debugging.\footnote{Percentage improvement is measured as $\frac{1-0.794}{1-0.737}$. Note that \emph{score} by itself gives the number of statements that need \emph{not} be examined.} This result is independent of the type of errors or program. \autoref{fig:effectiveness-by-benchmark} shows that dynamic slicing consistently outperforms statistical debugging for each benchmark, with slicing consistently localizing all faults ahead of statistical debugging.

\begin{result}
Overall, dynamic slicing was eight percentage points more effective than the best performing statistical debugging formula, i.e. \textnormal{\texttt{Kulczynski2}}.
\end{result}

For two-third of bugs (66\%, 243 out of 369 errors), dynamic slicing will find the fault earlier than the best performing statistical debugging formula. \autoref{fig:meandiff} shows a direct comparison of the scores computed for slicing and statistical debugging. Each scatter plot shows for each error the effectiveness score of statistical debugging on the x-axis and the effectiveness score of slicing on the y-axis. Errors plotted above the diagonal line are better localized using dynamic slicing. For all benchmarks, the majority of the points are above the diagonal line which indicates that slicing outperforms statistical debugging in most cases. We can see that dynamic slicing consistently outperforms statistical debugging across all benchmarks.

\begin{result}
For two-third (66\%) of bugs, dynamic slicing locates the fault earlier than the best performing statistical debugging formula, i.e. \textnormal{\texttt{Kulczynski2}}.
\end{result}

To compare the significance of dynamic slicing and statistical debugging, we compute the odds ratio and conduct a Mann-Whitney $U$\textit{-test} (\emph{cf. Slicing vs. Kulczynski2 in \autoref{tab:oddstable}}). The odds ratio is in favor of dynamic slicing ($\psi>1$) for all projects. In particular, slicing is 62\% more likely to find a faulty statement earlier than statistical debugging, this likelihood is also statistically significant according to the Mann-Whitney test. The \emph{statistically significant odds ratio} is explained by slicing being more effective than statistical debugging in most cases. For instance, slicing is more effective than statistical debugging for 50 out of 74 bugs in the \texttt{SIR} benchmark and for 18 out of 27 bugs in \Coreb.

\begin{result}
Dynamic slicing is significantly more likely to find a faulty statement earlier than statistical debugging. 
\end{result}

\begin{table}[!tp]\centering 
\sisetup{round-mode=places,round-precision=4}
\caption{Statistical Tests for all three Fault Localization Techniques: Odds ratio~$\psi$ (Mann-Whitney $U$\textit{-test}  p-values~($U$) are in brackets), odds ratio with statistically significant p-values determined by Mann-Whitney~($U$\textit{-test} ) are in \textbf{bold}
}
\begin{tabular}{|r |r r r|}\hline
\multicolumn{1}{|c|}{\textbf{Benchmark}}&\multicolumn{3}{c|}{\textbf{Odds Ratio}~$\psi$ (Mann whitney test score)} \\ 
& \multicolumn{1}{c}{Slicing} &\multicolumn{1}{c}{Slicing} &\multicolumn{1}{c|}{Hybrid-2}\\  
& \multicolumn{1}{c}{vs. \texttt{Kulczynski2}} &\multicolumn{1}{c}{vs. Hybrid-2} &\multicolumn{1}{c|}{vs. \texttt{Kulczynski2}}\\ \hline  
\texttt{SIR}                & \textbf{4.25} (\num{0.00001}) & 0.81 (\num{0.256828}) & \textbf{5.44} (\num{0.000}) \\
\texttt{IntroClass}                 & 2.16 (\num{0.108716}) & 0.68 (\num{0.2713}) & 0.68 (\num{0.2713})\\
\texttt{Codeflaws}                & 1.16 (\num{0.209351}) & \textbf{0.41} (\num{0.000}) & 1.05 (\num{0.393843})\\
\texttt{\Coreb}                & 2.06 (\num{0.090419}) & \textbf{0.06} (\num{0.000002}) &\textbf{16.74} (\num{0.000002})  \\ \hline
\textbf{All Bugs}  & \textbf{1.62} (\num{0.000595}) & \textbf{0.42} (\num{0.000}) & \textbf{1.69} (\num{0.000203})\\ \hline
\end{tabular}
\label{tab:oddstable}
\end{table}

\subsection{RQ4: Sensitiveness of the Hybrid Approach}
\label{sec:hybrid-sensitiveness}
\emph{How many suspicious statements (reported by statistical debugging, i.e. Kulczynski2) should a tool or developer inspect before switching to slicing?} 
We examine the sensitiveness of the hybrid approach to varying absolute values of $N$. We evaluate how the number of suspicious statements 
inspected 
before switching to slicing influences the effectiveness of the hybrid approach. In particular, we investigated the effect of $N$ values (2, 5, 10, 15, 20) on the performance of the hybrid approach, in order to determine the optimal number of suspicious statements to inspect before switching to slicing. 

\begin{figure}[!tp]\centering 
\begin{tabular}{@{}c@{}c@{}}
\includegraphics[width=0.5\columnwidth]{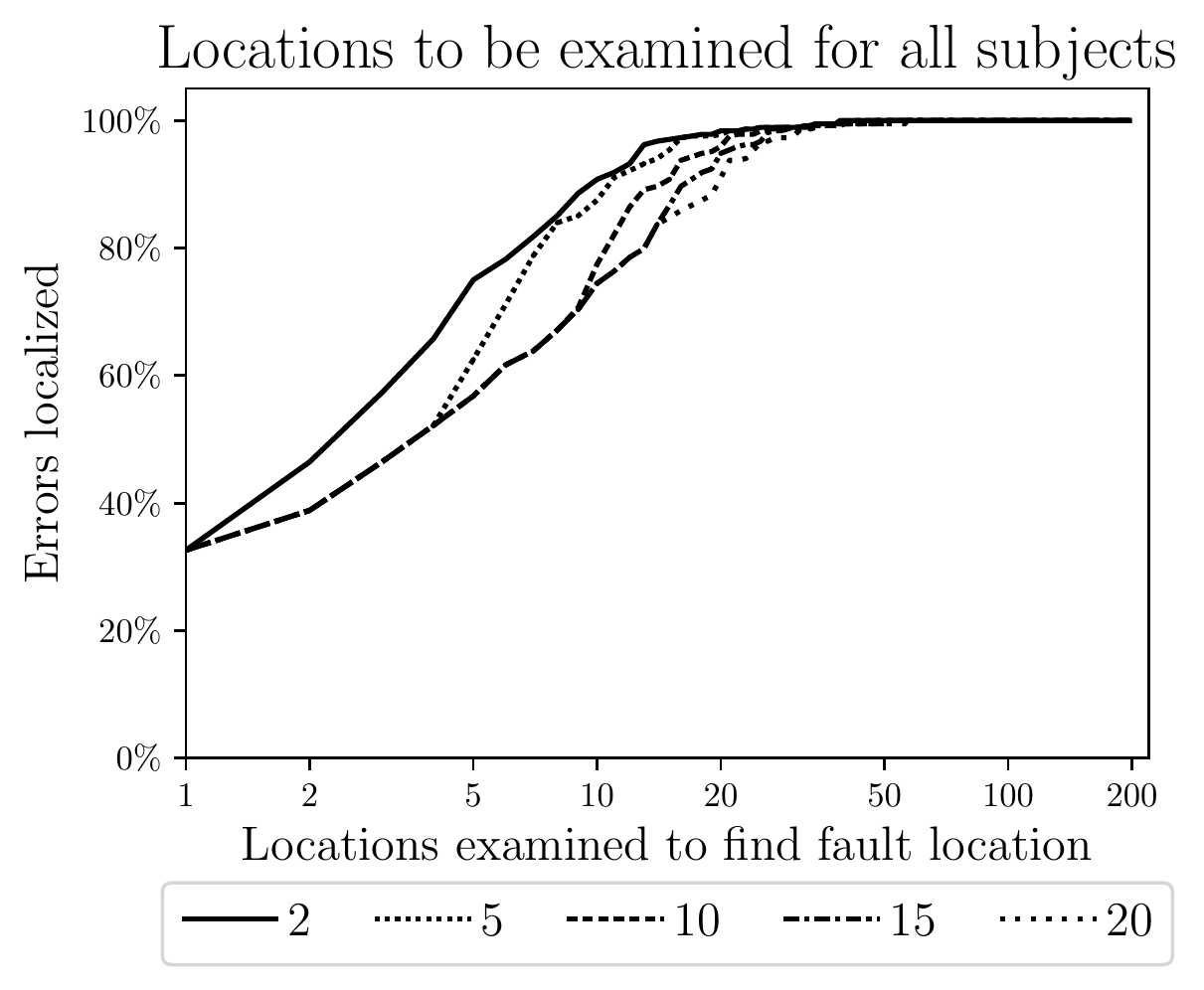} & 
\includegraphics[width=0.5\columnwidth]{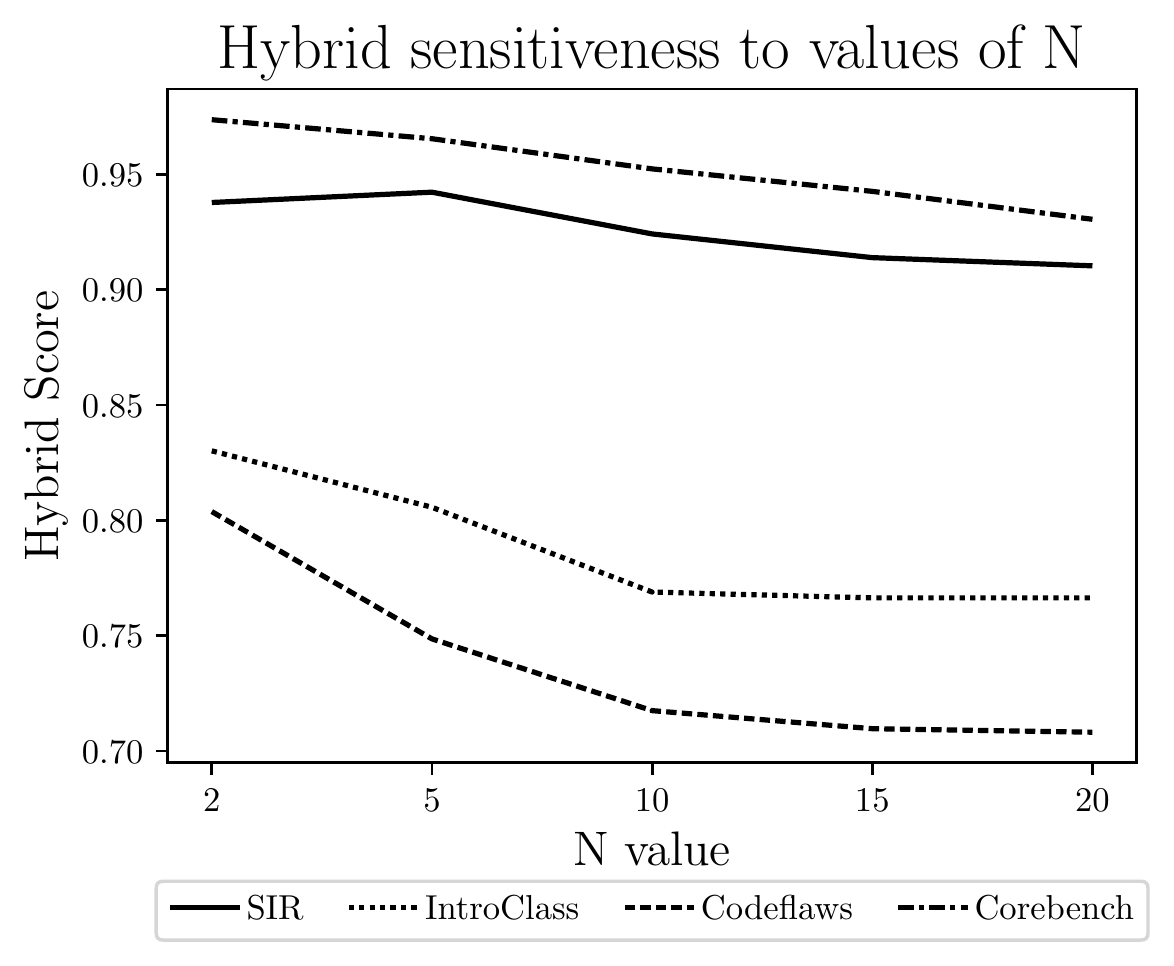} \\
\quad\quad\textbf{(a) Cumulative frequency} &\quad\quad\textbf{(b) Effectiveness Score }\\
\end{tabular} 
\caption{Hybrid sensitiveness to different values of $N\in\{2,5,10,15,20\}$ showing (a) the cumulative frequency of locations to be examined for all errors (left), and (b) the effectiveness score for each benchmark using the hybrid approach (right).}
\label{fig:hybrid_sensitiveness}
\end{figure}

\emph{A programmer that switches to slicing after investigating the top five most suspicious statements can localize more errors than if switching after investigating more suspicious statements.} \autoref{fig:hybrid_sensitiveness} shows the impact of other values of $N$ on the effectiveness of the hybrid approach. Note that the hybrid approach degenerates to dynamic slicing when $N=0$ and to statistical debugging when $N$ is large (e.g., program size). We see that Kochhar's suggestion of $N=5$ is a good value for our subjects, in particular inspecting at most five statements before switching to slicing outperforms both slicing and statistical debugging. As we see in \autoref{fig:comp-effectiveness}, the hybrid approach with $N=2$ and $N=5$ outperforms both slicing and statistical debugging (\texttt{Kulczynski2}). Hence, \emph{a developer is most effective 
if she inspects at most five most suspicious statements reported by statistical debugging before switching to slicing. }

A tool using our hybrid approach is most effective when inspecting only the top two most suspicious statements ($N=2$) reported by statistical debugging, before switching to slicing. Hence, we recommend the use of the hybrid approach (with $N=2$) for fault localization, and at most five suspicious statements should be inspected before switching to slicing.\footnote{Further evaluations of the \textit{hybrid approach} use the best values of $N$ (i.e. $N=2$ and $N=5$).}

\begin{result}
The hybrid approach is most effective when a programmer inspects at most two statements ($N=2$) before switching to slicing. 
\end{result}

\subsection{RQ5: Effectiveness of the Hybrid Approach}
\label{sec:hybrid-effectiveness} 
\begin{figure}[!tp]\centering 
\includegraphics[width=0.95\columnwidth]{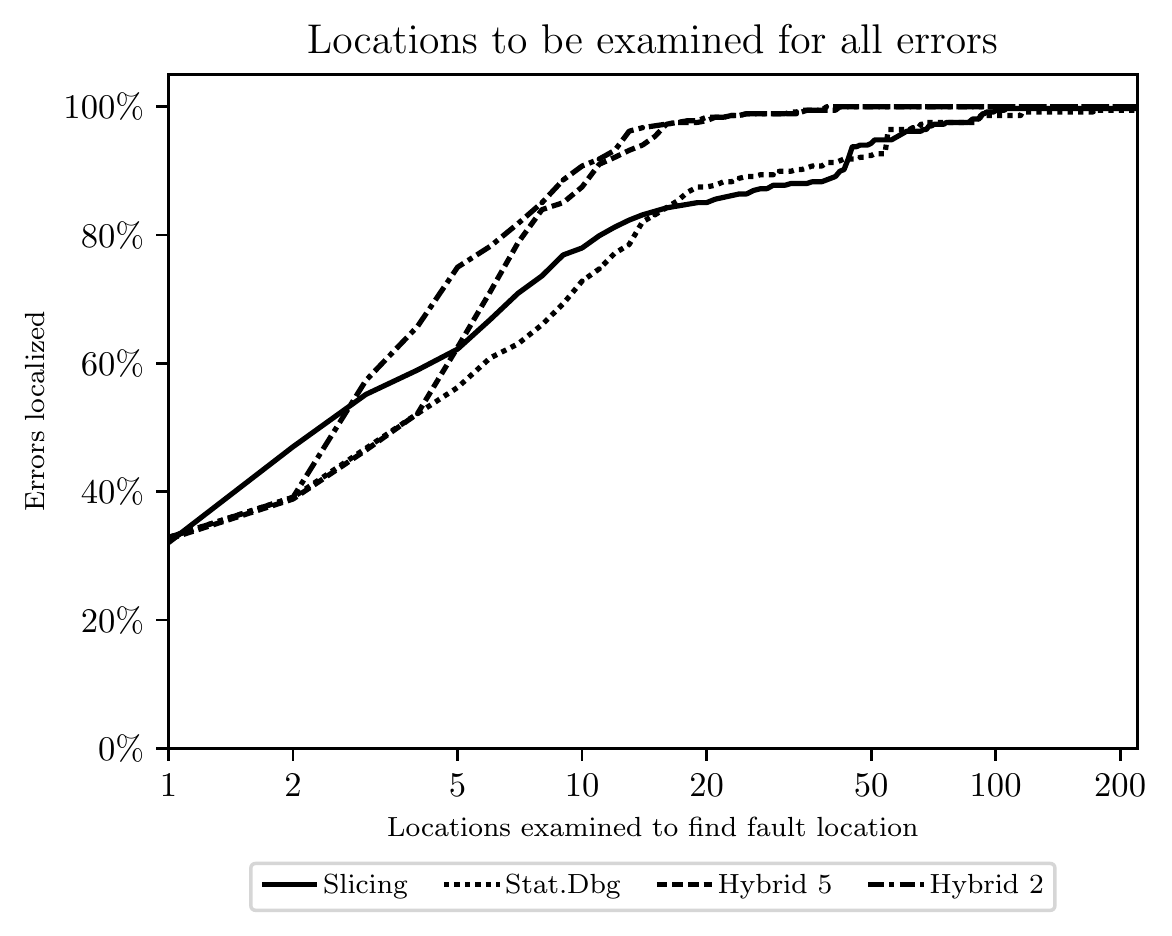}
\caption{Cumulative frequency of the locations to be examined for the hybrid approach, in comparison to statistical debugging (Kulczynski2) and dynamic slicing, for all (Single) Faults.
As expected, inspecting only the top two suspicious code locations, Hybrid-2  and statistical debugging perform similarly (localizing about 39\% of errors each); they are outperformed by dynamic slicing (47\% of errors localized). 
However, inspecting only the top five locations, Hybrid-2 clearly outperforms statistical debugging and dynamic slicing by localizing 75\% of errors, while slicing performs better than statistical debugging  (62\% vs. 56\% of errors).
}
\label{fig:comp-effectiveness}
\end{figure}

\emph{Which technique is the most effective in fault localization? Which technique is more likely to find fault locations earlier?} We now investigate the effectiveness of the hybrid approach, in comparison to slicing and statistical debugging. First, we examine the number of program statements that need to be inspected to localize all faults for each technique (\autoref{fig:comp-effectiveness}), as well as the absolute effectiveness score of each technique (\autoref{tab:slicing-effectiveness}, \autoref{tab:Kulczynski2-effectiveness} and \autoref{tab:Hybrid2-effectiveness}). Then, we evaluate the likelihood of each technique to find the fault locations earlier than the other two techniques (\autoref{tab:oddstable}).

Notably, if the programmer is willing to inspect no more than 20 statements, the hybrid approach will localize the fault location for almost all (98\%) of the bugs (\textit{cf. \autoref{tab:Hybrid2-effectiveness} and \autoref{fig:comp-effectiveness}}). In contrast, both statistical debugging and slicing can only localize almost all (98\%) faults after inspecting about five times as many statements, i.e. 100 LoC. 
In fact, if the programmer inspects 20 LoC, slicing and statistical debugging would find the fault location for about 85\% and 88\% of the bugs, respectively.

\begin{table}[!tp]\centering%
\scriptsize
\caption{Effectiveness of the Hybrid approach with $N=2$ (i.e., Hybrid-2)}
\begin{tabular}{| l | r | r | r | r | r | r | r | }
\hline
\multirow{4}{*}{\textbf{Benchmark}} &\multirow{4}{*}{\textbf{Score}}  &  \multicolumn{4}{c|}{\textbf{\% Errors Localized}}  \\ 
 & & \multicolumn{4}{c|}{if developer inspects \textbf{N}} \\ 
& & \multicolumn{4}{c|}{most suspicious LoC} \\ 
 &  &  \textbf{5} &  \textbf{10} &  \textbf{20} &  \textbf{30} \\
\hline
IntroClass  & 0.83 &  90.00 & 100 & 100 & 100 \\ 
 \hline
Codeflaws & 0.80 &  83.00 & 97.57 & 100 & 100  \\ 
\hline
\Coreb & 0.97 &  59.26 & 70.37 & 85.19 & 85.19 \\ 
\hline
\textbf{Real} & 0.83 & 81.29 & 95.24 & 98.64 & 98.64 \\  
\hline
\textbf{Artificial} (SIR)   & 0.94 &  50.00 & 72.97 & 97.30 & 100  \\ 
\hline
\textbf{Avg. (Bugs)} & 0.844 &   75.00 & 90.76 & 98.37 & 98.91 \\ 
\hline
\end{tabular}
\label{tab:Hybrid2-effectiveness}
\end{table}

\begin{result}
The hybrid approach can localize the fault location for almost all (98\%) of the bugs after inspecting no more than 20 LoC.
\end{result}

In absolute numbers, the hybrid approach is the most effective fault localization technique, followed by slicing, which is more effective than statistical debugging (see \autoref{tab:Hybrid2-effectiveness}, \autoref{fig:effectiveness-by-benchmark} and \autoref{fig:comp-effectiveness}). The hybrid approach ($N=2$) is about seven percentage points more effective than slicing, and about fifteen percentage points more effective than statistical debugging (\textit{cf. \autoref{tab:slicing-effectiveness}, \autoref{tab:Kulczynski2-effectiveness} and \autoref{tab:Hybrid2-effectiveness}}). Overall, it improves the performance of both slicing and statistical debugging. 
For instance, a programmer using the hybrid approach needs to examine about half (58\%) and three-quarter (75\%) of those statements that she would need to examine if she used slicing and statistical debugging, respectively.  

\begin{result}
The hybrid approach is significantly more effective than slicing and statistical debugging, respectively.
\end{result}

We compute the odds ratio and conduct a Mann-Whitney $U$\textit{-test}, in order to determine the significance of the hybrid approach. The odds ratio for all projects is strictly in favor of the hybrid approach ($\psi>1$ in \autoref{tab:oddstable}). Specifically, the hybrid approach is (69\%) more likely to find a faulty statement earlier than statistical debugging (\emph{cf. ``Hybrid-2 vs. Kulczynski2" in \autoref{tab:oddstable}}). Moreover, a programmer is (42\%) less likely to find the fault location early if she localizes with dynamic slicing instead of the hybrid approach (\emph{cf. ``Slicing vs. Hybrid-2" in \autoref{tab:oddstable}}). 

The \emph{statistically significant odds ratio} is explained by the hybrid approach being more effective than slicing and statistical debugging in most cases. \emph{The majority of bugs is best localized by the hybrid approach.} For more than half of the bugs (56\%, 208 out of 369 errors), the hybrid approach will find the fault earlier than both slicing and statistical debugging. In particular, for \Coreb, the hybrid approach is more effective than both techniques for 19 out of 27 bugs, as well as for 33 out of 74 bugs in \texttt{SIR}. 

\begin{result}
The hybrid approach is significantly more likely to find a faulty statement earlier than dynamic slicing and statistical debugging.
\end{result}

\subsection{RQ6: Real Errors vs. Artificial Errors}
In this section, we evaluate the effect of error type on the effectiveness of an AFL technique, in particular, the difference between evaluating AFL techniques on artificial errors (i.e., \texttt{SIR}\footnote{The \texttt{SIR} benchmark is the \emph{most used subject for the evaluation of AFL techniques}, especially statistical fault localization~\citep{tse}.}) versus real errors (i.e., \texttt{IntroClass}, \texttt{Codeflaws} and \texttt{\Coreb}). We examine the performance of each technique on each error type and portray the bias and differences in such evaluations. 
For real faults, we also evaluated the effect of program size on the effectiveness of an AFL technique. In particular, the difference in the performance of these AFL techniques on \textit{small programs} containing 16 to 18 executable LoC (i.e. \texttt{IntroClass} and \texttt{Codeflaws}) and \textit{large programs} with about 540 LoC (\texttt{\Coreb}), on average (\textit{see \autoref{tab:real-vs-artifical-faults}}). 
\autoref{fig:meandiff} and \autoref{fig:effectiveness-by-benchmark} highlight the difference between evaluating a fault localization technique on real or artificial errors. 
\autoref{tab:real-vs-artifical-faults} summarizes the difference in the effectiveness of each AFL technique when using real or artificial faults, as well as their performance on small or large programs. \autoref{tab:slicing-effectiveness}, \autoref{tab:Kulczynski2-effectiveness} and \autoref{tab:Hybrid2-effectiveness} also quantify the difference in the effectiveness of all three AFL techniques (i.e., dynamic slicing, statistical debugging and the hybrid approach, respectively) on real and artificial faults. 

\begin{table}[!tbp]\centering 
\sisetup{round-mode=places,round-precision=4}
\caption{
Effectiveness of AFL techniques on Real versus Artificial faults, as well as their effectiveness on small and large programs containing real faults (N/A =  Not applicable)
}
\begin{tabular}{|l| l |r r r |} \hline
 &  \multirow{3}{*}{\shortstack{\textbf{Program Size} \\ (Avg. \# LoC)}}  & \multicolumn{3}{c|}{\textbf{Percentage (\# LoC) of Statements}} \\   
\multicolumn{1}{|c|}{\textbf{Benchmark}}  &  &  \multicolumn{3}{c|}{\textbf{Inspected before locating fault}} \\  
&  &\multicolumn{1}{c}{Slicing} & \multicolumn{1}{c}{\texttt{Kulczynski2}} & \multicolumn{1}{c|}{Hybrid} \\ \hline  
\texttt{IntroClass} & small (15.5) & 17\% (3) & 24\% (4) & 17\% (3) \\
\texttt{Codeflaws} & small (17.7) & 23\% (4)  & 31\% (5) & 20\% (3) \\
\texttt{\Coreb} & large (540.4) & 15\% (80) & 21\% (112) & 3\% (14) \\ \hline
\textbf{Real} & N/A (200.3) & 18\% (29)  & 25\% (40) & 13\% (7) \\ \hline
\textbf{Artificial} (\texttt{SIR}) & N/A (148.1) & 21\% (31)  & 18\% (26) &  6\% (9) \\ \hline
\textbf{All Bugs} & N/A (193.5) & 21\% (40)  & 26\% (51) & 15\% (30) \\ \hline
\end{tabular}
\label{tab:real-vs-artifical-faults}
\end{table}

\emph{What is the most effective statistical debugging formula for artificial or real faults?} In our evaluation, the error type influences the effectiveness of a statistical debugging formula. \texttt{PattSim\_2} is the most effective statistical formula for artificial faults (\textit{score=0.85}), this is closely followed by \texttt{Ochiai} and \texttt{Naish\_1} with \textit{$score=0.83$} (\textit{see \autoref{tab:SBFL-Effectiveness}}). Meanwhile, for real faults, the most effective statistical formulas are \texttt{Kulczynski2} and \texttt{Tarantula} with scores 0.76, 0.70 and 0.79 for \texttt{IntroClass}, \texttt{Codeflaws} and \texttt{\Coreb}, respectively (\textit{see \autoref{tab:SBFL-Effectiveness}}). Notably, the most effective formula for artificial faults is not 
the most effective formula for real faults. This implies that the error type can influence the performance of an AFL technique. Thus, we recommended to always evaluate debugging aids using real faults.

\begin{result}
The performance of a statistical debugging formula depends on the error type: the most effective formula differs for artificial (\texttt{PattSim\_2}) and real faults  (\texttt{Kulczynski2} and \texttt{Tarantula}).
\end{result}

\emph{How does the effectiveness of statistical debugging compare to that of dynamic slicing, for artificial and real faults?} 
On one hand, dynamic slicing performs worse than statistical debugging on artificial faults (\textit{SIR}): A developer (or tool) has to inspect 21\% of the program to find the fault, in contrast to 18\% for \textit{Kulczynski2}, on average (\textit{see \autoref{tab:real-vs-artifical-faults}}). On the other hand, dynamic slicing performs better than statistical debugging on real faults (i.e., \texttt{IntroClass}, \texttt{Codeflaws} and \texttt{\Coreb}). For real errors, a developer has to inspect (7\%) less statements when using dynamic slicing (18\%) compared to slicing (25\%). Again, this shows that the error type has a significant influence on the effectiveness of an AFL technique. 

\begin{result}
Statistical debugging performs better on artificial faults, while dynamic slicing performs better on real faults.
\end{result}

\emph{What is the most effective AFL approach on artificial and real faults?}
The hybrid approach is the most effective AFL approach, outperforming both dynamic slicing and statistical debugging (\textit{see \autoref{tab:real-vs-artifical-faults}}). In particular,  depending on the error type, a developer or tool using the hybrid approach inspects one-third to less than three-quarter (0.3 to 0.7) of the statements inspected when using dynamic slicing or statistical debugging. This shows that the effectiveness of the hybrid approach is independent of error type. 

\begin{result}
The hybrid approach is the most effective approach, regardless of error type, i.e. artificial or real faults.
\end{result}

We observed that fault localization effectiveness on artificial errors does not predict results on real faults. In our evaluation, the performance of dynamic slicing and statistical debugging are different depending on the error type.  For instance, 
\autoref{tab:real-vs-artifical-faults} clearly shows that dynamic slicing performs better on real faults, while statistical debugging performs better on artificial faults. This result illustrates that the performance of an AFL technique on artificial faults is not predictive of its performance in practice. Hence, it is pertinent to evaluate AFL techniques on real faults rather than artificial faults, this is in line with the findings of previous studies~\citep{pearson2017evaluating}.

\begin{result}
The effectiveness of an AFL technique on artificial faults does not predict its effectiveness on real faults.
\end{result}

\emph{Does program size affect the effectiveness of individual statistical debugging formulas?
Does the most effective SBFL formula vary as program size varies, i.e. small versus large programs?} 
Among real faults, the most effective SBFL formula depends on the program size. 
Evidently, the most effective formula for small programs is not the most effective formula for large programs: For instance, \texttt{Tarantula} and \texttt{Kulczynski2} performed significantly better than \texttt{DStar} on small programs, but \texttt{DStar} was the most effective formula for large programs.\footnote{\autoref{tab:SBFL-Effectiveness} shows that \texttt{Tarantula} and \texttt{Kulczynski2} performed best on small programs with effectiveness scores 0.76 and 0.70 for \texttt{IntroClass} and \texttt{Codeflaws}, respectively. 
Despite the fact that \texttt{DStar} performs best on large programs (i.e. \Coreb) by slightly outperforming \texttt{Tarantula} and \texttt{Kulczynski2} (0.80 vs. 0.79); it performed significantly worse than \texttt{Tarantula} and \texttt{Kulczynski2} on small programs, with effectiveness score 0.62 and 0.56 for \texttt{IntroClass} and \texttt{Codeflaws}, respectively.
} 
Generally, the effectiveness of individual SBFL formulas varies as program size varies. Even though some SBFL formulas performed consistently well across program sizes (e.g. \texttt{Tarantula} and \texttt{Kulczynski2}), others are specialized for specific program sizes, e.g. \texttt{DStar} performs better on large programs (\Coreb). These results suggest that program size influences the effectiveness of individual statistical debugging techniques. 

\begin{result}
The effectiveness of individual SBFL formulas varies as program sizes varies: 
\texttt{Tarantula} and \texttt{Kulczynski2} are the most effective formulas for small programs, but \texttt{DStar} is the most effective formula for large programs.
\end{result}

\emph{
Does the comparative effectiveness of our AFL techniques (i.e., Hybrid vs. Slicing vs. SBFL)  vary as program sizes varies?}
For real faults, we investigated if there is a difference in the comparative effectiveness of our AFL techniques on small or large programs. Among real faults, the most effective technique is the same across program sizes (\textit{see \autoref{fig:meandiff} and \autoref{tab:real-vs-artifical-faults}}): Consistently, the hybrid approach performs best and slicing outperforms statistical debugging, regardless of program size. This observation holds across program sizes, for all three AFL techniques (\textit{see \autoref{fig:effectiveness-by-benchmark}}). These results suggest that program size does not influence the comparative effectiveness of these techniques. In fact, the comparative effectiveness of these AFL techniques is predictable across program sizes; the hybrid approach performs best, followed by slicing then statistical debugging (i.e., \texttt{Kulczynski2}). 

\begin{result}
For real faults, the comparative effectiveness of our AFL techniques is predictable across program sizes; the hybrid approach performs best, followed by slicing then statistical debugging.
\end{result}

\begin{table}[!htp]\centering
\caption{
Effectiveness of all AFL techniques on Single and Multiple Faults for \texttt{SIR} and \texttt{IntroClass} benchmarks. Single Fault Scores are in italics and bracketed, i.e. (\textit{Single}), while Multiple Fault Scores are in normal text. For multiple faults, the best scores for each (sub)category are in \textbf{bold}; higher scores are better. For instance, \texttt{Tarantula} is the best performing (popular) statistical debugging formula for \emph{all programs with multiple faults with score $0.8269$}, on average. 
} 
\begin{tabular}{| c | l | c | c | c | c |}
\hline
 \multirow{3}{*}{\shortstack{\textbf{AFL}\\ \textbf{Technique}}}  & \multirow{3}{*}{\shortstack{\textbf{Formula/}\\ \textbf{Approach}}} & \multirow{3}{*}{\shortstack{\texttt{SIR}\\MULT \\ (\textit{Single)}}} & \multirow{3}{*}{\shortstack{\texttt{IntroClass}\\MULT \\(\textit{Single)}}} & \multicolumn{2}{c|}{\textbf{Average} (Mean)} \\ 
& & &  &\multirow{2}{*}{\textbf{Programs}}   &\multirow{2}{*}{\textbf{Bugs}}    \\
& & &  &  &  \\
\hline
\multirow{6}{*}{\shortstack{Popular \\SBFL}}
 & \multirow{2}{*}{\texttt{Tarantula}} & \textbf{0.8214 }&  \textbf{0.8324} & \textbf{0.8269}  & \textbf{0.7878} \\ 
   & &  (\textit{0.6907}) & (\textit{0.7464}) & (\textit{0.7186}) & (\textit{0.6266}) \\    \cline{2-6}
 & \multirow{2}{*}{\texttt{Ochiai}} & 0.7796 & 0.7747  & 0.7772  & 0.7560 \\ 
   &  &  (\textit{0.7337}) & (\textit{0.7464}) & (\textit{0.7401}) & (\textit{0.6514}) \\    \cline{2-6}
 & \multirow{2}{*}{\texttt{Jaccard}} & 0.7720 & 0.7747 & 0.7734  & 0.7532 \\ 
  &  & (\textit{0.7029}) & (\textit{0.7464}) & (\textit{0.7247}) & (\textit{0.6342}) \\    \cline{2-6}
 \hline
\multirow{14}{*}{\shortstack{Human \\Generated \\SBFL}} 
 & \multirow{2}{*}{\texttt{Naish\_1}} &  0.7215 & 0.7638 & 0.7426 & 0.7136 \\ 
   &  &  (\textit{0.7399}) & (\textit{0.7448}) & (\textit{0.7423}) & (\textit{0.6682}) \\    \cline{2-6}
 & \multirow{2}{*}{\texttt{Naish\_2}} &  0.7484 & \textbf{0.7747} & 0.7615 & 0.7404 \\ 
   &  &  (\textit{0.7207}) & (\textit{0.7464}) & (\textit{0.7336}) & (\textit{0.6596}) \\    \cline{2-6}
 & \multirow{2}{*}{\texttt{Russel\_Rao}} & 0.7350 & 0.7586 & 0.7468 & 0.7185\\ 
   &  &  (\textit{0.6088}) & (\textit{0.6394}) & (\textit{0.6241}) & (\textit{0.6051}) \\    \cline{2-6}
 & \multirow{2}{*}{\texttt{Binary}} &  0.7125 & 0.7553 & 0.7339 & 0.6975 \\  
   &  & (\textit{0.6283}) & (\textit{0.6378}) & (\textit{0.633}) & (\textit{0.6129}) \\    \cline{2-6}
 & \multirow{2}{*}{\texttt{Wong\_1}} & 0.7350 & 0.7586 & 0.7468 & 0.7185\\ 
   &  & (\textit{0.6088}) & (\textit{0.6394}) & (\textit{0.6241}) & (\textit{0.6051})  \\    \cline{2-6}
  & \multirow{2}{*}{\texttt{$D^2$}} & \textbf{ 0.7718} & \textbf{0.7747} & 0.\textbf{7732} & 0.\textbf{7553}  \\  
    &  & (\textit{0.7345}) & (\textit{0.7464}) & (\textit{0.7404}) & (\textit{0.6523}) \\    \cline{2-6}
  &\multirow{2}{*}{\texttt{$D^3$}} & 0.7680 & \textbf{0.7747} & 0.7714 & \textbf{0.7553}\\ 
    &  &  (\textit{0.7484}) & (\textit{0.7464}) & (\textit{0.7474}) & (\textit{0.6611})  \\    \cline{2-6}
\hline
\multirow{8}{*}{\shortstack{GP \\Evolved \\SBFL}} 
 & \multirow{2}{*}{\texttt{GP\_02}} & \textbf{0.7633} & \textbf{0.7747} & \textbf{0.7690 }& \textbf{0.7498}  \\  
     &  &  (\textit{0.6725}) & (\textit{0.7157}) & (\textit{0.6941}) & (\textit{0.6133}) \\    \cline{2-6}
 & \multirow{2}{*}{\texttt{GP\_03}} & 0.7473 & \textbf{0.7747} & 0.7610 & 0.7402 \\  
     &  &  (\textit{0.6813}) & (\textit{0.6169}) & (\textit{0.6491}) & (\textit{0.6285})  \\    \cline{2-6}
 &\multirow{2}{*}{\texttt{GP\_13}} &0.7456 & \textbf{0.7747} & 0.7602 & 0.7398 \\ 
     &  &  (\textit{0.7211}) & (\textit{0.7464}) & (\textit{0.7338}) & (\textit{0.6606})  \\     \cline{2-6}
 & \multirow{2}{*}{\texttt{GP\_19}} & 0.7629 & 0.7558 & 0.7593 & 0.7279 \\ 
     &  & (\textit{0.4673}) & (\textit{0.6237}) & (\textit{0.5455}) & (\textit{0.4876})  \\     \cline{2-6}
 \hline
\multirow{8}{*}{\shortstack{Single Bug \\Optimal\\SBFL}}
 & \multirow{2}{*}{\texttt{PattSim\_2}} & 0.7608 & 0.7747 & 0.7677 & 0.7301 \\ 
     &  & (\textit{0.7537}) & (\textit{0.6544}) & (\textit{0.704}) & (\textit{0.6506}) \\     \cline{2-6}
 & \multirow{2}{*}{\texttt{lex\_Ochiai}} & 0.7532 & 0.7747 & 0.7640 & 0.7396  \\  
     &  &  (\textit{0.7356}) & (\textit{0.7464}) & (\textit{0.741}) & (\textit{0.6646})  \\    \cline{2-6} 
 & \multirow{2}{*}{\texttt{m9185}} & \textbf{0.8183 }& \textbf{0.8315} &\textbf{ 0.8249} & \textbf{0.7664} \\   
     &  & (\textit{0.7570}) & (\textit{0.7225}) & (\textit{0.7397}) & (\textit{0.6646})  \\ 
        \cline{2-6}
 & \multirow{2}{*}{\texttt{Kulczynski2} }& 0.7885 & 0.7747 & 0.7816 & 0.7588 \\ 
     &  & (\textit{0.7572}) & (\textit{0.7464}) & (\textit{0.7518}) & (\textit{0.6689}) \\ 
 \hline		
\multirow{2}{*}{\shortstack{Program\\ Slicing}}  & \multirow{2}{*}{\shortstack{\texttt{Dynamic} \\\texttt{Slicing}}}   & \multirow{1}{*}{0.8357} & \multirow{1}{*}{0.5487} & \multirow{1}{*}{0.6922} & \multirow{1}{*}{0.7840} \\ 
& & (\textit{0.7935})  & (\textit{0.8602}) & (\textit{0.8269})  & (\textit{0.7535}) \\ 
 \hline		
\multirow{4}{*}{\shortstack{Hybrid\\ Approach}}  & \multirow{2}{*}{\texttt{Hybrid-2}} & \textbf{0.9627 }& \textbf{0.9057} & \textbf{0.9342} & \textbf{0.9457}  \\ 
    &  &  (\textit{0.9358}) & (\textit{0.888}) & (\textit{0.9119}) & (\textit{0.9237})\\ 
      \cline{2-6}
  & \multirow{2}{*}{\texttt{Hybrid-5}} &  0.9505 & 0.8406 & 0.8955 & 0.9206 \\ 
     &  & (\textit{0.9397}) & (\textit{0.814}) & (\textit{0.8768}) & (\textit{0.8974}) \\

 \hline		
\end{tabular}
\label{tab:mult-SBFL-Effectiveness}
\end{table}

\begin{figure}[!tp]\centering 
\begin{tabular}{@{}c@{}c@{}}
\includegraphics[width=0.5\columnwidth]{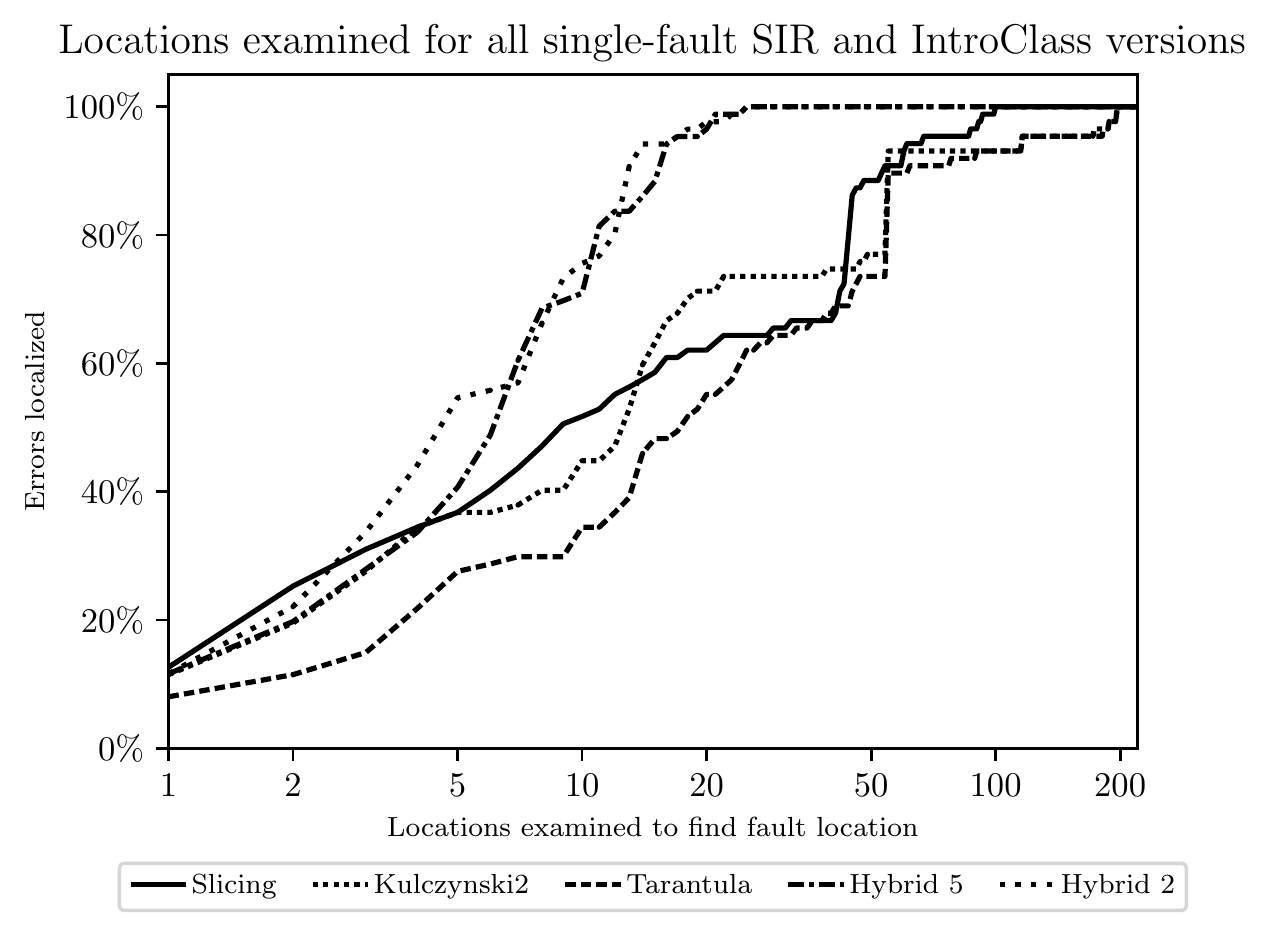} &
\includegraphics[width=0.5\columnwidth]{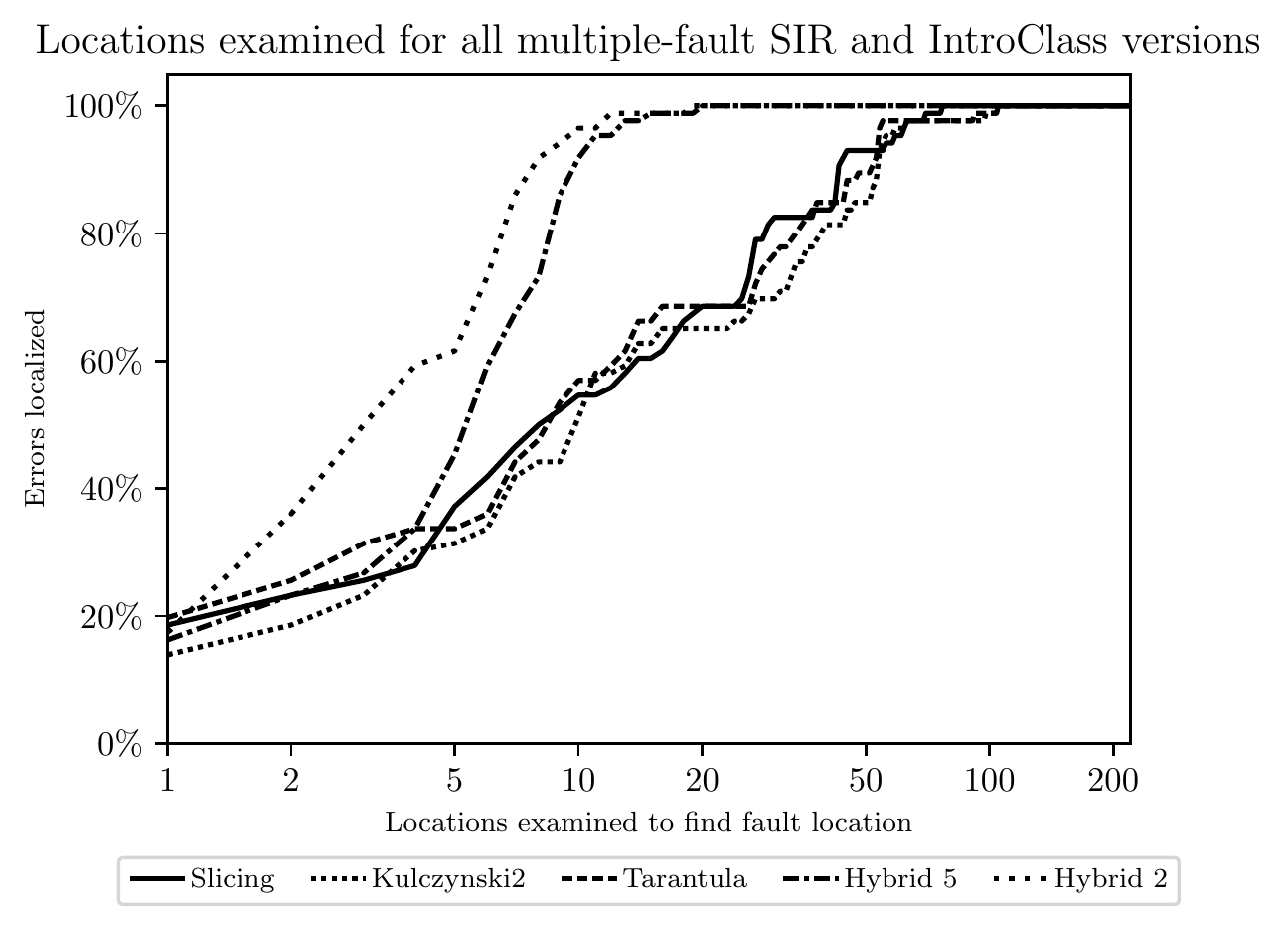} \\
\quad\quad\textbf{(a) Single Faults} &\quad\quad\textbf{(b) Multiple Faults}\\
\end{tabular} 
\caption{
Cumulative frequency of the locations to be examined for (a) single-fault and (b) multiple-fault versions of the \texttt{SIR} and \texttt{IntroClass}, using the hybrid approach, statistical debugging (\texttt{Kulczynski2} and \texttt{Tarantula}) and dynamic slicing
}
\label{fig:mult-comp-effectiveness}
\end{figure}

\subsection{RQ7: Single Fault vs. Multiple Faults}
\begin{figure}[!tp]\centering 
\begin{tabular}{@{}c@{}c@{}}
\includegraphics[width=0.5\columnwidth]{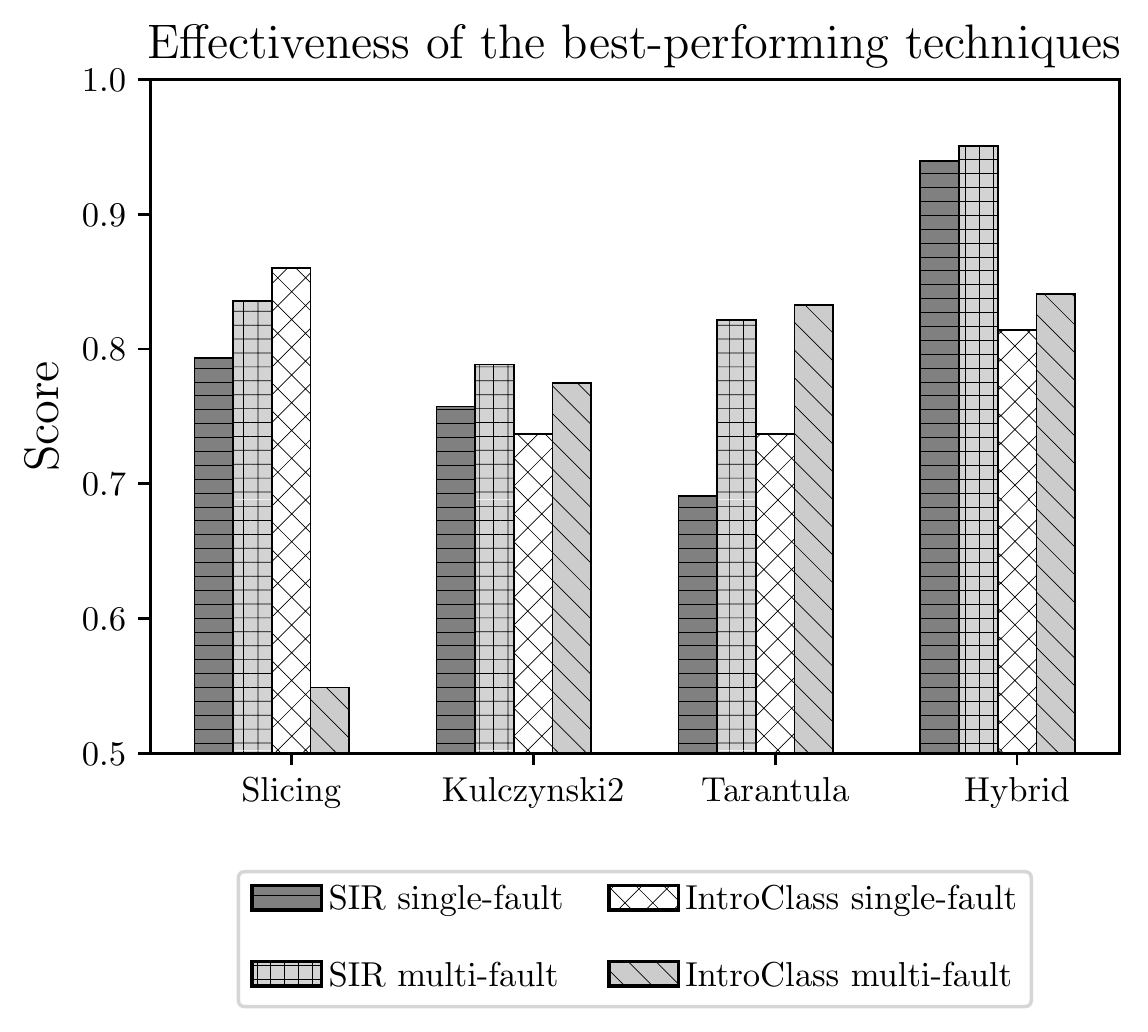} &
\includegraphics[width=0.5\columnwidth]{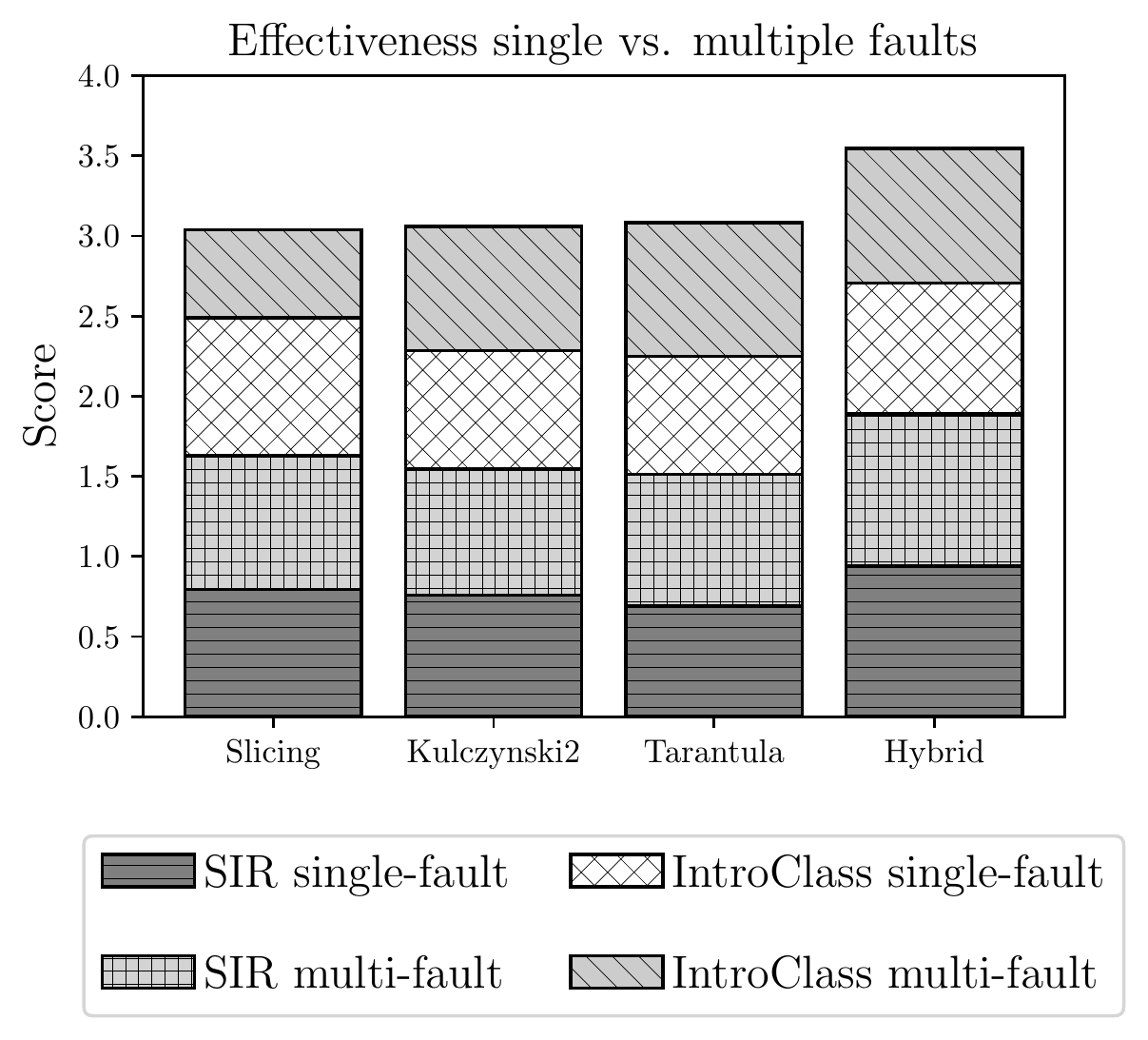} 
\\
\quad\quad\textbf{(a) Each Benchmark} &\quad\quad\textbf{(b) Cumulative Score}\\
\end{tabular} 
\caption{Effectiveness of each technique for Single and Multiple Fault(s) in \texttt{SIR} and \texttt{IntroClass}: (a) Scores for each benchmark and (b) Scores for both benchmarks}
\label{fig:bar-charts-comp-effectiveness}
\end{figure}

\begin{figure}[!tp]\centering 
\includegraphics[width=0.95\columnwidth]{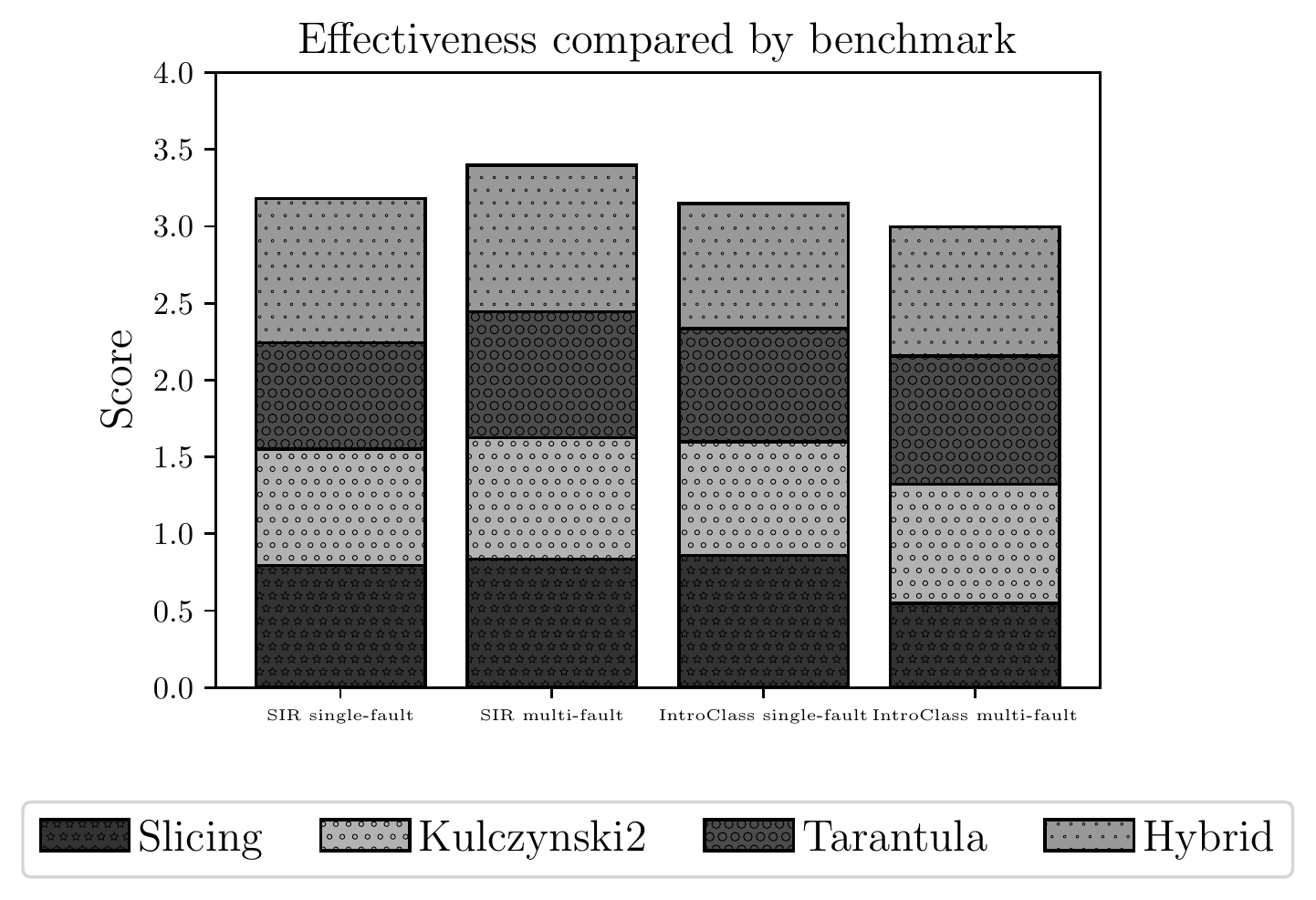}
\caption{
Effectiveness of each technique on Single and Multiple Faults compared by benchmark, i.e., \texttt{SIR} and \texttt{IntroClass}}
\label{fig:benchmark-comp-effectiveness}
\end{figure}

In  this section, we compare the effectiveness of all three AFL techniques on programs with \emph{multiple faults}. Then, we examine the effect of multiple faults on the performance of each technique and the difference between evaluating an AFL technique on single or multiple fault(s). In this experiment, we employ the original single-fault versions of the \texttt{SIR} and \texttt{IntroClass} benchmarks, as well as the multiple-fault versions of the same benchmarks, called \texttt{SIR-MULT} and \texttt{IntroClass-MULT}, respectively. \autoref{tab:mult-SBFL-Effectiveness} highlights the results for single and multiple fault(s) for all AFL techniques, including statistical debugging, hybrid and dynamic slicing. \autoref{fig:mult-comp-effectiveness}, \autoref{fig:bar-charts-comp-effectiveness} and \autoref{fig:benchmark-comp-effectiveness} illustrate the difference in the performance of each technique when given programs with a single fault or multiple faults.

\emph{What is the most effective statistical debugging formula for multiple faults?} 
In our evaluation, the most effective SBFL formula for multiple faults is \texttt{Tarantula} (0.8269), from the \textit{popular} SBFL family. It outperforms the other SBFL formulas (\emph{cf. \autoref{tab:mult-SBFL-Effectiveness} and \autoref{fig:bar-charts-comp-effectiveness}}). For the other statistical debugging families, 
the most effective formula for multiple faults are \texttt{DStar} ($D^2$ and $D^3$), \texttt{GP02} and \texttt{m9185} for the \textit{popular}, \textit{human-generated} and \textit{genetically evolved} families, respectively (\emph{cf. \autoref{tab:mult-SBFL-Effectiveness}}). The performance of \texttt{Tarantula} is closely followed by that of the single-bug optimal formulas \texttt{m9185} (0.8249). However, the difference in the performance of \texttt{m9185} and \texttt{Tarantula} is not statistically significant, i.e. $\psi<1$ (\emph{odds ratio $\psi=0.14$, Mann-Whitney $U$\textit{-test} p-value $U=0$}). Notably, the most effective single-bug optimal formulas (i.e. \texttt{m9185}) outperformed the human-generated and genetically evolved formulas (\emph{cf. \autoref{tab:mult-SBFL-Effectiveness}}). This illustrates that \textit{single bug optimal} formulas are also effective for multiple faults, despite being specialized for single faults.

\begin{result}
\texttt{Tarantula} is the most effective statistical debugging formula for multiple faults; it outperforms all other statistical debugging formulas.
\end{result}

\emph{For multiple faults, how does the effectiveness of statistical debugging (\texttt{Tara-\\ntula}) compare to that of dynamic slicing and hybrid?} \texttt{Tarantula} performs better than dynamic slicing ($0.8269$ vs. $0.6922$) for multiple faults, our results show that the effectiveness of slicing is 16\% worse than that of \texttt{Tarantula} on multiple faults in the \texttt{SIR-MULT} and \texttt{IntroClass-MULT} programs (\emph{see \autoref{tab:mult-SBFL-Effectiveness}}). 
This is despite the fact that dynamic slicing ($0.8269$) outperforms \texttt{Tarantula} ($0.7186$) by 15\% on single fault programs (i.e., in \texttt{SIR} and \texttt{IntroCla\-ss} benchmarks). 
Indeed, there is a 13\% decrease in the performance of dynamic slicing on multiple faults. 
This is evident in \autoref{fig:bar-charts-comp-effectiveness} (a) where the performance of dynamic slicing drops for multiple faults for \texttt{IntroClass-MULT}. 
This shows that it is beneficial for an AFL technique to employ coverage data from (numerous) failing test cases when diagnosing 
programs with multiple faults. As expected, it is more difficult for dynamic slicing to diagnose multiple faults: Since a dynamic slice is constructed for only a single failing test case, it is difficult to account for the effect of multiple faults. Overall, the performance of the hybrid approach remains superior to that of dynamic slicing and statistical debugging, regardless of the number of faults present in the program (\emph{cf. \autoref{tab:mult-SBFL-Effectiveness}, \autoref{fig:mult-comp-effectiveness} and \autoref{fig:bar-charts-comp-effectiveness}}).

\begin{result}
Statistical debugging performs better on multiple faults: \texttt{Tarantula} is 19\% more effective than dynamic slicing on multiple faults.
\end{result}

\emph{Given single or multiple faults, does the effectiveness of an AFL technique improve or worsen?} 
\autoref{fig:bar-charts-comp-effectiveness} (a) illustrates the difference in the performance of all techniques for single and multiple faults. 
Results show that all techniques (except dynamic slicing) perform better on multiple faults in comparison to single faults, improvements range from two to 11 percentage points. Notably, \texttt{Tarantula}'s performance improved by 11\% on multiple faults. Meanwhile, other approaches improved by two to three percentage points, in particular, the hybrid approach, \texttt{Kulczynski2} and \texttt{DStar} ($D^2$ and $D^3$). This illustrates that most AFL approaches -- especially SBFL -- perform better on multiple faults than single faults (\textit{see \autoref{fig:bar-charts-comp-effectiveness} (a) and \autoref{fig:benchmark-comp-effectiveness}}). 

\begin{result}
Statistical debugging is better suited for diagnosing multiple faults, while dynamic slicing is more effective at localizing single faults.
\end{result}

Generally, we found that the performance of a technique on programs containing single faults does not predict its performance on multiple faults. For instance, although \texttt{Kulczynski2} outperformed the other statistical formulas for single faults ($0.7518$), it is outperformed by \texttt{m9185} for multiple faults ($0.7816$ vs. $0.8249$) (\emph{cf. \autoref{tab:mult-SBFL-Effectiveness}}). %
This result is also evident from \autoref{fig:bar-charts-comp-effectiveness} (a) and \autoref{tab:mult-SBFL-Effectiveness}, where dynamic slicing outperforms statistical debugging (\texttt{Kulczynski2}) for \texttt{SIR} single faults ($0.8269$ vs. $0.7518$), but statistical debugging (\textit{m9185} and \texttt{Kulczynski2}) clearly outperform dynamic slicing for multiple faults. These results suggest that the number of faults in the program influences the effectiveness of an AFL technique.

\begin{result}
The effectiveness of an AFL technique on single faults does not predict its effectiveness on multiple faults.
\end{result}

\section{Threats to Validity}
\label{sec:threats}
We discuss the threats to validity for this fault localization study within the framework of \cite{framework}.

\subsection{External Validity}
\noindent
External validity refers to the extent to which the reported results can be generalized to other objects which are not included in the study. The most immediate threats to external validity are the following:
\begin{itemize}
\item 
\emph{EV.1) Heterogeneity of Probands}. The quality of the test suites provided by the object of analysis may vary greatly which hampers the assessment of accuracy for practical purposes. However, in our study the test suites are well-stocked and maintained. All projects are open source C programs which are subject to common measures of quality control, such as code review and providing a test case with bug fixes and feature additions.

\item 
\emph{EV.2) Faulty Versions and Fault Injection}. For studies involving artificially injected faults, it is important to control the type and number of injected faults. Test cases become subject to accidental fault injection. Some failures may be spurious. However, in our study we also use real errors that were introduced (unintentionally) by real developers. Failing test cases are guaranteed to fail because of the error.

\item 
\emph{EV.3) Language Idiosyncrasies}. Indeed, our objects contain well-main\-tained open-source C projects with real errors typical for such projects. However, for instance  errors in projects written in other languages, like Java, or in commercially developed software may be of different kind and complexity. Hence, we cannot claim generality for all languages and suggest reproducing our experiments for real errors in projects written in other languages as well.

\item 
\emph{EV.4) Test Suites}. The size of a test suite can influence the performance of an AFL technique. Testing strategies that reduce or increase test suite size such as test reduction or test generation methods (i.e. removing tests or generating new tests) have been shown to improve the performance of some AFL techniques~\citep{yang2017better, yu2008empirical}. We mitigate the effect of test suite size by employing projects with varying test suite sizes (ranging from tens to  tens of thousands of test cases) as provided by our subject programs. In our evaluation, we do not generate additional tests or remove any tests from the test suite provided by the benchmarks, in order to simulate the typical debugging scenario for the software project.

\item 
\emph{EV.5) Missing Statements in Slices}. Although, there is a risk of discarding faulty statements during program slicing, dynamic slicing rarely miss faulty statements during fault localization. \cite{reis2019demystifying} found that dynamic slicing reports the faulty statement 
in the top-ten most suspicious statement 91\% of the time. We further mitigate this risk by first inspecting statements in the dynamic slice before inspecting other executable statements. Thus, dynamic slicing (eventually) finds the faulty statement for all bugs in our evaluation. 

\end{itemize}

\subsection{Construct Validity}
\noindent
Construct validity refers to the degree to which a test measures what it claims to be measuring. The most immediate threats to construct validity are the following: 
\begin{itemize}
\item 
\emph{CV.1) Measure of Effectiveness}. Conforming to the standard~\citep{tse}, we measure fault localization effectiveness as ranking-based relative wasted effort. The technique that ranks the faulty statement higher is considered more effective. Parnin and Orso  find that ``programmers will stop inspecting statements, and transition to traditional debugging, if they do not get promising results within the first few statements they inspect''~\citep{parnin}. However, \cite{framework} insist that one may question the \emph{usefulness} of fault locators, but measures of ranking-based relative wasted effort are certainly necessary for evaluating their performance, particularly in the absence of the subjective user as the evaluator.

\item 
\emph{CV.2) Implementation Flaws}. Tools that we used for the evaluation process may be inaccurate. Despite all care taken, our implementation of the 18 studied statistical fault localization techniques, or of \adslicing, or of the empirical evaluation may be flawed or subject to random factors. However, we make all implementations and experimental results available online for public scrutiny.
\end{itemize}

\section{Future Work}
\label{sec:future-work}
Fault localization based on dependencies still has much room for improvement. For instance:

\begin{description}
\item[\textbf{Cognitive load}.]  In our investigation, we did not consider or model the \emph{cognitive load} it takes to understand the role of individual statements in context.  Since following dependencies in a program is much more likely to stay within same or similar contexts than statistical debugging, where the ranked suspicious lines can be strewn arbitrarily over the code, we would expect dependency-based techniques to take a lead here. The seminal study of Parnin and Orso~\citep{parnin} found that ranked lists of statements are hardly helping human programmers---let us find out which techniques work best for humans.

\item[\textbf{Alternate search techniques}.]  Furthermore, there would be other search strategies along dependencies (for instance, starting with the input, and progressing forward through a program; starting at some suspicious or recently changed location; or moving along coarse-grained functions first, and fine-grained lines later) that may be even more efficient both in terms of nodes visited as well as from the assumed cognitive load.  Again, all this calls for more human studies in debugging. 

\item[\textbf{Experimental techniques}.] such as delta debugging~\citep{dd} offer another means to reduce the cognitive load---by systematically narrowing down the conditions under which a failure occurs.  The work of Burger and Zeller~\citep{jinsi} on minimization of calling sequences with delta debugging showed dramatic improvements over dynamic slicing, reducing ``the search space to 13.7 \% of the dynamic slice or 0.22 \% of the source code''.  In a recent human study, delta debugging ``statistically significantly increased programmers efficiency in failure detection, fault localization and fault correction.''~\citep{hammoudi}.

\item[\textbf{Symbolic techniques}.]   Finally, following dependencies is still one of the simplest methods to exploit program semantics.  Applying symbolic execution and constraint solving would narrow down the set of possible faults.  Model-based debugging~\citep{wotawa} was one of the first to apply this idea in practice; the more recent \textsc{bugassist} work of Jose and Majumdar ``quickly and precisely isolates a few lines of code whose change eliminates the error.''~\citep{jose}.
\end{description}

All these techniques would profit from wider evaluation and assessment; however, they  can also be joined and combined; for instance, one could start with suspicious statements as indicated by statistical fault localization, follow dependencies from there, and skip influences deemed impossible by symbolic analysis.  What we need, though, is true defects which we can use to compare the techniques with---and a willingness to actually compare state of the art techniques, as we do in this paper.

\section{Related Work}
\label{sec:related-work}
\subsection{Evaluation of Fault Localization Techniques}
The effectiveness of various fault localization approaches have been studied by several colleagues, see \cite{tse}. Most papers investigated the effects of program, test and bug features on the effectiveness of statistical debugging. \cite{statEval2} examined the effects of the number of passing and failing test cases on the effectiveness of statistical debugging, they established that the suspiciousness scores stabilize starting from an average six~(6) failing and twenty~(20) passing test cases. \cite{pearson2017evaluating} evaluated the difference between evaluating fault localization techniques on real faults versus artificial faults, using two main techniques, namely statistical debugging and mutation-based fault localization. Notably, their evaluation results shows that results on artificial faults
do not predict results on real faults for both techniques, and a hybrid technique is significantly better than both techniques. 
\cite{keller2017critical} and \cite{heiden2019evaluation} evaluated the effectiveness of statistical fault localization on real world large-scale software systems. The authors found that, for realistic large-scale programs, the accuracy of statistical debugging is not suitable for human developers. In fact, the authors emphasize the obvious need to improve statistical debugging with contextual information such as information from the bug report or from version history of the code lines. In contrast to our work, none of these papers evaluated the fault localization effectiveness of program slicing, nor compare the effectiveness of slicing to that of statistical debugging. 

A few approaches have evaluated the effectiveness of dynamic slices in fault localization~\citep{zhang2007study,zhang2005experimental,pmd-slicer}. In particular, \cite{zhang2005experimental} evaluated the effectiveness of three variants of dynamic slicing algorithms, namely data slicing, full (dynamic) slicing, and relevant slicing. 
Recently, \cite{pmd-slicer} proposed the concept of \textit{potential memory-address dependence} (PMD) to improve the accuracy of dynamic slicing. 
Traditional dynamic dependence graphs (DDG) do not account for PMDs since they are not actual data or control dependencies in the program~\citep{pmd-slicer}. In particular, PMD-slicer determines the potential memory dependencies in a program and represents them on the DDG. This allows developers to detect faults that are due to program assignments that modify the wrong code location. 
Like our study, these papers also found that (variants of) program slicing considerably reduces the number of program statements that need to be examined to locate faulty statements. 
However, in contrast to our study, these papers 
have not empirically compared the performance of dynamic slicing to that of statistical debugging.

\subsection{Improvements of Statistical Fault Localization}
Several authors have proposed approaches to improve statistical fault localization. Most approaches are focused on reducing the program spectra (i.e. the code coverage information) fed to statistical debugging, sometimes by using delta debugging~\citep{christi2018reduce}, program slicing~\citep{alves2011fault,lei2012effective,guo2018empirical}, test generation~\citep{liu2017improving}, test prioritization~\citep{zhang2017boosting} or machine learning~\citep{zou2019empirical, b2016learning}. In particular, some techniques apply program slicing to reduce the program spectra fed to statistical debugging formulas~\citep{shu2017fault, alves2011fault,liu2016simulink,alves2011fault,lei2012effective,guo2018empirical}. The popular page rank algorithm has been used to boost statistical debugging effectiveness by estimating the contributions of different tests to re-compute program spectral~\citep{zhang2017boosting}.
Machine learning algorithms (such as learning to rank) have also been used to improve the effectiveness of statistical debugging~\citep{b2016learning,zou2019empirical}. Besides, BARINEL employs a combination of bayesian reasoning and and statistical debugging to improve fault localization effectiveness, especially for programs with multiple faults~\citep{abreu2009spectrum}. BARINEL combines statistical debugging and \textit{model-based diagnosis} (MBD), i.e., logic reasoning over a behavioral model to deduce multiple-fault candidates: The goal is to overcome the high computational complexity of typical MBD. Search-based test generation has also been combined with SBFL, in order to improve the performance of statistical debugging for Simulink models~\citep{liu2017improving}. However, none of these papers localize faults by following control and data dependencies in the program, i.e. they do not directly use program slicing as a fault localization technique. 

\cite{zou2019empirical} found that the combination of fault localization techniques improves over individual techniques, the authors recommend that future fault localization techniques should be evaluated in the combined setting. 
\cite{wotawa2010fault} proposed a combination of model-based debugging and program slicing for fault localization, called \emph{slicing hitting set computation} (SHSC). 
In contrast to our work, SHSC combines slices of faulty variables, which causes undesirable high ranking of statements executed in many test cases~\citep{hofer2012spectrum}. To address this, Hofer and Wotawa also proposed SENDYS -- a combination of statistical debugging and SHSC to improve the ranking of faulty statements~\citep{hofer2012spectrum}. 
The focus of this work is to provide fault locations at a finer granularity than program blocks. In contrast to dynamic slicing, SENDYS analyzes the execution information from both passing and failing test cases and uses statistical debugging results as a-priori fault probabilities of single statements in SHSC~\citep{hofer2012spectrum}. 

\section{Conclusion and Consequences}
\label{sec:conclusion}

As it comes to debugging, \emph{dynamic slicing remains the technique of choice for programmers.}  Suspicious statements, as produced by statistical debugging, can provide good starting points for an investigation; but beyond the top-ranked statements, following dependencies is much more likely to be effective.  As it comes to teaching debugging, as well as for interactive debugging tools, we therefore recommend that following dependencies should remain the primary method of fault localization---it is a safe and robust technique that will get programmers towards the goal.  

For automated repair techniques, the picture is different.  Since current approaches benefit from a small set of suspicious locations, focusing on a small set of top ranked locations, as produced by statistical debugging, remains the strategy of choice.  Still, automated repair tools could benefit from static and dynamic dependencies just as human debuggers.

While easy to deploy, the techniques discussed in this paper should by no means be considered the best of fault localization techniques.  \emph{Experimental techniques} which reduce inputs~\citep{dd,hammoudi}, or executions~\citep{jinsi} may dramatically improve fault localization by focusing on relevant parts of the execution. \emph{Grammar-based techniques} that debug inputs~\citep{ddmax}, generalize inputs~\citep{ddset} or determine the circumstances of failure~\citep{alhazen} may provide contextual information for developers during debugging by focusing on input features that explain failures. \emph{Symbolic techniques} also show a great potential---such as the technique of Jose and Majumdar, which ``quickly and precisely isolates a few lines of code whose change eliminates the error''~\citep{jose}.  The key challenge of automated fault localization will be to bring the best of the available techniques together in ways that are \emph{applicable} to a wide range of programs and \emph{useful} for real programmers, who must fix their bugs by the end of the day.

\bigskip
\noindent
\noindent
\textbf{Additional material.} All of our scripts, tools, benchmarks, and results are freely available as an artifact, in order to support scrutiny, evaluation, reproduction, and extension:
\begin{center}
	\url{https://tinyurl.com/HybridFaultLocalization}
\end{center}

\begin{acknowledgements}
We thank the anonymous reviewers for their helpful comments. This work was funded by Deutsche Forschungsgemeinschaft, Project ``Extracting and Mining of Probabilistic Event Structures from Software Systems (EMPRESS)''.
\end{acknowledgements}

%
%


\bibliographystyle{spbasic}
\bibliography{emse-debugging}

%
%

\end{document}